\documentstyle[emulateapj]{article}

\def\lsim{\mathrel{\rlap{\lower 4pt 
\hbox{\hskip 1pt $\sim$}}\raise 1pt \hbox {$<$}}} 
\def\gsim{\mathrel{\rlap{\lower 4pt 
\hbox{\hskip 1pt $\sim$}}\raise 1pt \hbox {$>$}}}

\newcommand{\msun}{M_{\odot}}

\lefthead{Maeda \& Nomoto}
\righthead{Bipolar Supernovae Explosions}

\begin{document}

\submitted{To appear in the Astrophysical Journal}

\title{Bipolar Supernova Explosions: Nucleosynthesis \& 
Implication on Abundances in Extremely Metal-Poor Stars}

\author{Keiichi Maeda\altaffilmark{1}, \& 
Ken'ichi Nomoto\altaffilmark{1,2} 
}

\altaffiltext{1}{Department of Astronomy, School of Science, 
University of Tokyo, Hongo 7-3-1, Bunkyo-ku, Tokyo 113-0033, Japan: 
maeda@astron.s.u-tokyo.ac.jp}
\altaffiltext{2}{Research Center for the Early Universe, School of Science, 
University of Tokyo, Hongo 7-3-1, Bunkyo-ku, Tokyo 113-0033, Japan} 

\begin{abstract}
Hydrodynamics and explosive nucleosynthesis in bipolar supernova 
explosions are examined to account for some peculiar properties of hypernovae 
as well as peculiar abundance patterns of metal-poor stars. 
The explosion is assumed to be driven by bipolar jets which are powered 
by accretion onto a central remnant. 
The energy injection rate by the jets is assumed to be proportional 
to the accretion rate, i.e., 
$\dot E_{\rm jet} = \alpha \dot M c^2$. 
We explore the features of the explosions with varying progenitors' masses and 
jet properties. 
The outcomes are different from conventional spherical models. 
(1) In the bipolar models, Fe-rich materials are ejected at high velocities 
along the jet axis, while O-rich materials occupy 
the central region whose density becomes very high 
as a consequence of continuous accretion from the side. 
This configuration can explain 
some peculiar features in 
the light curves and the nebular spectra of hypernovae. 
(2) Production of $^{56}$Ni 
tends to be smaller than in spherical thermal bomb models. 
To account for a large amount of $^{56}$Ni observed in hypernovae, 
the jets should be initiated when 
the compact remnant mass is still smaller than $2-3\msun$, 
or the jets should be very massive and slow. 
(3) Ejected isotopes are distributed as follows in order of decreasing velocities: 
$^{64}$Zn, $^{59}$Co, $^{56}$Fe, $^{44}$Ti, and $^{4}$He 
at the highest velocities, 
$^{55}$Mn, $^{52}$Cr, $^{32}$S, and $^{28}$Si 
at the intermediate velocities, 
and $^{24}$Mg, $^{16}$O 
at the lowest velocities. 
(4) The abundance ratios (Zn, Co)/Fe are enhanced while the ratios 
(Mn, Cr)/Fe are suppressed.  
This can account for 
the abundance pattern of extremely metal-poor stars. 
These agreements between the models and observations suggest that 
hypernovae are driven by bipolar jets 
and have significantly contributed to the early Galactic chemical evolution. 
\end{abstract}




\section{Introduction}

The basic explosion mechanisms of core-collapse (Type II, Ib, Ic) supernovae 
(SNe II, Ib, Ic) have not been clarified (e.g., Janka et al. 2003), 
and the essential importance 
of asphericity has been suggested from observations, including 
a direct HST image of SN1987A (Wang et al. 2002), 
and large optical linear polarization in SNe Ib/Ic (Wang et al. 2001).  

The discovery of a new class of supernovae, 'hypernovae', which have 
been suggested to have distinctly 
large kinetic energies (hereafter $E$) 
as compared with previously known supernovae 
(Galama et al. 1998; Iwamoto et al. 1998), brought us a new light on this issue. 
As will be discussed below, 
some peculiar properties of hypernovae have indicated a large deviation 
from spherical symmetry. 
Here we present bipolar explosion models for hypernovae 
to account for those properties. 
We further apply our nucleosynthesis results to 
account for some peculiar abundance patterns of metal-poor stars, 
which have not been predicted by spherical explosion models. 

\subsection{Properties of Hypernovae}

In 1998, the unusually bright SN Ic 1998bw with very broad 
spectral features was discovered 
in the small error box of the $\gamma$-ray burst GRB980425 (Galama et al. 1998). 
The large expansion velocity combined with relatively slow evolution 
in the optical spectra and the light curve 
(when compared with those of SN 1994I for example) led to the 
conclusion that the supernova was fairly energetic with $E_{51} \sim 30$ 
(where $E_{51} = E/10^{51}$ergs) 
and had very massive C+O ejecta $M_{\rm ej} \sim 10\msun$ 
(Iwamoto et al. 1998; Woosley, Eastman, \& Schmidt, 1999). 
It was most likely an end product of a star 
with the zero-age main-sequence mass 
$M_{\rm ZAMS} \sim 40\msun$. 
Iwamoto et al. (1998) used the term 'hypernova' to describe 
such a hyper-energetic supernova with $E_{51} \gsim 10$. 
We follow this terminology hereafter in this paper. 

Beginning with SN1998bw, there is a growing list of hypernovae. 
It now includes SNe Ic 1997ef (Iwamoto et al. 2000; Mazzali, Iwamoto, \& Nomoto 2000), 
1999as (Knop et al. 1999; Hatano et al. 2001), 
2002ap (Mazzali et al. 2002; Kinugasa et al. 2002), 
Type IIn supernovae (SNe IIn) 1997cy (Germany et al. 2000; Turatto et al. 2000) 
and 1999E (Rigon et al. 2003). 
Still more SNe have been proposed as hypernova candidates 
(e.g., Nomoto et al. 2001). 
For reviews of hypernovae, see Nomoto et al. (2001, 2003).

With the growing number of samples, 
interesting features have been uncovered (e.g., Nomoto et al. 2003). 
Figure~\ref{f1} summarizes $E$ and $M$($^{56}$Ni) 
(the ejected mass of $^{56}$Ni) against $M_{\rm ZAMS}$ 
for several core-collapse supernovae/hypernovae 
derived by the previous studies. 
\begin{itemize}
\item[(1)]
For progenitors with $M_{\rm ZAMS} \lsim 20\msun$, $E$ and $M$($^{56}$Ni) 
are fairly homogeneous with the representative values of $E_{51} \sim 1$ 
and $M(^{56}{\rm Ni}) \sim 0.07\msun$. 
\item[(2)]
There seems to be a transition in those properties around 
$M_{\rm ZAMS} = 20 - 25\msun$. 
In the regime where $M_{\rm ZAMS} > 20 -25\msun$, 
there exist bright energetic hypernovae 
at one extreme and faint low-energy Type II supernovae 
SN II 1997D (Turatto et al. 1998) and SN II 1999br (Hamuy 2003; Zampieri et al. 2003) 
on the other. There may be a variety of SNe 
which express intermediate cases between those two branches (Hamuy 2003). 

\item[(3)]
In the 'hypernova branch', $E$ and $M$ ($^{56}$Ni) are larger for 
larger $M_{\rm ZAMS}$.  
\end{itemize}
The apparent transition leads to the hypothesis that the explosion mechanism 
of stars with zero-age main-sequence masses $M_{\rm ZAMS} > 20 - 25\msun$ 
might be different from that of less massive stars ($M_{\rm ZAMS} \lsim 20\msun$), 
especially for hypernovae given their large energies. (For faint SNe, 
the small $E$ may simply be attributed to a large gravitational binding energy. 
Unless the explosion mechanism provides a much larger energy than a typical SNe, 
a more massive star will experience more siginificant fallback. 
See Woosley \& Weaver (1995).)

\subsection{Indications of Asphericity in Hypernovae}

Subsequent studies have indeed uncovered the discrepancies between 
observations of hypernovae and a conventional spherical model 
(a spherical thermal bomb model: Aufderheide, Baron, \& Thielemann, 1991). 
The model had well described observations of 
normal core collapse supernovae, e.g., SN1994I 
(e.g., Iwamoto et al. 1994). 
The discrepancies raised by these studies are summarized as follows: 
\begin{itemize}
\item[(1)]
Late time spectra of SN1998bw showed that the FeII] blended lines were 
broader than the OI] 6300 lines, indicating Fe was ejected with larger velocities 
than O along the line of sight (Mazzali et al. 2001). 
Maeda et al. (2002) suggested that the explosion was highly aspherical 
and bipolar by demonstrating that such an explosion 
naturally explains this property. 
\item[(2)]
Modeling the light curves of SNe~1998bw (Nakamura et al. 2001a), 
1997ef (Iwamoto et al. 2000; Mazzali et al. 2000), and 
2002ap (Yoshii et al. 2003) showed the deviation of 
the observed curves from the synthetic curves based on the 
spherical explosion models after $\sim 50$ days. 
All the model curves are much fainter than the observed curves 
at these epochs. Maeda et al. (2003) have shown that this failure 
is resulted from too low densities at low velocities 
in the spherical hydrodynamic models. 
By introducing a dense inner component, 
which is not expected in spherical hydrodynamic models, 
they have reproduced 
the light curves of SNe~1998bw, 1997ef, and 2002ap 
at the whole epochs observed. 
\item[(3)]
The spectroscopic modeling of SN1997ef also suggested the 
existence of a dense core. 
It also requires that the density distribution at an outer envelope is flatter than 
obtained by spherical thermal bomb models (Mazzali et al. 2000). 
\item[(4)]
On spectropolarimetries, SNe 1998bw and 2002ap had linear polarization of 
$\sim 1$\% (Iwamoto et al. 1998) and 
$\sim 2$\% (Kawabata et al. 2002; Leonard et al. 2002; 
Wang et al. 2003), respectively. 
Because local polarization should be totally cancelled out 
in strictly spherically symmetrical geometry, 
existence of net polarization is interpreted as evidence of asphericity in ejecta.  
Kawabata et al. (2002) suggested that the combination 
of the mildly distorted ejecta and the jet material at 
$v \gsim 0.115 c$ could explain their spectropolarimetric data of SN 2002ap. 
\end{itemize}
These features of hypernovae, combined with their large $E$, have suggested 
that hypernovae are bipolar explosions.

\subsection{Bipolar Explosion Models}

A bipolar explosion could arise in a core-collapse supernova through 
several mechanisms, 
if rotation in a pre-explosion core is sufficiently large. 
In the neutrino driven delayed explosion scenario, the rotation would make 
a newly born neutron star a 
pancake configuration. This prefers strong neutrino heating 
along the rotational axis (Yamada \& Sato, 1994). 
In the extreme case, the core would become a black hole surrounded 
by a hyper accreting torus. 
It has been suggested that such a system can generate an outflow 
with the energy comparable to those in hypernovae through neutrino annihilation 
and/or viscous dissipation (Woosley 1993; MacFadyen \& Woosley 1999).  
Another possible mechanism is a MHD driven explosion, where 
the magnetic field amplified by rotation 
up to $\sim 10^{15} - 10^{16}$G would make a strong outflow along 
the axis of magnetic field lines. 
The central object can be either a neutron star (LeBlanc \& Wilson 1970; 
Wheeler et al. 2000) or a black hole (Brown et al. 2000). 
The latter provides further two options for the energy source: 
the binding energy of the accretion disk and 
the rotational energy of the black hole through Blanford \& Znajek 
mechanism (Blandford \& Znajek 1977). 

Such an explosion should show different characteristics 
in hydrodynamics and nucleosynthesis. 
The nature of the central energy-generating system is quite complicated 
and involves physics which is not well understood yet. 
The outcome, however, is relatively simple. It is the interaction 
between the jets and the stellar envelope. 
Therefore, once the property of the jets is specified, 
subsequent hydrodynamical evolution and nucleosynthesis can be examined in detail. 
In fact, the inverse approach is possible. First assume the jet property, 
then examine the outcome of the explosion. 
The results can be compared with observation. 
The comparison finally puts constraints on the properties of the jets, 
and thus the energy generation mechanism. 

This strategy has been taken by several authors in studying 
the outcome and observational consequences of 
jet-driven supernova explosions. 
Khokhlov et al. (1999) examined hydrodynamical interaction 
between the jets and the surrounding stellar mantle for one 
representative case with $M_{\rm ZAMS} = 15\msun$ and a fixed jet property. 
MacFadyen, Woosley, \& Heger (2001) investigated the effects of jets from 
a central compact remnant in the context of a failed supernova explosion of 
a $25 \msun$ progenitor star. 
They studied the interaction by varying the jet parameters.  
However, neither of the above hydrodynamical studies computed nuclosynthesis. 

On the other hand, 
Nagataki (2000) and Maeda et al. (2002) examined 
nucleosynthesis in aspherical supernova/hypernova explosions, using 
aspherical energy inputs at the center of stars with 
$M_{\rm ZAMS} = 20$ and $40\msun$, respectively. 
In their studies, however, 
the energy injection by the jets was simplified as compared to that of 
Khokhlov et al. (1999) and MacFadyen et al. (2001), 
as their models represented aspherical prompt explosions 
which apply only to the case 
where the time scale of the energy generation is much shorter 
than the hydrodynamical time scale. 

There are two possible sites for nucleosynthesis associated with jet-driven 
supernova explosions: the stellar materials heated by the jets and 
the materials in the jets themselves. 
As stellar materials falls onto a central remnant, they cool 
via photodisintegration and neutrino emissions to form an accretion disk. 
A fraction of the accreted materials is likely ejected from the accretion 
disk, rather than is accreted onto the central remnant 
(Narayan, Piran, and Kumar 2001). 
This accretion disk wind likely escapes from the central region along the rotational 
axis and shows collimation, i.e., a jetted wind. 

How much fraction of the accreting gas is ejected depends on the accretion rate 
and the typical radius where the accretion disk forms (Narayan et al. 2001). 
They are physically related to the angular momentum distribution and 
the viscosity in the progenitor star (MacFadyen 2003). 
Unfortunately, both are rather uncertain. 
In the present study, we assume that only a small fraction of the accreting 
gas is ejected as the jets from the central region 
($10$ or $50 \%$: denoted as $\mu$ in section 2), so that nucleosynthesis 
in the heated stellar materials is more prominent than in the jet (wind) 
materials themselves. 

The purpose of this paper is first to investigate in more detail 
the outcome (hydrodynamics and nucleosynthesis) of supernova explosions 
driven by bipolar jets 
for such massive stars as $M_{\rm ZAMS} \gsim 25\msun$.  
We model the jet properties in terms of the accretion rate, 
which itself is affected by the jet properties through the 
hydrodynamical interaction. 
This approach makes it possible to calculate the self-regulated interaction, 
and investigate 
how the jet properties affect the outcome by modeling the interaction 
with varying parameters. 
With the detailed nucleosynthesis yields, 
we discuss their possible influence on 
the early Galactic chemical evolution. 

We note the two main assumptions in the present study: 
the parameterized constant jet properties and the small mass of the jet 
materials. 
In reality, the jets could be highly variable. 
The variability will affect the detailed 
structure in the jet materials (e.g., Aloy et al. 2000), 
though we believe that the constant jet properties express 
the typical behavior and are adequate for the purpose of this paper. 
The assumption that only a small fraction of the accreting gas is ejected 
as the jets (i.e., the small mass of the jets) puts more important limitation. 
Changing the mass of the jets will change the evolution of the mass of the central 
remnant, may change the hydrodynamic interaction between the jets and the 
stellar materials. Moreover, in the case of very massive jets, 
if they are realized, nucleosynthesis products in the jets will 
become prominent than those 
in the heated stellar materials (Pruet, Woosley, \& Hoffman 2003: 
MacFadyen 2003). 
This is an interesting possibility, 
while we postpone to study such a very massive jets to future works.

In Section 2, we describe our models of bipolar supernova explosions in detail. 
Results are shown in Section 3, which is divided into three subsections. 
Section 3.1 gives the results of hydrodynamics, discussing how the outcome 
(e.g., the degree of asphericity and the final mass of the 
central compact remnant) depends on the jet properties. 
Section 3.2 focuses on the production of $^{56}$Ni. 
Section 3.3 shows the results of detailed nucleosynthesis, 
where we put emphasis on the difference between our models and previous yields. 
In Section 4, we examine their contribution to the early Galactic chemical evolution 
by comparing our yields with abundances in extremely metal poor stars. 
Section 5 closes this paper with conclusions and discussion.

\section{Methods \& Models}
The main ingredient of our models is a pair of jets penetrating into a stellar mantle. 
At the beginning of each calculation, the central part ($M_r \leq M_{\rm REM 0}$) 
of a progenitor star is displaced by a point mass 
with transmitted boundary condition 
at the interface. 
Without detailed knowledge of how the central object generates energy, 
we take $M_{\rm REM 0}$ as a parameter which expresses the mass of the central remnant 
when it begins to produce the jets (strictly speaking, when the jets emerge 
out of the central region). 

The jets are injected at the inner boundary 
along the $z$-axis with the opening half-angle $\theta_{\rm jet}$. 
For the property of the jets, we adopt the formalism similar to 
MacFadyen et al (2001). 
We must specify two thermodynamical variables and one hydrodynamical 
variable to determine 
the property of the jets at their emergence. For these, 
we take the density, the momentum, and the ratio $f$ 
of the internal energy to the total energy in the jets. 
We set $f = 0.01$, 
so that the internal energy in the jets is negligible. 
Thus the jets are 'momentum-driven'. 
To connect the jet properties consistently to the flow around the central region, 
we assume that the energy and mass fluxes of the jets are 
proportional to the mass accretion rate. 
In this formalism, the jet properties are expressed as follows: 
\begin{eqnarray}
	\dot M_{\rm jet} & = & \mu \dot M_{\rm accretion} 
= \rho_{\rm jet} v_{\rm jet} A_{\rm jet} \ , \nonumber\\
	\dot E_{\rm jet} & = & \alpha \dot M_{\rm accretion} 
 =  \left( e_{\rm thermal} + \frac{1}{2} \rho_{\rm jet} v_{\rm jet}^2 \right) 
v_{\rm jet} A_{\rm jet} \ , \\ 
	f & = & \frac{\dot E_{\rm thermal}}{\dot E_{\rm jet}} 
= 0.01 \ . \nonumber 
\end{eqnarray}
Here $A_{\rm jet}$ is the area where the jets emerge, i.e., 
$A_{\rm jet} = 4 \pi R_0^2 (1-\cos(\theta_{\rm jet}))$ with 
$R_0$ the radius of the inner boundary (typically $\sim 10^8$cm). 
Equation (1) leads to the following explicit expression for the jet properties: 

\begin{eqnarray}
	\rho_{\rm jet} & = & 
\frac{\mu \dot M_{\rm accretion}}{v_{\rm jet} A_{\rm jet}} \nonumber \ , \\
	e_{\rm jet} & = & 
e_{\rm thermal} + \frac{1}{2} \rho_{\rm jet} v_{\rm jet}^2 
 =  \frac{\alpha \dot M_{\rm accretion} c^2}{v_{\rm jet} A_{\rm jet}} \ , \\
	v_{\rm jet} & = & 
\left( \frac{2 \alpha (1-f)}{\mu} \right)^{\frac{1}{2}}c 
\sim \left( \frac{2 \alpha}{\mu} \right)^{\frac{1}{2}} c \ . \nonumber
\end{eqnarray}

Generally, $\mu$ and $\alpha$ are the functions of $\dot M_{\rm accretion}$. 
For simplicity, we take them as constants if 
$\dot M_{\rm accretion} \ge \dot M_{\rm critical}$. 
When the accretion rate falls down below $\dot M_{\rm critical}$, 
the injection of jets is turned off. 
$\dot M_{\rm critical}$ expresses the critical accretion rate at which 
the central accretion 
flow changes its characteristics dramatically. 
Such transition would occur in a hyper-accreting disk by the strong dependence 
of neutrino cooling on the mass accretion rate (Popham, Woosley, \& Fryer, 1999). 
We set $\dot M_{\rm critical} = 0.2 \msun$s$^{-1}$. 
For the relation between $\alpha$ and $\mu$, 
we set $\mu = 10 \alpha$, so that 
$v_{\rm jet} \sim 1.3 \times 10^{10}$cm s$^{-1}$ for all the models.

Our models are summarized in the first five columns of Table 1. 
For the progenitor models, we take the $16\msun$ and $8\msun$ 
He cores of stars with $M_{\rm ZAMS} =40\msun$ and $25\msun$, respectively, 
from the solar metalicity models of 
Nomoto \& Hashimoto (1988). 
The H-rich envelope in the progenitor star is removed and the remaining 
He star is mapped onto the numerical grids with the outer radii $R_{*} = 3.7$ 
and $6.7 \times 10^{10}$cm for $40\msun$ and $25\msun$, respectively.

With the parameterization described above, 
we follow hydrodynamics of the explosion. 
At each time step, we compute $\dot M_{\rm accretion}$ at the inner boundary, 
and then update the jet properties and the mass 
of the central remnant ($M_{\rm REM}$) at the next step.  
To determine the final central remnant mass and the composition in the ejected 
materials, it is necessary to follow the hydrodynamics until 
$\dot M_{\rm accretion}$ becomes very small and the ejecta 
reaches homologous expansion. This is done by expanding the numerical grids and 
adding the exponential atmosphere above the surface of the He star when the leading 
edge of the jets reaches the outer boundary. The hydrodynamical evolution is 
followed up to $40 \sim 100$ seconds in most runs, so that the expansion becomes 
nearly homologous.

Roe's scheme based on Riemann solver is applied to a 
two-dimensional Eulerian hydrodynamical code (Hachisu et al. 1992, 1994). 
Typically $100 \times 30$ zones are used in spherical polar 
coordinate descritisized logarithmically in the radial direction. 
This relatively coarse zoning is due to the small time steps 
imposed at the inner boundary as a consequence of large $v_{\rm jet}$. 
A typical time step is 
$\Delta t = R_0 \Delta \theta / v_{\rm jet} 
\sim (0.5 \pi/30 \times 10^8$cm$)/(1.3 \times 10^{10}$cm s$^{-1}) 
\sim 4 \times 10^{-4}$ seconds. We have set the CFL number to be 0.1 so that 
the calculations are stable, which makes 
$\Delta t \sim 4 \times 10^{-5}$ seconds. As mentioned above, 
we have to follow the evolution up to $\sim 100$ seconds, so that 
more than $10^6$ time steps are needed to run one model. 
Also, we need to run many models (about 10 models: see Table 1) 
to investigate the outcome with 
various jet properties. 
In addition, the number of points to calculate nucleosynthesis 
should be proportional to the number of the meshes used 
in the hydrodynamic calculation, 
which puts another need for computational time for a higher resolution run. 

Despite the practical difficulties mentioned above, the coarse zoning is 
physically little justifiable. We have performed a convergence test 
by running model 40A with lower and higher resolutions in Appendix A. 
We have found that the results do not change significantly, 
if the number of the meshes is greater than $100 \times 30$ 
(see Appendix A for the details).

We employ equation of state including radiation 
and e$^{+}$e$^{-}$ pairs in an approximated 
analytical form (Freiburghaus et al. 1999). 
The integrated form of Poisson equation (Hachisu 1986) handles with both 
self gravity of a stellar mantle and gravity of a central point mass 
(in Newtonian limit). 
In the hydrodynamical calculations, the energies generated by nuclear reactions 
are omitted. This is justified by the following reason: 
We have used He star models, so that the original composition of the 
materials which has burned into $^{56}$Ni is heavier than $^{4}$He. 
The energy generated by the reactions is therefore at most $3 \times 10^{51}$ergs 
$/$($M$($^{56}$Ni)$/\msun$), and even smaller in a typical case because most 
of the materials burned into $^{56}$Ni are heavier than $^{16}$O. Because 
$M$($^{56}$Ni) is of the order of $0.1\msun$, This energy is about $10 - 20$ \% for 
the models with $E_{51} \sim 1$ and much smaller for the models with $E_{51} 
\gsim 10$. 

With the hydrodynamical calculation, we trace thermodynamical 
histories of individual Lagrangian elements. 
3000 particles are used for stellar mantle. 
At the base of the jet, new particles flow into the numerical domain, 
and are also traced. 
These histories are used to calculate nucleosynthesis as post-processing. 
The reaction network includes 222 isotopes up to $^{71}$Ge 
(Hix, \& Thielemann 1996, 1999).

The initial compositions of the particles in the post-processing are taken as follows: 
\begin{itemize}
\item[(1)]
For those which initially reside above the Si layer in the progenitor star, 
we use the original composition of Nomoto \& Hashimoto (1998). 
\item[(2)]

For those in the iron core, the initial composition, especially $Y_e$ 
(i.e., electron molecule number), is uncertain. 
Guided by the recent investigation on how $Y_e$ affects the resultant yields 
(Umeda \& Nomoto 2002), 
we take $Y_e = 0.5$ in this region. 

\item[(3)]
The initial composition of the jet material should wait for detailed 
study on the flow around the central remnant. 
Recently Pruet et al. (2003) reported based on the 
one-dimensional accretion disk model 
that $Y_e$ in the bulk of the disk wind is close to 0.5. 
In this study we take $Y_{e} = 0.5$. 
\end{itemize}
Because of the large uncertainty in the initial composition of the jet material, 
we restrict ourselves in this paper to parameter space where the 
mass of the isotopes newly synthesized in the shocked stellar mantles 
overwhelms the mass contained in the jets.  
We postpone the whole survey, such as very massive jets, to future works. 
This should require to follow the accretion process to determined 
the initial composition and will involve the physics neglected 
in the present study, e.g., angular momentum transport, nuclear energy generation, 
and cooling by photodisintegration and neutrino emissions, as well as 
highly resolved numerical simulation around the central remnant. 

\section{Results}
We perform hydrodynamic calculations for the models in Table 1. 
The models are named with the numbers describing 
progenitors' zero-age main-sequence masses ($M_{\rm ZAMS}$). 
For $M_{\rm ZAMS} = 40\msun$, we examine 
the parameter dependences in detail. 
\begin{itemize}
\item[(1)]
Models 40A and 25A are regarded as standard runs in this paper. 
\item[(2)]
Models 40B and 25B express models with loosely collimated jets. 
\item[(3)]
Model 40C corresponds to very efficient jets with $5 \%$ efficiency 
in converting the accreted energy into the jets' energy. 
\item[(4)]
Model 40D is calculated in order to examine dependence 
on the initial central remnant's mass ($M_{\rm REM 0}$), 
where the jets are initiated with $M_{\rm REM 0} = 3\msun$, 
twice the mass in the standard run. 
\item[(5)]
Models 40SH, 40SL, and 25S express spherical prompt explosion models, 
which are calculated to compare the bipolar models with 
the conventional spherical models 
using the same numerical code. 
In their names, S means Spherical and 
H and L mean High and Low energies. 
\end{itemize}

\subsection{Hydrodynamics}
The hydrodynamical evolution of the interaction between the bipolar jets 
and the stellar mantle of model 40A is shown in Figure~\ref{f2}, 
and on an expanded scale in Figure~\ref{f3}. 
As the jets run through the stellar mantle, they make configuration which 
is typical of interaction between jets and surrounding matter 
(e.g., Khokhlov et al. 1999). 
At the leading edge of the jets is a working surface 
which divides the jet matter and the ambient stellar matter. 
At the working surface, energy is deposited to the stellar material 
which is pushed sideways to form bow shocks. 
The shocked material forms the high-pressure cocoon surrounding the jets. 
It expands laterally to push the stellar material further 
at the bow shock. 
The cocoon also pushes the jet material toward the $z$-axis 
to help the jet even more collimated (MacFadyen et al. 2001), 
with itself flowing continuously into behind the jets. 
After the jets lose the power supply at their bases, the internal structure 
in the jets and the shocked material tends to be smoothed out, making finally 
the bipolar explosion.

Figure~\ref{f3} exhibits the complicated structure of the interaction 
between the jets and the surrounding matter in the central region. 
The outflow generated by the jets and the inflow caused by gravity compete 
with each other during the jet propagation. 
Strong outflow occurs along the z-axis, while 
matter continues to accrete onto the central region from the side. 
The working surface works as the 'jet channel', 
through which the energy is transported efficiently into the outer layers. 
In the central region, the energy is deposited less effectively. 
Thus after the jets are turned off, 
the central part of the star still continues to accrete. 

This leads to important results on the final density structure. 
In Figure~\ref{f4}, we plot the final density distribution of 
model 40A along the z-axis and r-axis, as well as 
that of models 40SH and 40SL. 
The same plot for the $25\msun$ star is also given. 
We note two points which are completely deferent from spherical (thermal bomb) models. 
\begin{itemize}
\item[(1)]
In the outer envelope, we find higher velocities in the $z$-axis 
than in the $r$-axis as expected. 
What is unexpected is that 
the density distribution is flatter than in spherical prompt models. 
The matter along the $z$-axis is dominated by the jet material. 
Thus the density structure is determined by the jet properties rather 
than the pre-explosion structure. 
The latter affects the final density structure only through 
the hydrodynamical interaction. 
Its effect, however, is of less significance 
than the jet property itself. 

The density distribution at high velocities ($v \gsim 10^{9}$ cm s$^{-1}$) 
is responsible for spectral evolution in early, photospheric phases. 
Analyzing supernova photospheric spectra 
indicates the structure in this region.  
For SNe 1997ef and 2002ap, Mazzali et al. (2000, 2002) derived
$\rho \propto v^{-4}$ at $v \gsim 3 \times 10^{9}$ cm s$^{-1}$. 
They stressed that this was not reproduced by spherical models, 
and might indicate asphericity in the hypernova explosions. 
The densities in this region is roughly expressed as $\rho \propto v^{-8}$ 
for the spherical model 40SH, which is consistent with the previous studies 
(e.g., Nakamura et al. 2001b). 
Model 40A results in the density structure in the same region fitted by 
$\rho \propto v^{-3.6}$. This indeed well agrees with the structure derived by 
the spectral modeling. 

The caveat is that the detailed 
density structure at the highest velocities along the $z$-axis 
depends on how fine structures of the jets are resolved, i.e., on the 
numerical resolution (see appendix). In appendix, we show it is the case but 
the overall slope does not depend on the resolution sensitively. 
Especially, model 40A results in much flatter density distribution than 
spherical models 40SH and 40SL irrespective of the assumed numerical resolution.  

\item[(2)]
In the central region, a high density core is formed. It should be noted 
that such a high density region is found not only along the $r$-axis, 
but also along the $z$-axis. It is really a 'central' core. 
This is because the jet deposits its energy mostly in the 
outer layers (i.e., through the working surface), 
while in the central region the energy deposition is inefficient. 
Thus the central region continues to fall back onto the center. 
This produces the central dense core, 
which agrees with the suggestion 
that a dense core is the feature of a hypernova explosion 
according to the spectral analyses of SNe 1998bw and 1997ef 
(Mazzali et al. 2000, 2001: Maeda et al. 2002).  

As Nakamura et al. (2001a) showed that their low energetic models 
($E_{51}=1$ and $7$) 
have ejecta dense enough to compare to the density required to explain 
the brightness of SN~1998bw at late epochs, and as model 40A has 
denser core than the spherical models 40SH ($E_{51} = 10$) 
and 40SL ($E_{51} = 1$), we expect that the dense core in model 40A 
is capable to reproduce the 
brightness of SN~1998bw at the late phases. 
The densities below $\sim 10^{8}$ cm s$^{-1}$ is enhanced 
in model 40A roughly by a factor of 100 and 10 
compared to the spherical models 40SH and 40SL, respectively. 
This is close to the value 
seen in the central region 
in the light curve modeling for 
SNe 1998bw, 1997ef, and 2002ap by Maeda et al. (2003), in which we 
re-examined the light curves of these hypernovae and got good 
fits to them at the whole epochs both at the peak and late phases. 
\end{itemize}

As noted above, the density structure is determined by 
the parameterized jet properties which are put by hand. 
Figure~\ref{f5} shows the density structure resulted 
from different jet properties. 
The jet models in general have different density structures from 
spherical prompt models, and moreover, have flatter density structures 
at the highest velocities than spherical models. Thus we suggest 
that the flatter density structure is a typical and distinguished 
feature of the 
jet-driven explosions as compared to spherical prompt models. 
Of course, the detailed structure, e.g., the slope 
($\rho \propto v^{-3.6}$ for a particular model 40A), 
does depend on the assumed jet properties 
(Fig.~\ref{f5}) and on time variability of the central source 
which produces the jets. 
The deference will be useful to distinguish the models in future 
by comparing the observed and multi dimensional synthetic spectra 
(e.g., Thomas et al. 2002).

The competition between the outflow and the inflow determines the 
hydrodynamical and nucleosynthetic outcome in the present models. 
We note that the final remnant's mass 
(i.e., $M_{\rm REM}$) is determined self-consistently 
by the hydrodynamical evolution, in contrast to the conventional spherical models 
in which it is treated as a parameter. 

How the hydrodynamical results depend on the parameters is summarized in Table 1, 
where we list some important quantities in characterizing dynamics, 
i.e., $E_{\rm total}$, $M_{\rm REM}$, and $M$($^{56}$Ni) 
(the last we discuss in the next subsection). 
Note that $E_{\rm total}$ is the total energy 
(almost equal to the kinetic energy) at the end of the calculation 
when the ejecta reaches nearly the homologous expansion, 
thus being basically the energy deposited by the jets reduced 
by the gravitational binding energy of the ejecta. 
In Figure~\ref{f6}, we show the evolution of the mass of the 
central compact remnant, $M_{\rm REM}$. 
The results are summarized as follows.
\begin{enumerate}
\item[(1)]
For a given set of the jet parameters, the explosion of the 
$40\msun$ star is more energetic than the $25\msun$ star. 
This is because the gravity is stronger in a more massive star, 
making an accretion rate higher 
and thus the explosion more energetic. 
At the same time, a higher accretion rate leads to a more massive central remnant. 
\item[(2)]
The jet with a wider opening angle $\theta_{\rm jet}$ induces a 
less energetic explosion. 
The larger $\theta_{\rm jet}$ is, more efficiently the energy is transported from 
the jets to the surrounding stellar envelopes. 
The bow shock running through the envelope tries to push material outward. 
$\dot M_{\rm accretion}$, therefore, falls off rapidly in models 40B and 25B 
(larger $\theta_{\rm jet}$) than 
in models 40A and 25A (smaller $\theta_{\rm jet}$). 
Small $\dot M_{\rm accretion}$ leads to a less 
energetic explosion. 
\item[(3)]
The final remnant's mass depends on $\theta_{\rm jet}$. 
It is related to the progenitors' mass as follows: 
\begin{enumerate}
\item[(3-1)]
For the $40\msun$ star, model 40B (large $\theta_{\rm jet}$) 
forms a more massive remnant than 40A. 
Figure~\ref{f6} shows that 
the decrease in the accretion rate after turning off the jets 
is slower in 40B than 40A. When the jets are turned off ($\sim 1$ sec in 40B), 
model 40B still has a mantle of more than $14\msun$. 
The energy injected by the jets ($E_{\rm jet}$) till that time is 
$\sim 6 \times 10^{51}$ergs, 
while the binding energy of the whole star is $\sim 7 \times 10^{51}$ergs.  
This binding energy is comparable to and indeed larger than $E_{\rm jet}$. 
However, it does not mean that the explosion must be failed. 
Because the energy is mainly deposited in the outer layers along the z axis, matter 
continues to accrete in the central region. It reduces the final mass of 
the ejecta and thus the energy required to set the outword motion 
against gravitation. 
It can (and does in 40B) induce the explosion. 
In model 40A, at the time the jets are turned off, $E_{\rm jet}$ greatly exceeds 
the binding energy. 
\item[(3-2)]
For the $25\msun$ star, model 25B forms a smaller mass remnant than model 25A. 
At turning off the jets, even smaller $E_{\rm jet}$ ($\sim 3 \times10^{51}$ ergs) 
in model 25B is enough 
to eject the remaining stellar mantle without further accretion, 
because of the much smaller binding energy ($\sim 2 \times 10^{51}$ ergs) 
of the $25\msun$ star than the $40\msun$ star.

\end{enumerate}
\item[(4)]
For larger efficiency of the central engine, a more energetic explosion takes place. 
Also the central remnant is less massive, because 
$E_{\rm jet}$ is much larger than the binding energy and the lateral expansion 
keeps the accretion rate small. 
\end{enumerate}

\subsection{Production of $^{56}$Ni}

Regarding nucleosynthesis, we first direct our attention to the 
production of $^{56}$Ni. 
The ejected mass of $^{56}$Ni, $M$($^{56}$Ni), is listed in Table 1. 
The models presented here differ from spherical thermal bomb models mainly 
in two respects. 
\begin{enumerate}
\item[(1)]
The time scale for the energy release from the central remnants 
can be comparable to the hydrodynamical time scale for the jet propagation. 
\item[(2)]
The explosion proceeds mainly along the z-axis. 
\end{enumerate}

We first show how the time scale of the explosion affects the final 
$^{56}$Ni mass produced at the shocked stellar material. 
$M$($^{56}$Ni) can be estimated as follows, which generalizes 
the equation (2) of Nakamura et al. (2001b) for 
a prompt explosion model. 
In the following derivation, the energy input rate ($\dot E_{\rm jet}$) 
is assumed to be constant for simplicity. 

Let $t_{\rm f}$ be the time when the jets lose the power supply at their bases, 
i.e., $\dot E_{\rm jet} \times t_{\rm f} = E_{\rm jet}$. 
The region having experienced the shock wave is dominated by radiation, and contains 
the energy which is injected till the time $t$. 
Assuming spherical symmetry, the following approximate relation holds. 

\begin{equation}
  \frac{4 \pi}{3} R^3 a T^4 = \left\{ 
     \begin{array}{ll}
        \dot E_{\rm jet} t & \quad \mbox{for $t<t_{\rm f}$} \ , \\
        \dot E_{\rm jet} t_{\rm f} 
= E_{\rm jet} & \quad \mbox{for $t \ge t_{\rm f}$} \ . 
     \end{array}\right.
\end{equation}

The position of the shock may be approximated as $R = V t$.  Using the approximations 
of $ V \sim \sqrt{\dot E_{\rm jet} \min(t,t_{\rm f}) / M_{\rm sw}}$ for 
the shock velocity and $M_{\rm sw} = (4 \pi /3) R^3 \bar \rho$ 
for the mass of the swept up material, 
we get the following equations for the post-shock temperature when 
the shock arrives at $R$ (here we take $\bar \rho = 10^6$g cm$^{-3}$). 

\begin{equation}
  T_9 \sim \left\{ 
     \begin{array}{ll}
        6 \dot E_{51}^{1/6} R_8^{-1/3} & \quad \mbox{for $t<t_{\rm f}$} \ , \\
        20 E_{52}^{1/4} R_8^{-3/4} & \quad \mbox{for $t \ge t_{\rm f}$} \ . 
     \end{array}\right.
\end{equation}
Here $X_n = X/10^{n}$ in cgs unit. 
$T_9$ is plotted as a function of the shock radius $R$ in Figure~\ref{f7}. 
For given $\dot E_{\rm jet}$ and $E_{\rm jet}$ (or, equivalently, $t_{\rm f}$), 
one gets two curves in the $M_r-T_9$ ($R-T_9$) plot, 
at $t < t_{\rm f}$ and $t \ge t_{\rm f}$, respectively. 
$R(t_{\rm f})$ is given as the intersection of the two curves and estimated as 

\begin{equation}
  R_8 (t=t_{\rm f}) \sim 30 \dot E_{51}^{1/5} (t_f/10 s)^{3/5} 
\end{equation}
by setting $t=t_{\rm f}$ in equation (3) and in the expression for $R$. 
In the case $t_{\rm f}$ is very small, the relations above reduce to those 
in a prompt explosion model. 

With Figure~\ref{f7}, $M$($^{56}$Ni) can be estimated as the mass contained 
in the regions with $T_9 > 5$. 
In the prompt explosion with $E_{51} = 10$ and $M_{\rm ZAMS} = 40\msun$, 
$T_9 > 5$ is satisfied up to 
the enclosed mass $M_r \sim 3\msun$, among which $\sim 1\msun$ is reprocessed 
into $^{56}$Ni and the rest falls back onto the central remnant. 
This is consistent with the result of model 40SH within a factor of a few. 

The noticeable difference from the prompt models arises when $t_{\rm f}$ is large. 
Let $\dot E_{51} = 1$ and $t_{\rm f} = 10$ sec so that $E_{51} = 10$ 
being the same as in model 40SH. 
In this case, however, the region with $T_9 > 5$ is small 
(up to the enclosed mass $M_r \sim 1.8\msun$), 
so that the production of $^{56}$Ni is smaller than 40SH, 
as also found by Nagataki et al. (2003). 
The reason is simple. At a given time $t$, the total energy injected 
till that time is smaller in this kind of models than in prompt explosion models. 
At later times, matter already begins to expand, 
so that continuous energy deposition yields lower temperature because of 
larger ejecta volume. 

Let us look at the other case where $t_{\rm f}$ is sufficiently small, 
e.g., $t_{\rm f} = 1$ sec. 
This is shorter than the typical hydrodynamical time scale, 
as it takes the jets $\gsim 4$ s to propagate through the entire stellar radius. 
For a given final explosion energy, e.g., $E_{51}=10$ (thus $\dot E_{51}=10$), 
the energy injection is finished before the post-shock temperature 
falls off below $T_9 = 5$. Therefore, the 
expected $M$($^{56}$Ni) is similar to the corresponding prompt model.

Deviation from spherical symmetry makes the production of $^{56}$Ni less efficient 
for given $E$. 
Given the opening half-angle $\theta_{\rm jet} = 15^{\circ}$, 
the solid angle subtended by the jets is 
$\Omega = 2 \times 2 \pi (1 - \cos\theta_{\rm jet}) \sim 4 \pi /30$. 
This in effect corresponds to the enhancement of the isotropic energy by a factor 
($4 \pi / \Omega$) in this direction, i.e., 
\begin{equation}
  \frac{\Omega}{4 \pi} \frac{4 \pi}{3} R^3 a T^4 \sim 
        \dot E_{\rm jet} t \ . 
\end{equation}
For $\dot E_{51} =10$ s$^{-1}$ with $t_{\rm f} =1$ s, 
matter up to $M_{r} \sim 5\msun$, i.e., 
up to larger $R$ than the spherical case, is heated to $T_9 > 5$. 
For $M_{\rm cut} = 1.5\msun$, 
the ejected $M$($^{56}$Ni) is about $0.1\msun$
(after reduced by the factor of $\Omega / 4 \pi$). 
It roughly explains $M$($^{56}$Ni) in model 40A. 
This is much smaller than $M$($^{56}$Ni) $ \gsim 1.0\msun$ 
in the spherical model with the same parameters. 
The effect of asphericity, therefore, is to reduce $M$($^{56}$Ni) 
for given $E_{\rm jet}$. 
The reason is as follows: 
though the region with $T_9 > 5$ is elongated to the outer, lower density regions 
in a specific direction in the aspherical models, 
the mass in such outer regions is small. 
The mass of the region with $T_9 > 5$ in 
the bipolar models, 
therefore, is reduced by a factor 
$\sim$ ($\Omega / 4 \pi$). 

Also the initial mass of the compact remnant, $M_{\rm REM 0}$, 
greatly affects 
the $^{56}$Ni production. Model 40D produces 
very little amount of $^{56}$Ni. 
The reason for this can be seen in Figure~\ref{f7}. 
With $E_{51} \sim 10$, the region with $T_9 > 5$ 
extends to $M_r \sim 5\msun$ along the $z$-axis 
as discusses for model 40A. 
Because $M_{\rm REM0}$ is as large as $3\msun$ in model 40D, 
$M$($^{56}$Ni) is estimated to be smaller than $0.1\msun$. 
Thus, with $M_{\rm REM0} > 3\msun$, 
even a hypernova-like explosion produces much smaller $M$($^{56}$Ni) than 
typical supernovae. In other words, $M$($^{56}$Ni) may not be 
connected with $E_{\rm jet}$.

\begin{figure}
\begin{center}
	\begin{minipage}[t]{0.5\textwidth}
		\epsscale{1.0}
		\plotone{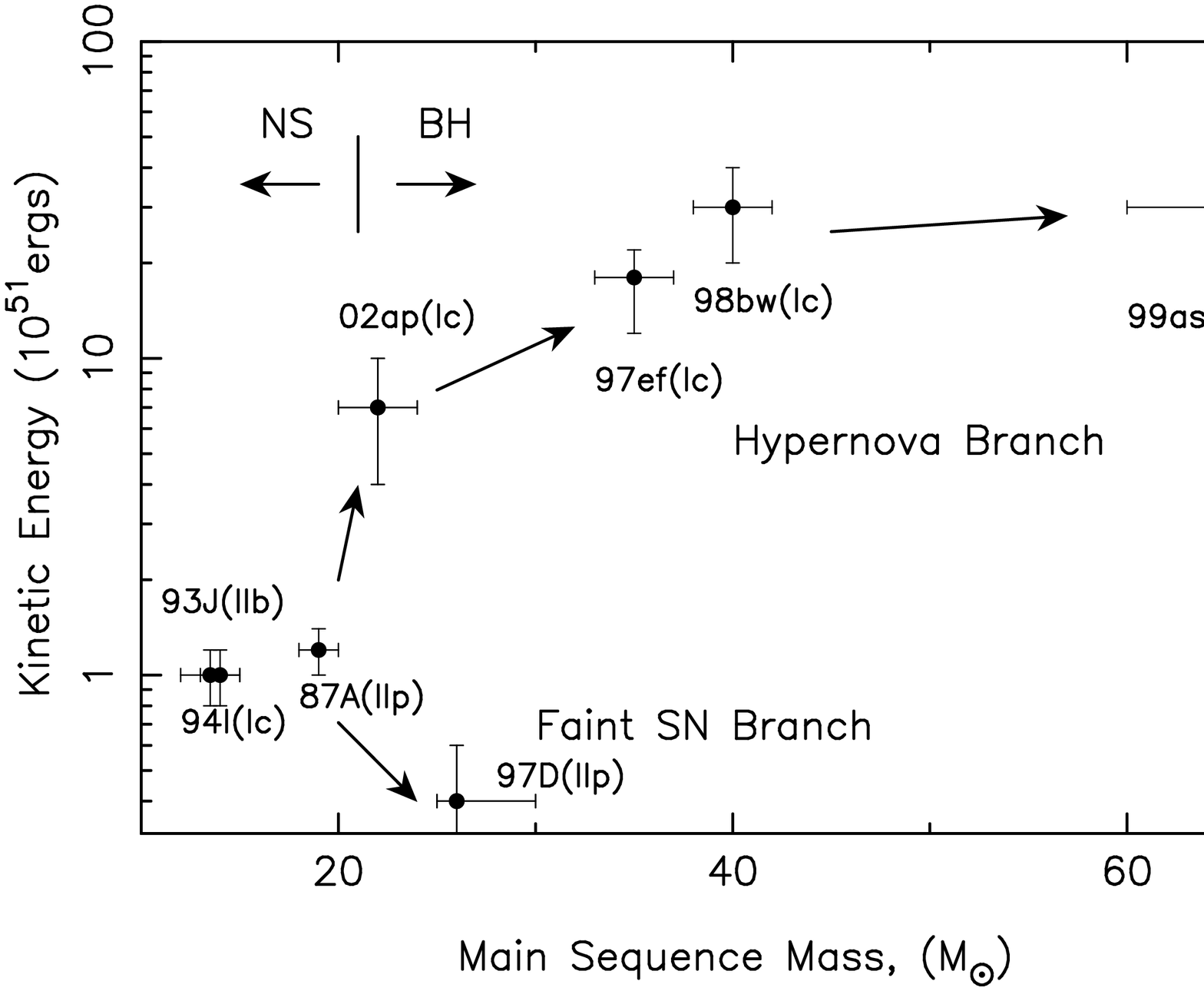}
	\end{minipage}
	\begin{minipage}[t]{0.5\textwidth}
		\epsscale{1.0}
		\plotone{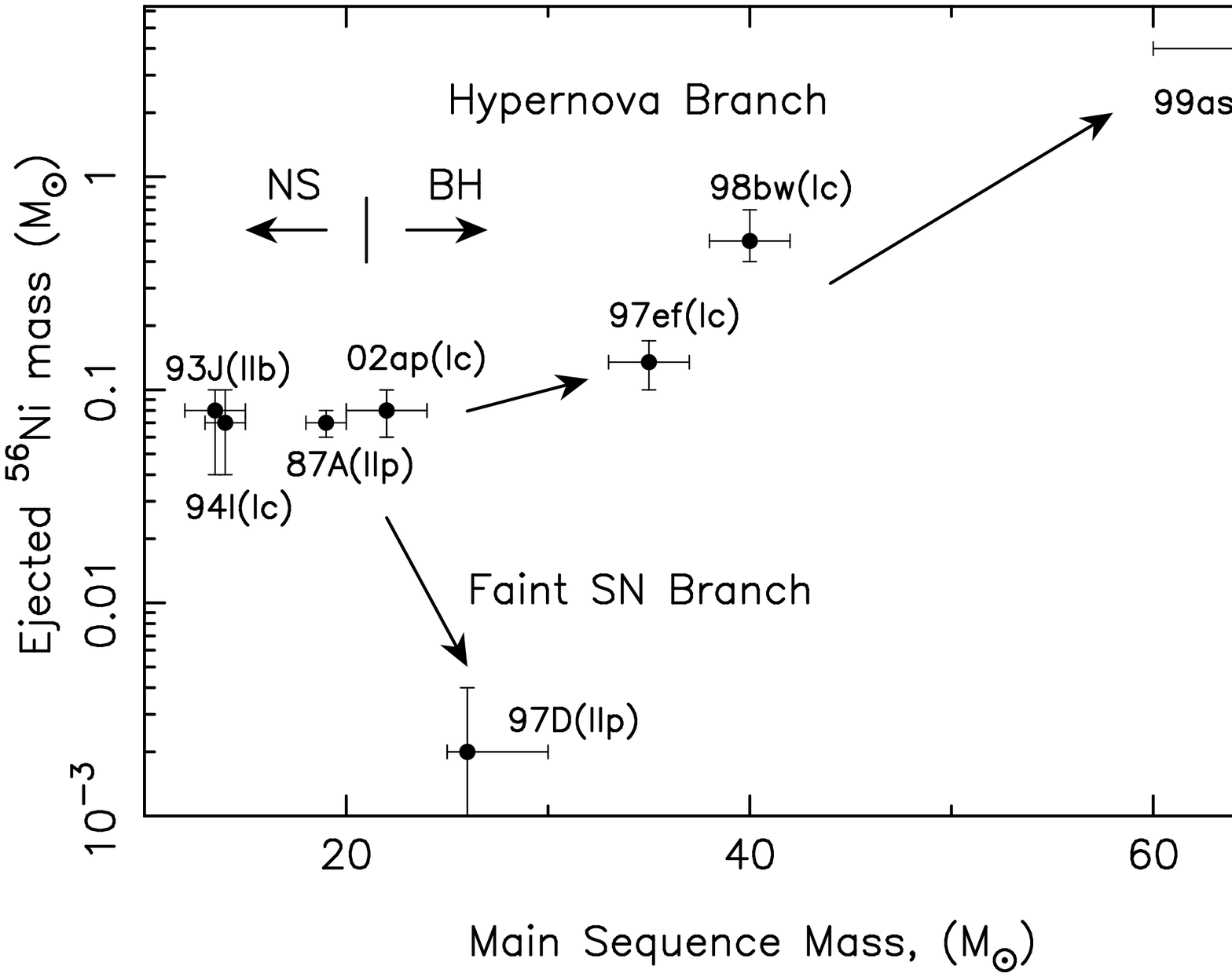}
	\end{minipage}
\end{center}
\caption{(a) Explosion Energies and (b) ejected $^{56}$Ni masses against 
main-sequence mass of the progenitors for core-collapse supernovae/hypernovae, 
as estimated for SN~II 1987A (Shigeyama \& Nomoto 1990); 
SN~IIb 1993J (Shigeyama et al. 1994); 
SN~Ic 1994I (Iwamoto et al. 1994); 
SN~II 1997D (Turatto et al. 1998); 
SN~Ic 1997ef (Mazzali et al. 2000);  
SN~Ic 1998bw (Iwamoto et al. 1998); 
SN~Ic 1999as (Hatano et al. 2001); 
SN~Ic 2002ap (Mazzali et al. 2002). 
\label{f1}}
\end{figure}

\begin{figure}
\begin{center}
	\begin{minipage}[t]{0.35\textwidth}
		\epsscale{1.0}
		\plotone{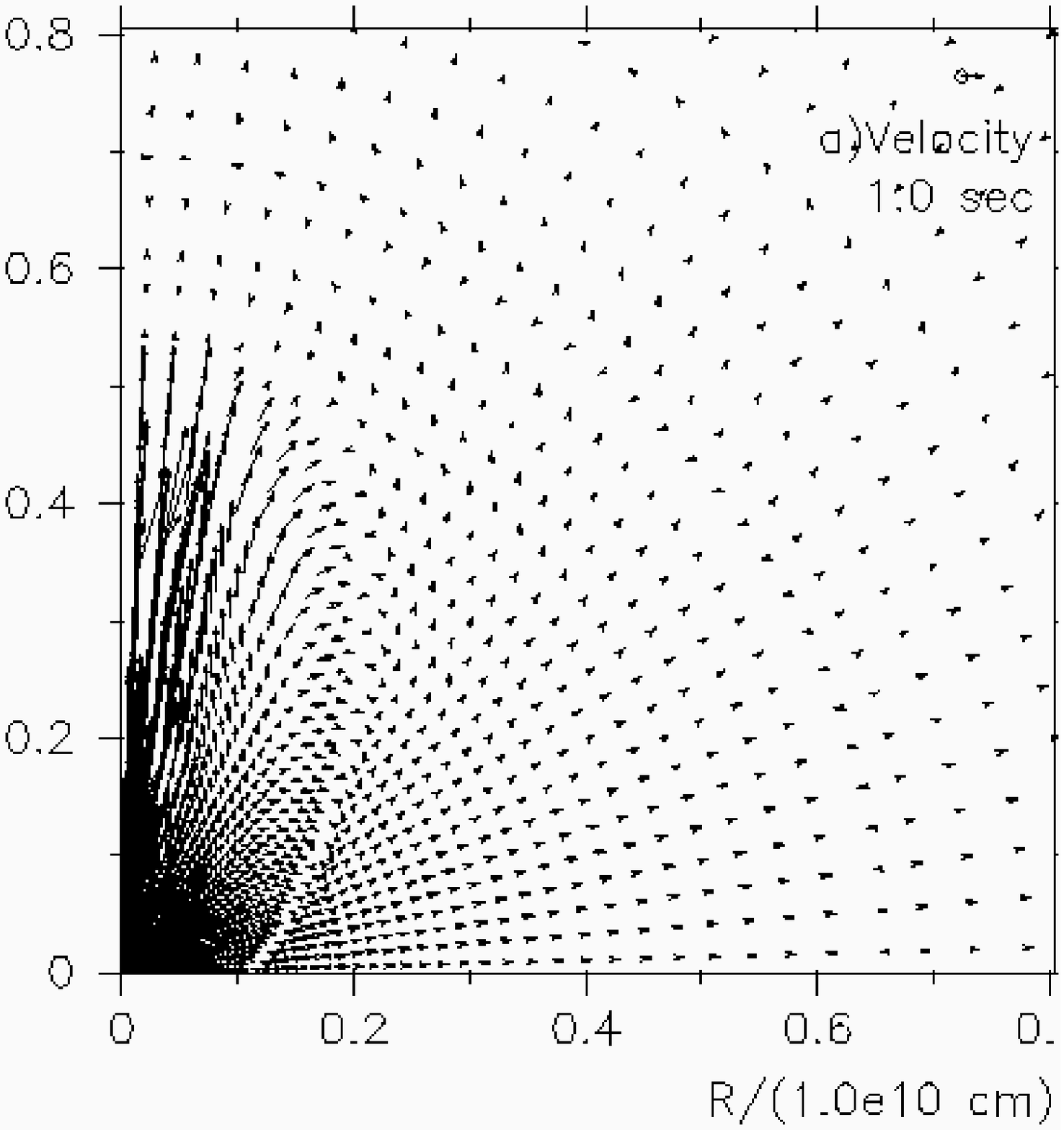}
	\end{minipage}
	\begin{minipage}[t]{0.35\textwidth}
		\epsscale{0.95}
		\plotone{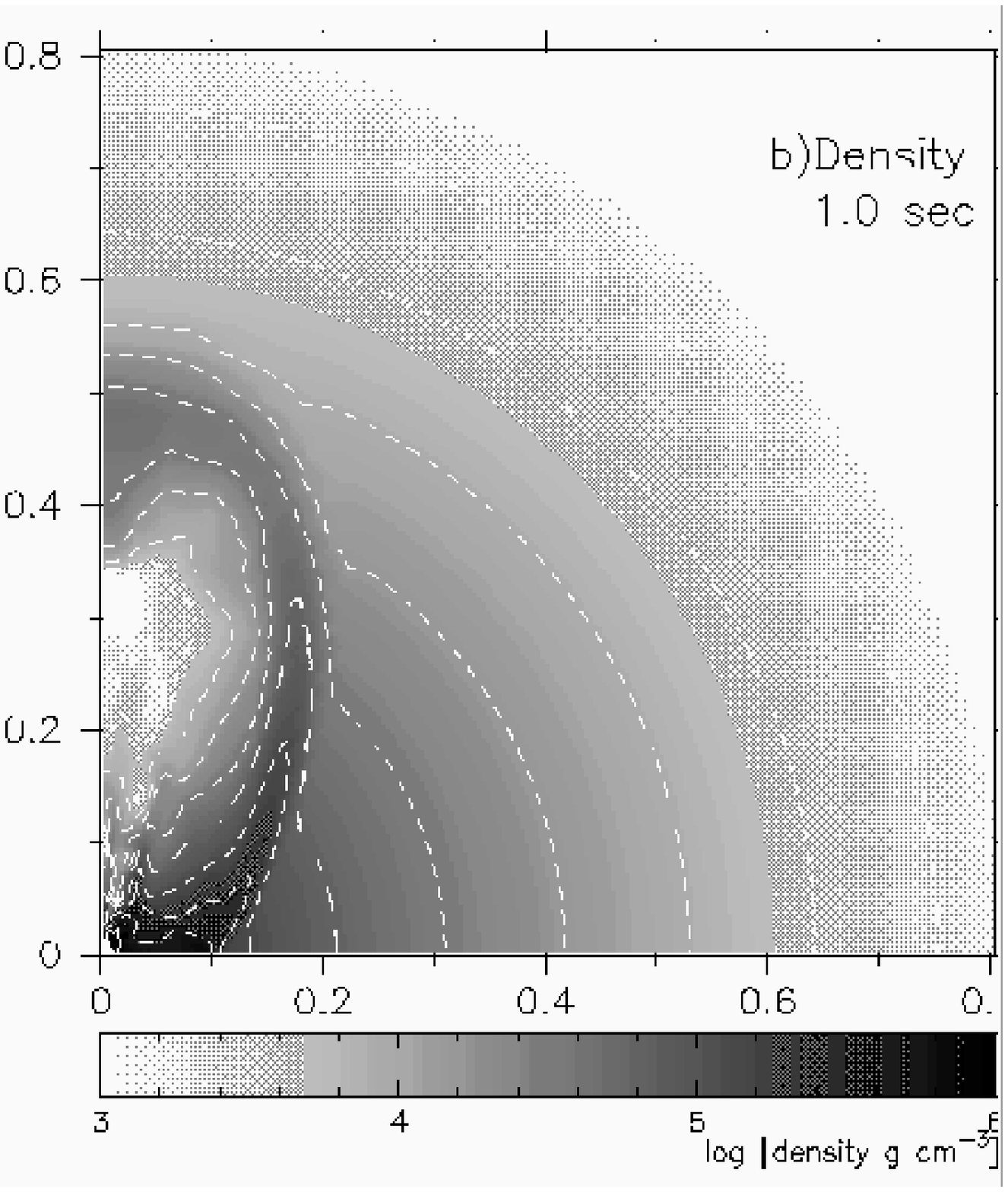}
	\end{minipage}\\
	\begin{minipage}[t]{0.35\textwidth}
		\epsscale{1.0}
		\plotone{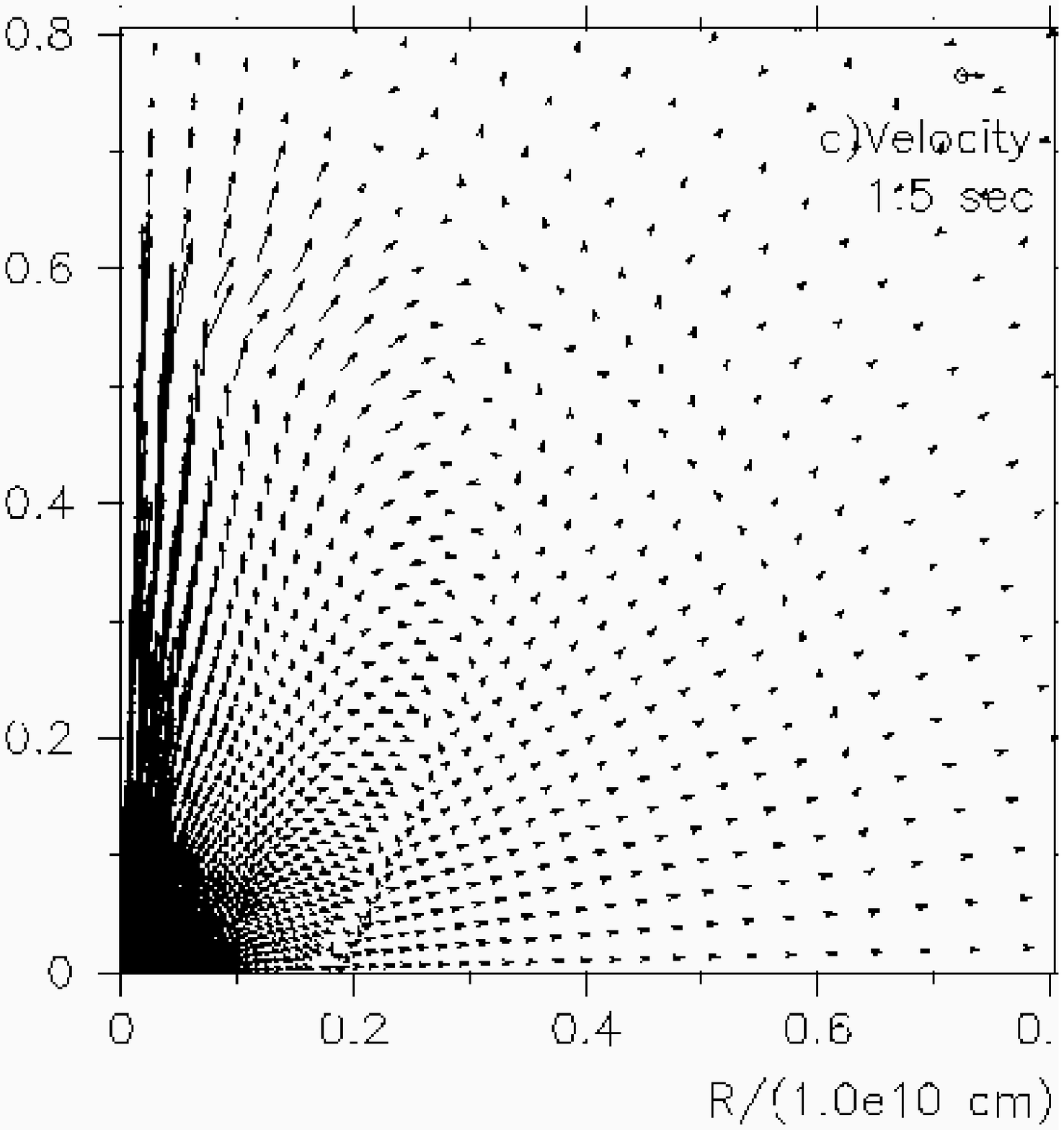}
	\end{minipage}
	\begin{minipage}[t]{0.35\textwidth}
		\epsscale{0.95}
		\plotone{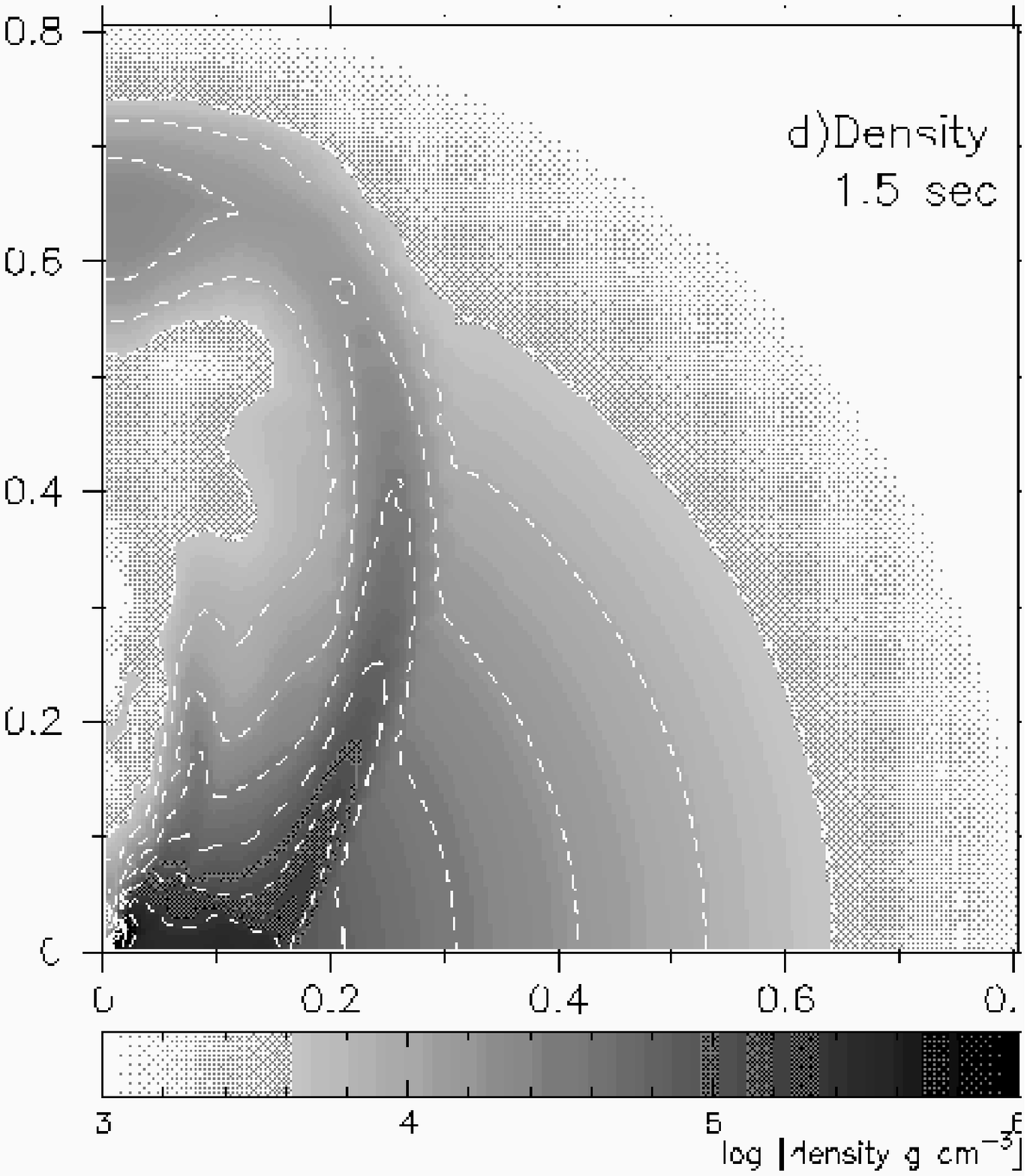}
	\end{minipage}\\
	\begin{minipage}[t]{0.35\textwidth}
		\epsscale{1.0}
		\plotone{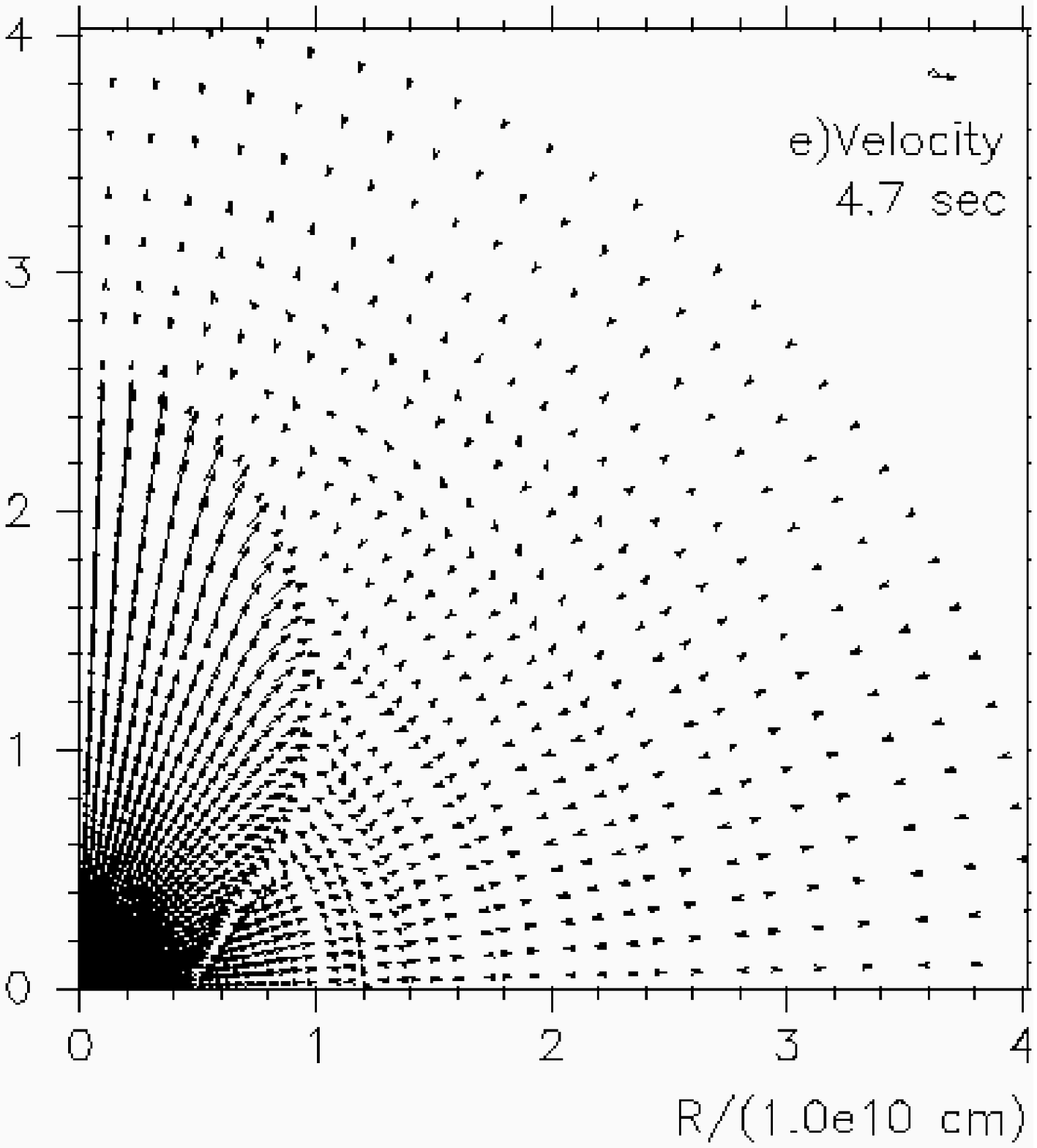}
	\end{minipage}
	\begin{minipage}[t]{0.35\textwidth}
		\epsscale{0.95}
		\plotone{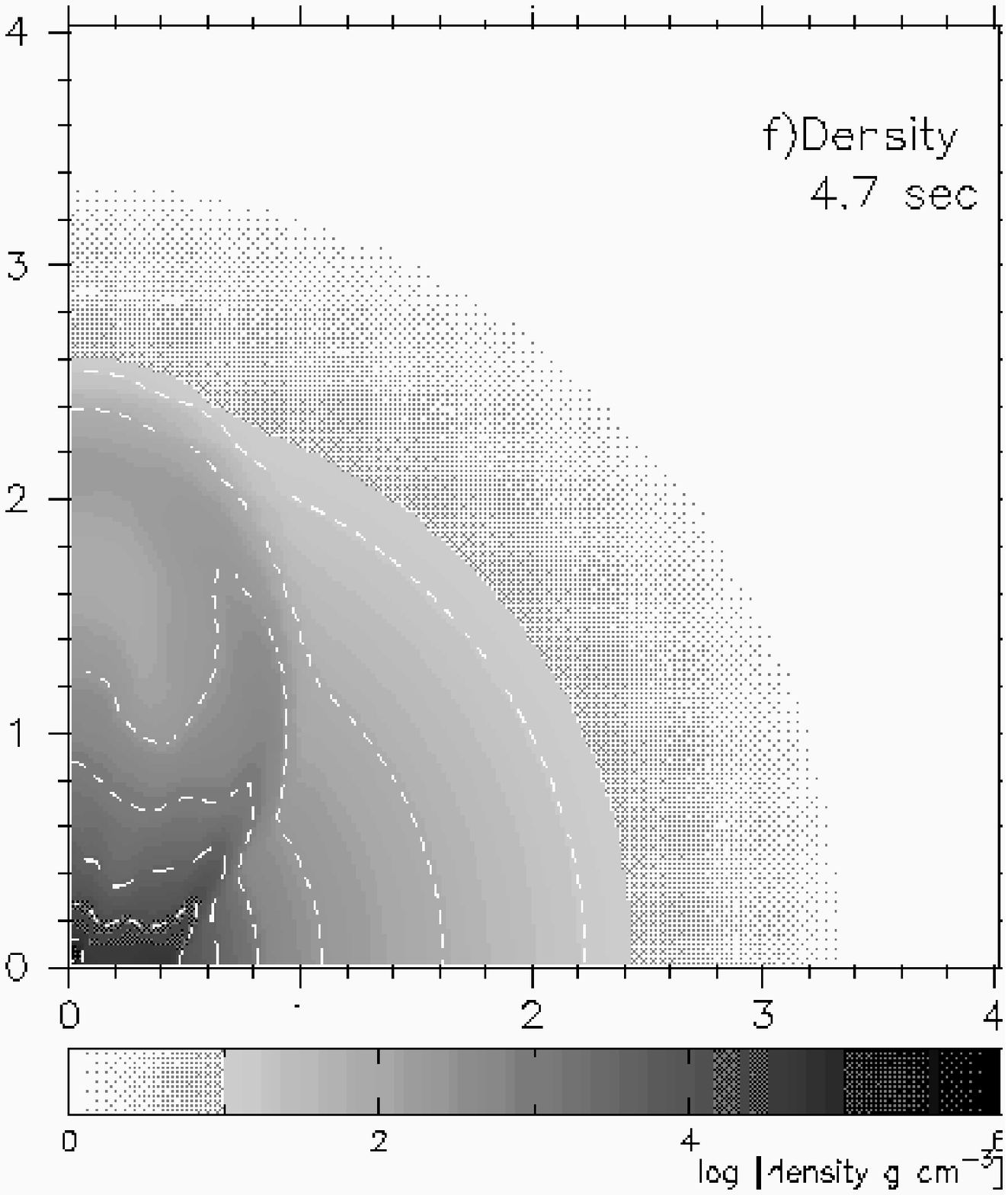}
	\end{minipage}
\end{center}
\caption{(a,c,e) Velocity and (b,d,f) density distributions of model 40A, 
at 1.0s (a,b), 1.5s (c,d), and 4.7s (e,f) after the initiation of the jets. 
(a,c,e) The arrows with a circle on its base shown on the upper right of 
each figure represent $2 \times 10^{9}$cm s$^{-1}$. 
(b,d,e) The interval of contour is (max-min)/10 in log scale. 
\label{f2}}

\end{figure}

\begin{figure}
\begin{center}
	\begin{minipage}[t]{0.35\textwidth}
		\epsscale{1.0}
		\plotone{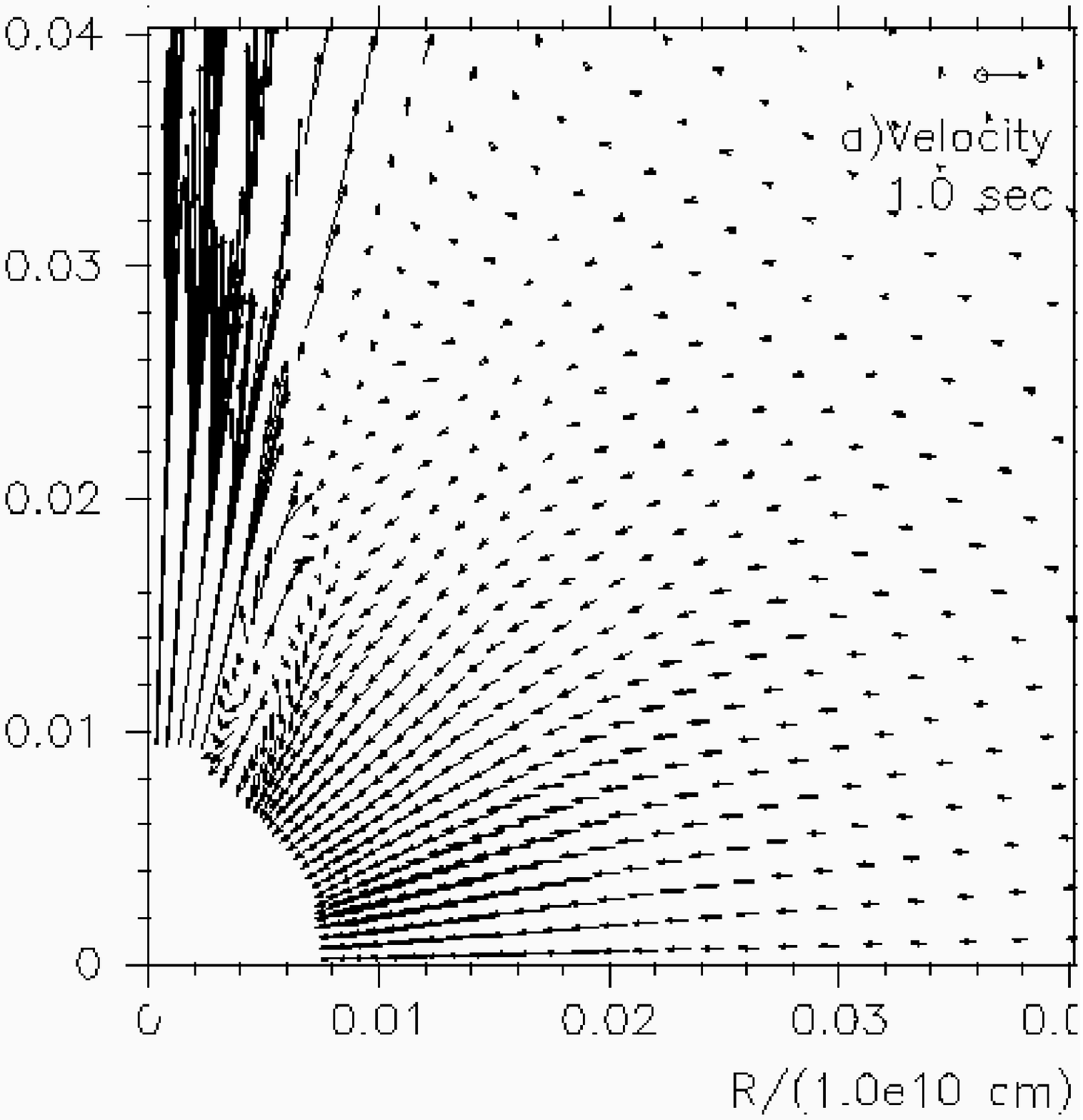}
	\end{minipage}
	\begin{minipage}[t]{0.35\textwidth}
		\epsscale{0.95}
		\plotone{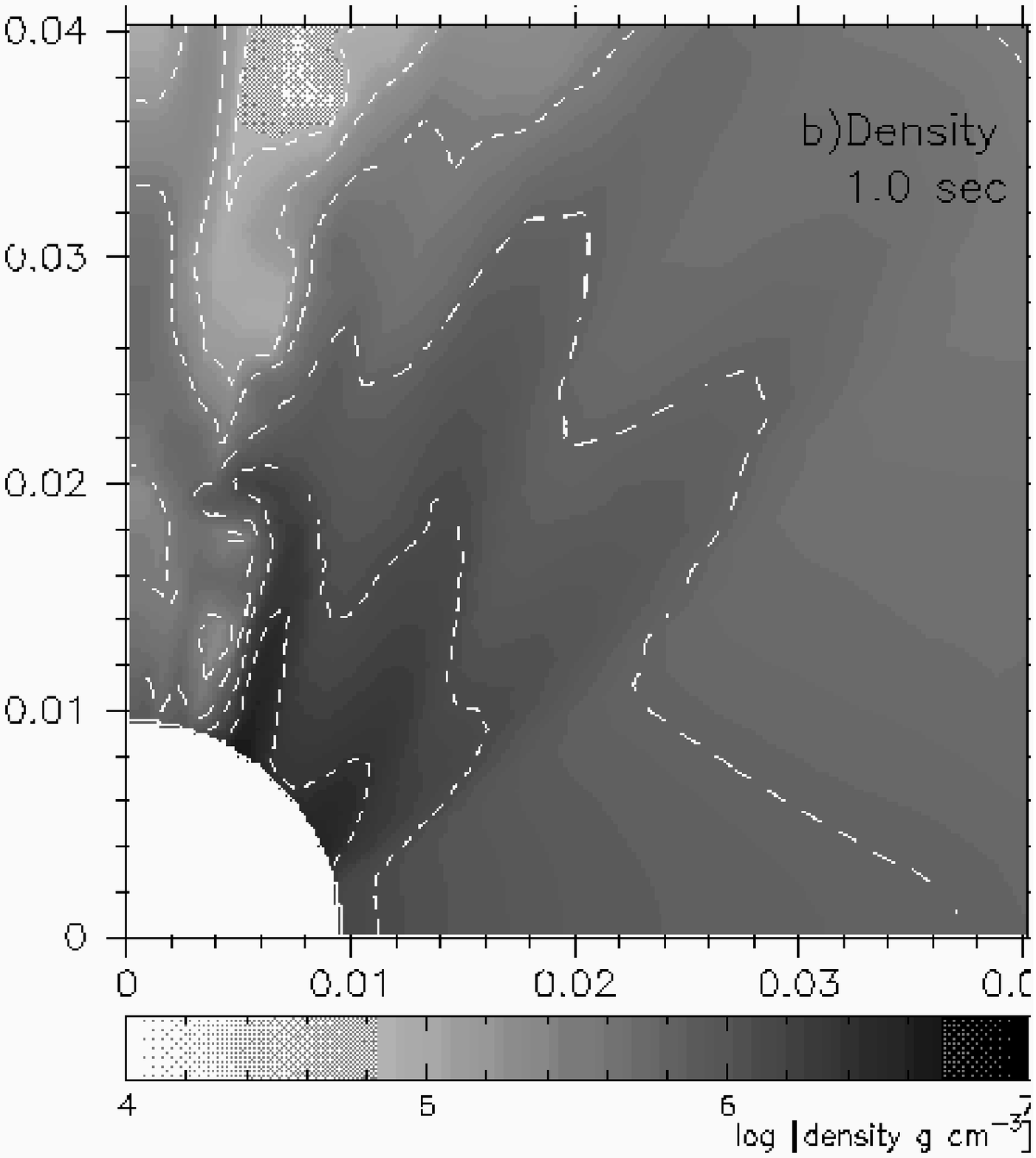}
	\end{minipage}\\
	\begin{minipage}[t]{0.35\textwidth}
		\epsscale{1.0}
		\plotone{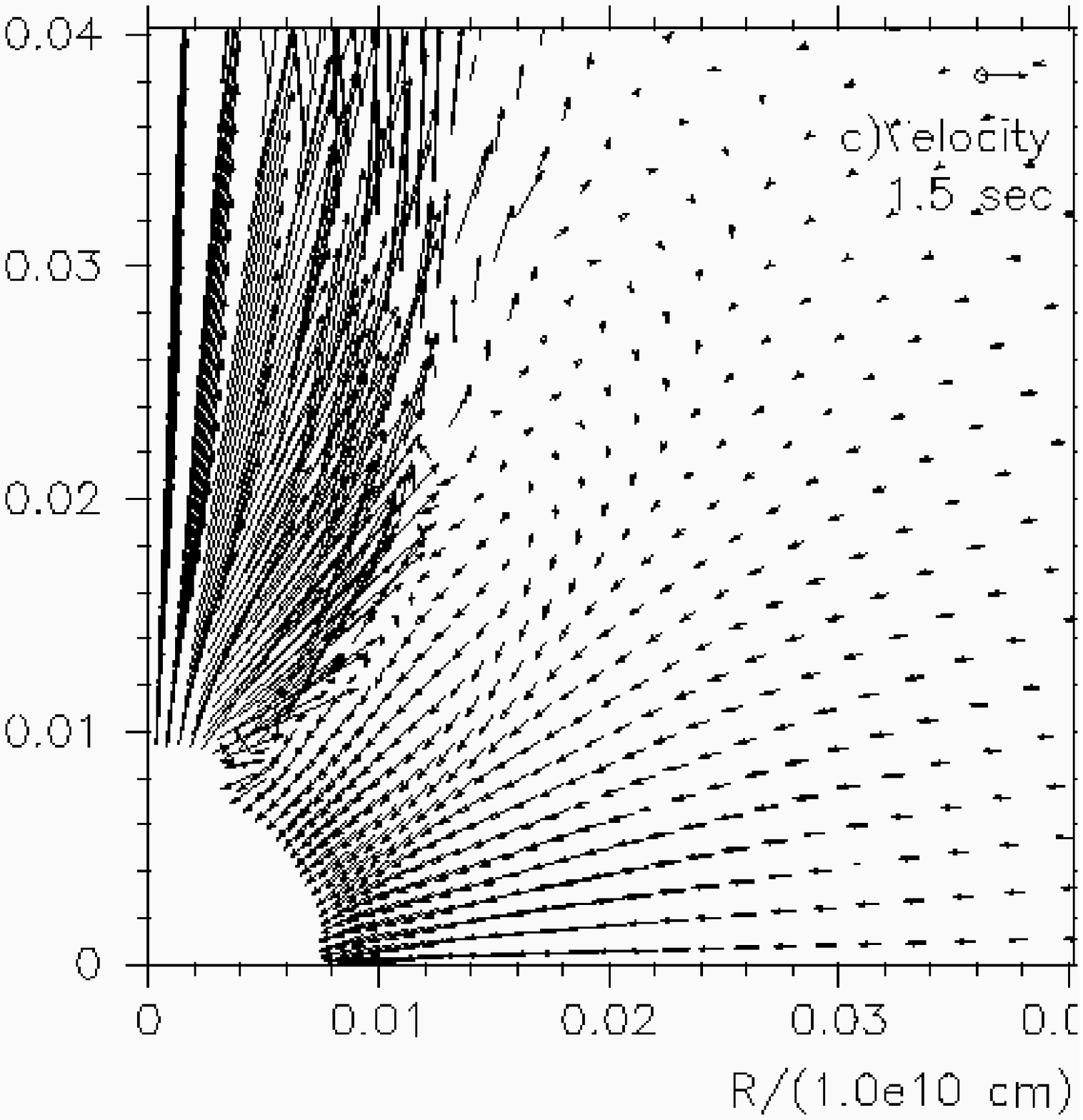}
	\end{minipage}
	\begin{minipage}[t]{0.35\textwidth}
		\epsscale{0.95}
		\plotone{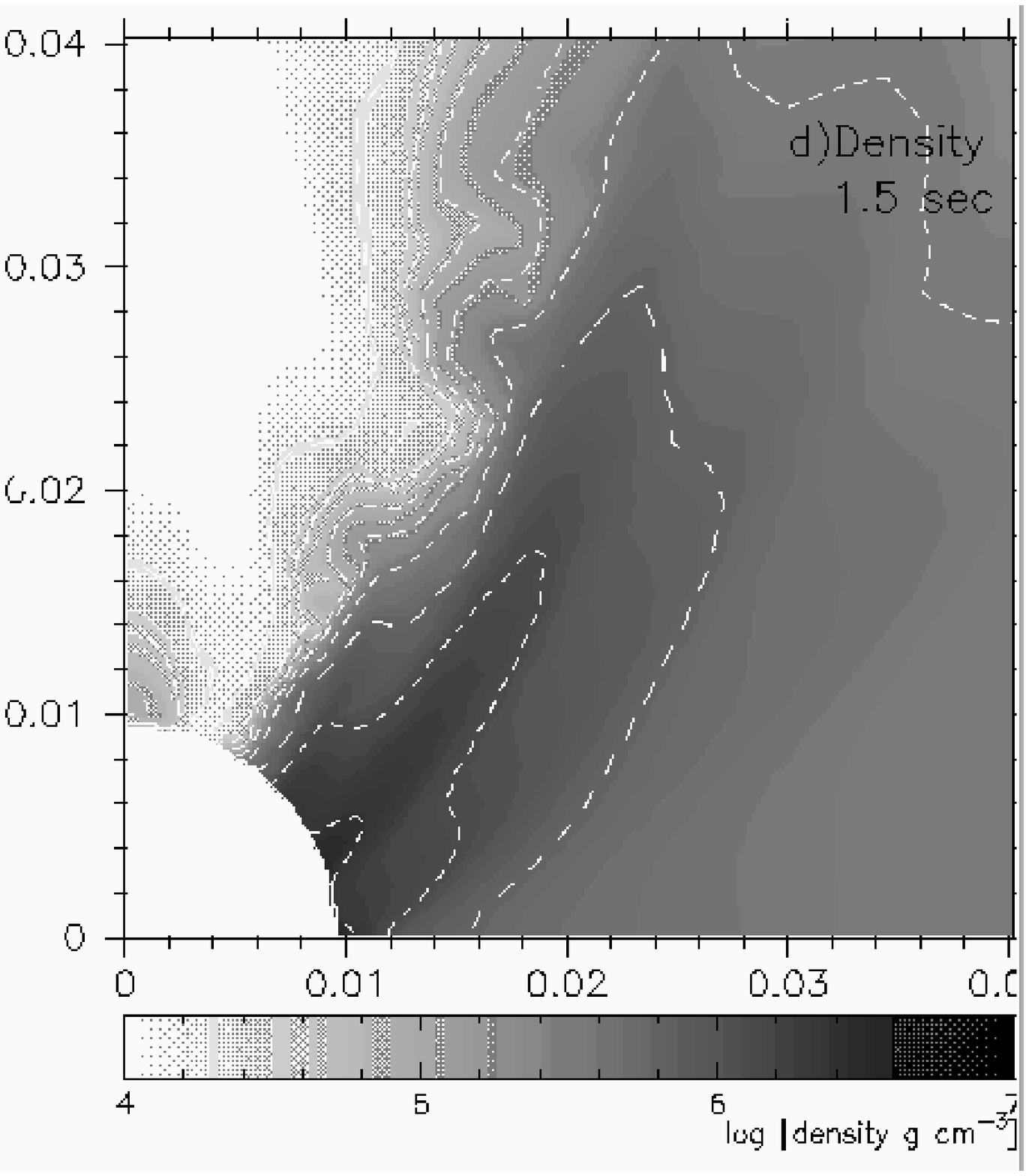}
	\end{minipage}\\
	\begin{minipage}[t]{0.35\textwidth}
		\epsscale{1.0}
		\plotone{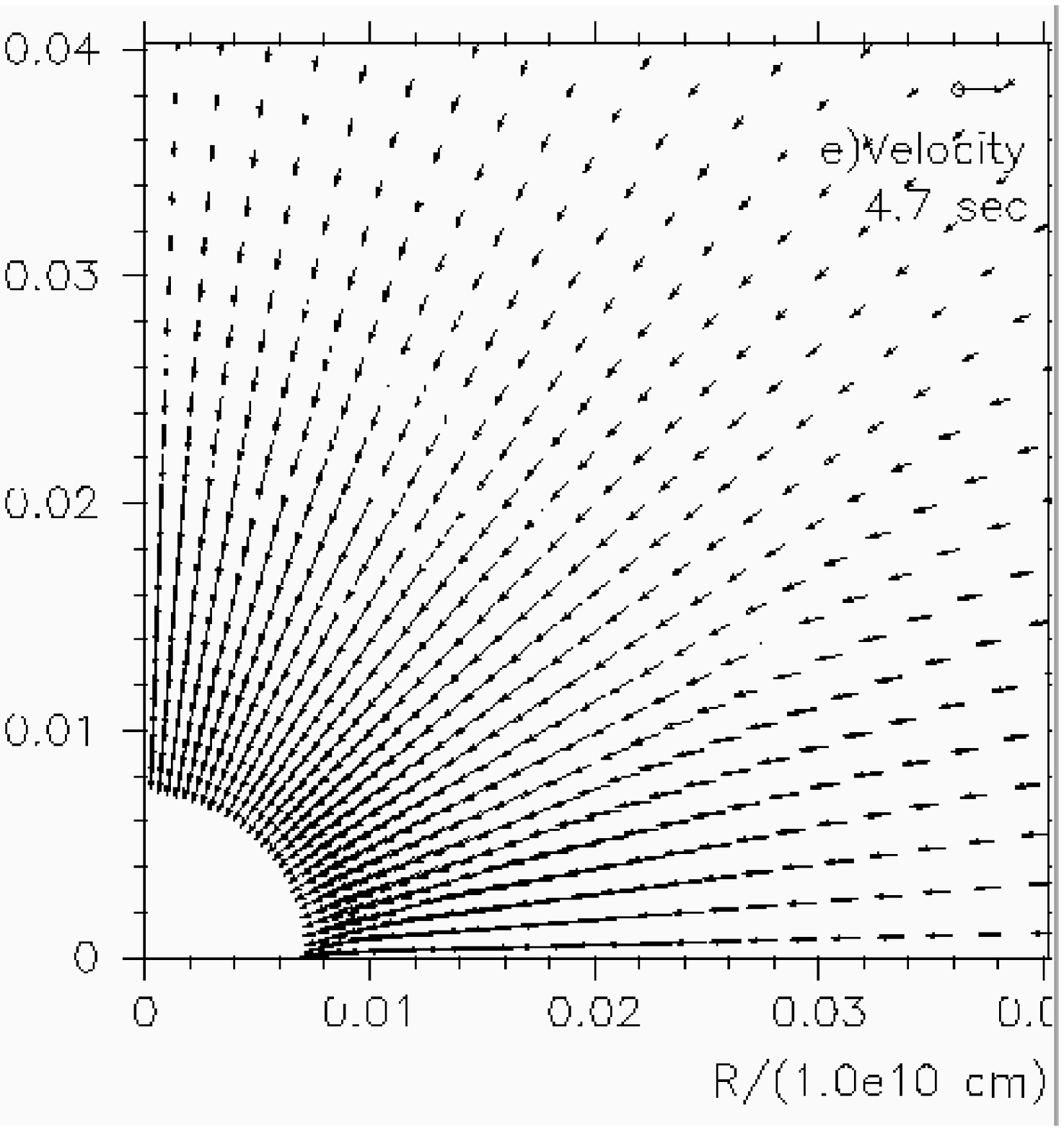}
	\end{minipage}
	\begin{minipage}[t]{0.35\textwidth}
		\epsscale{0.95}
		\plotone{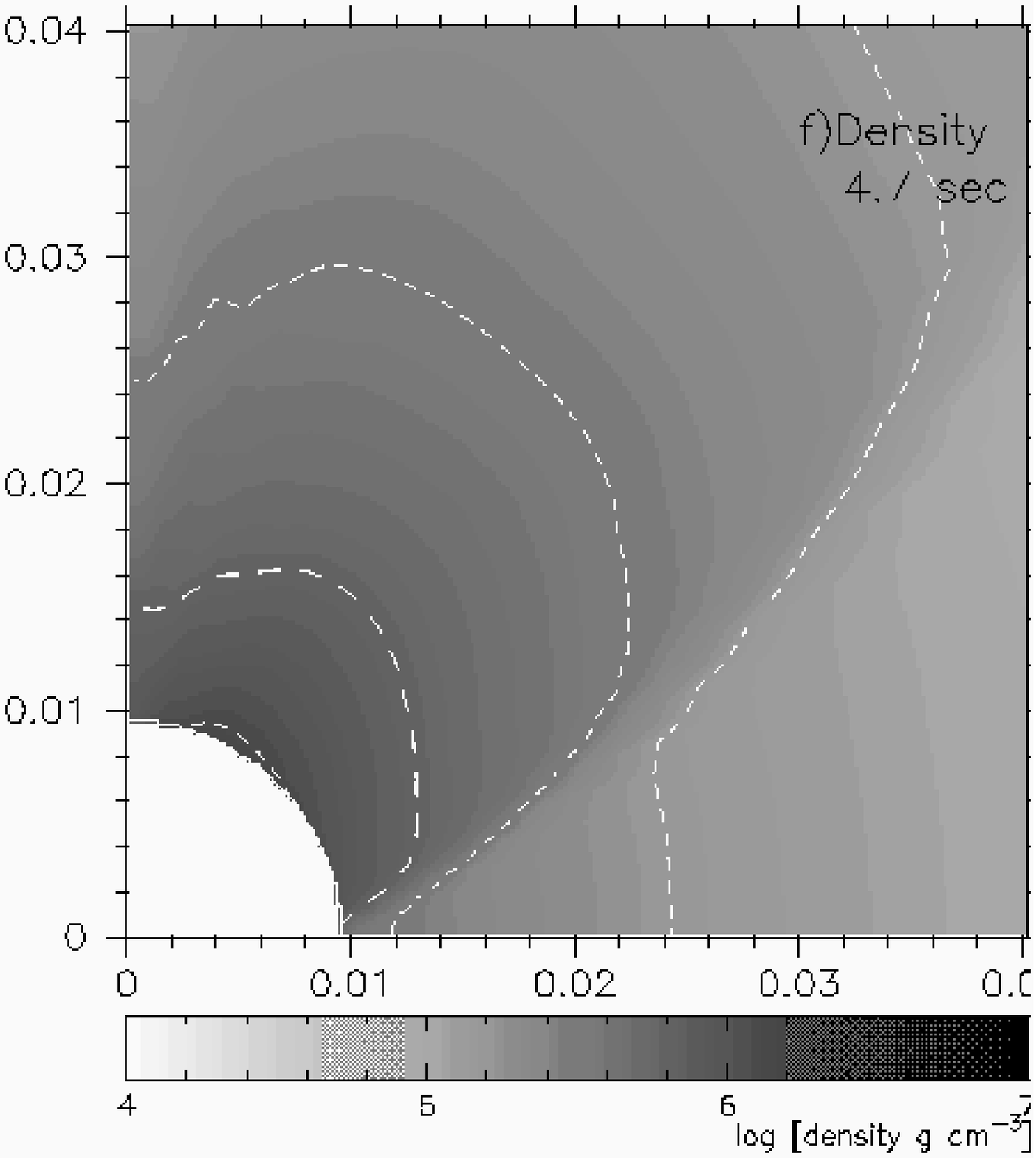}
	\end{minipage}
\end{center}
\caption{The same as Figure 2, but on an expanded scale. 
\label{f3}}
\end{figure}

\begin{figure}
\begin{center}
	\begin{minipage}[t]{0.4\textwidth}
		\epsscale{1.0}
		\plotone{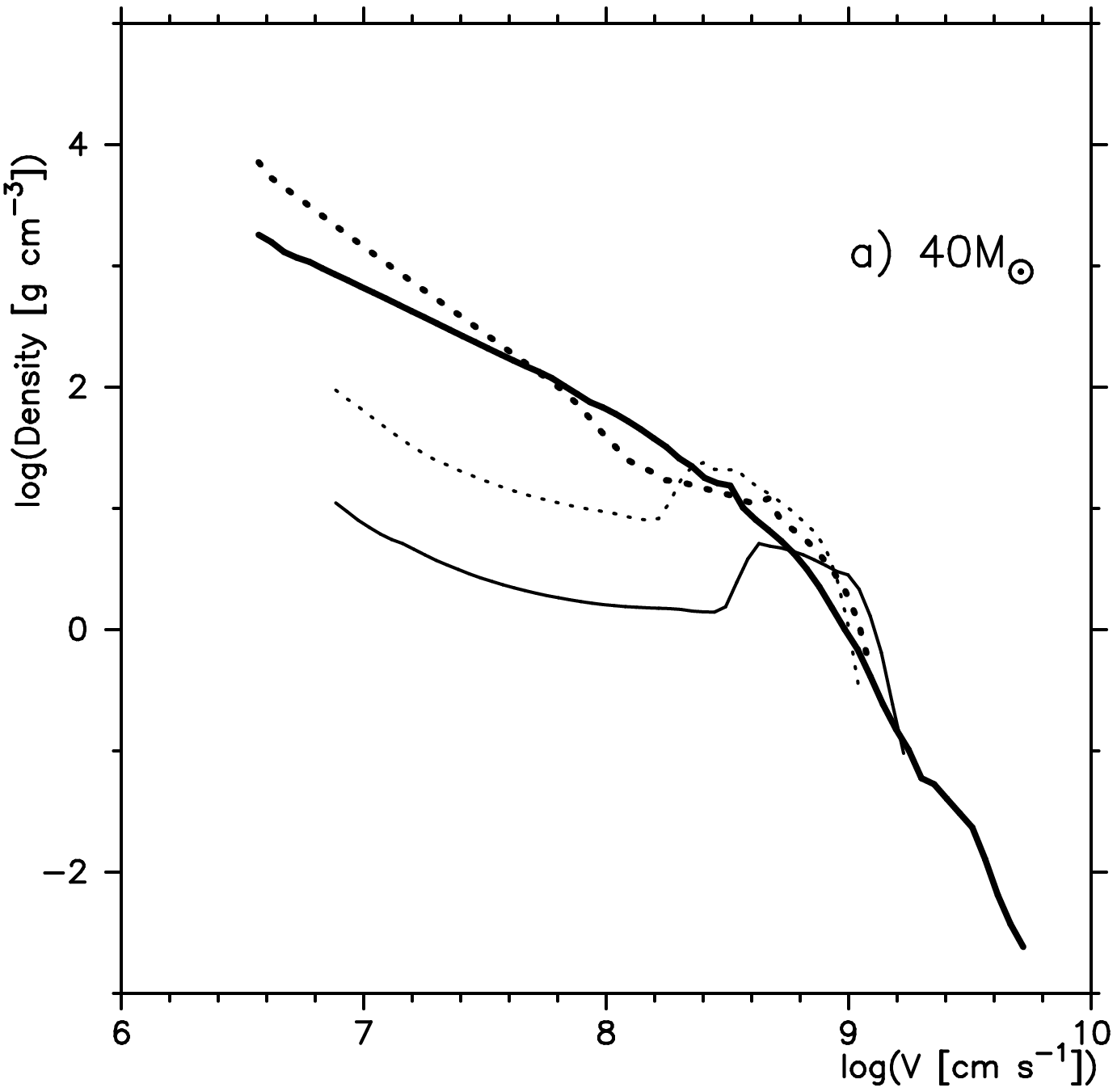}
	\end{minipage}
	\begin{minipage}[t]{0.4\textwidth}
		\epsscale{1.0}
		\plotone{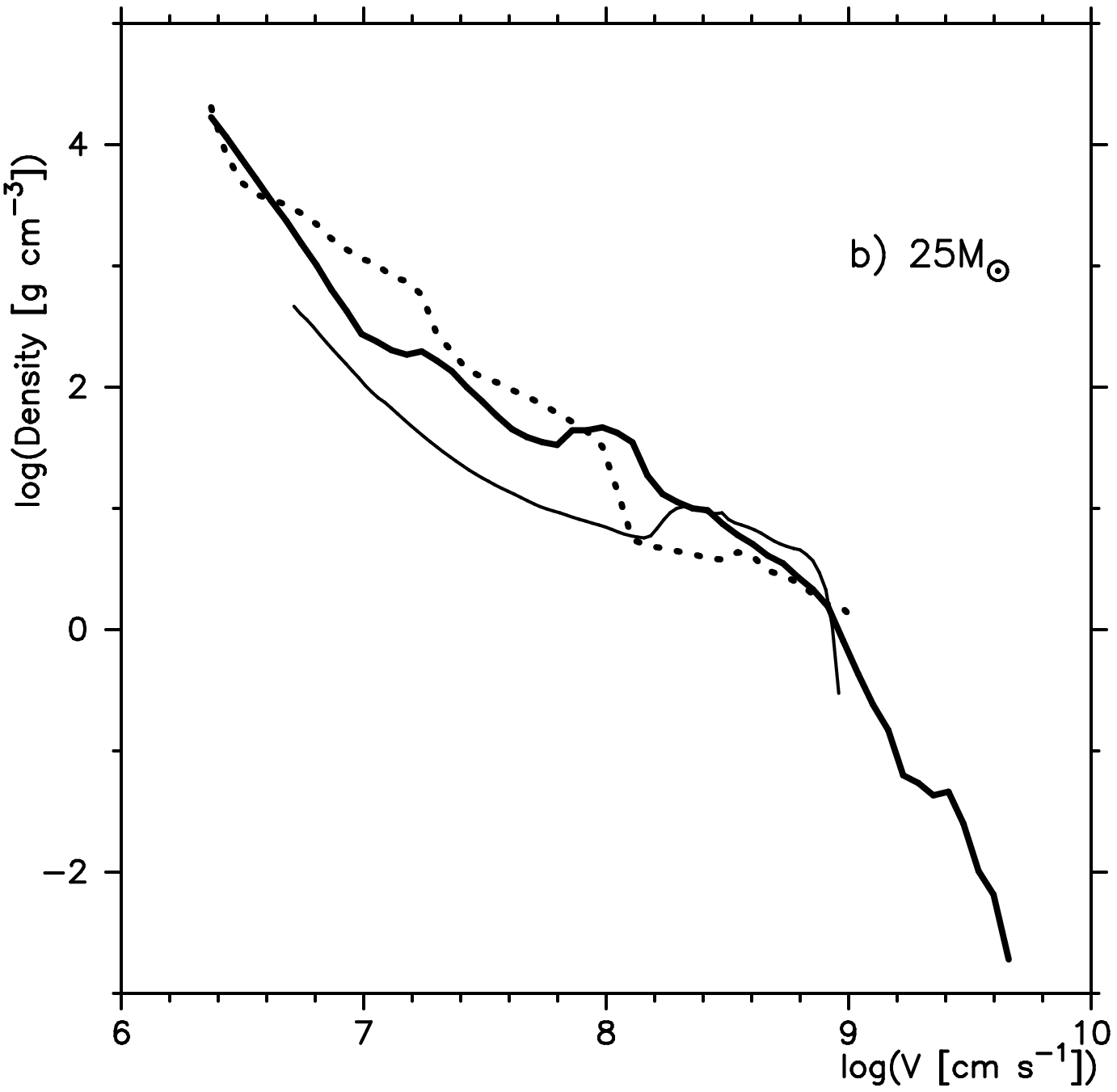}
	\end{minipage}
\end{center}
\caption{Density structure at 100 sec. 
(a) Along the $z$-axis of 40A (thick-solid) 
and the $r$-axis of 40A (thick-dotted), 40SH (thin-solid), and 40SL (thin-dotted). 
(b) Along the $z$-axis of 25A (thick-solid) and 
the $r$-axis of 25A (thick-dotted), and 25S (thin-solid). 
\label{f4}}
\end{figure}

\begin{figure}
\begin{center}
	\begin{minipage}[t]{0.4\textwidth}
		\epsscale{1.0}
		\plotone{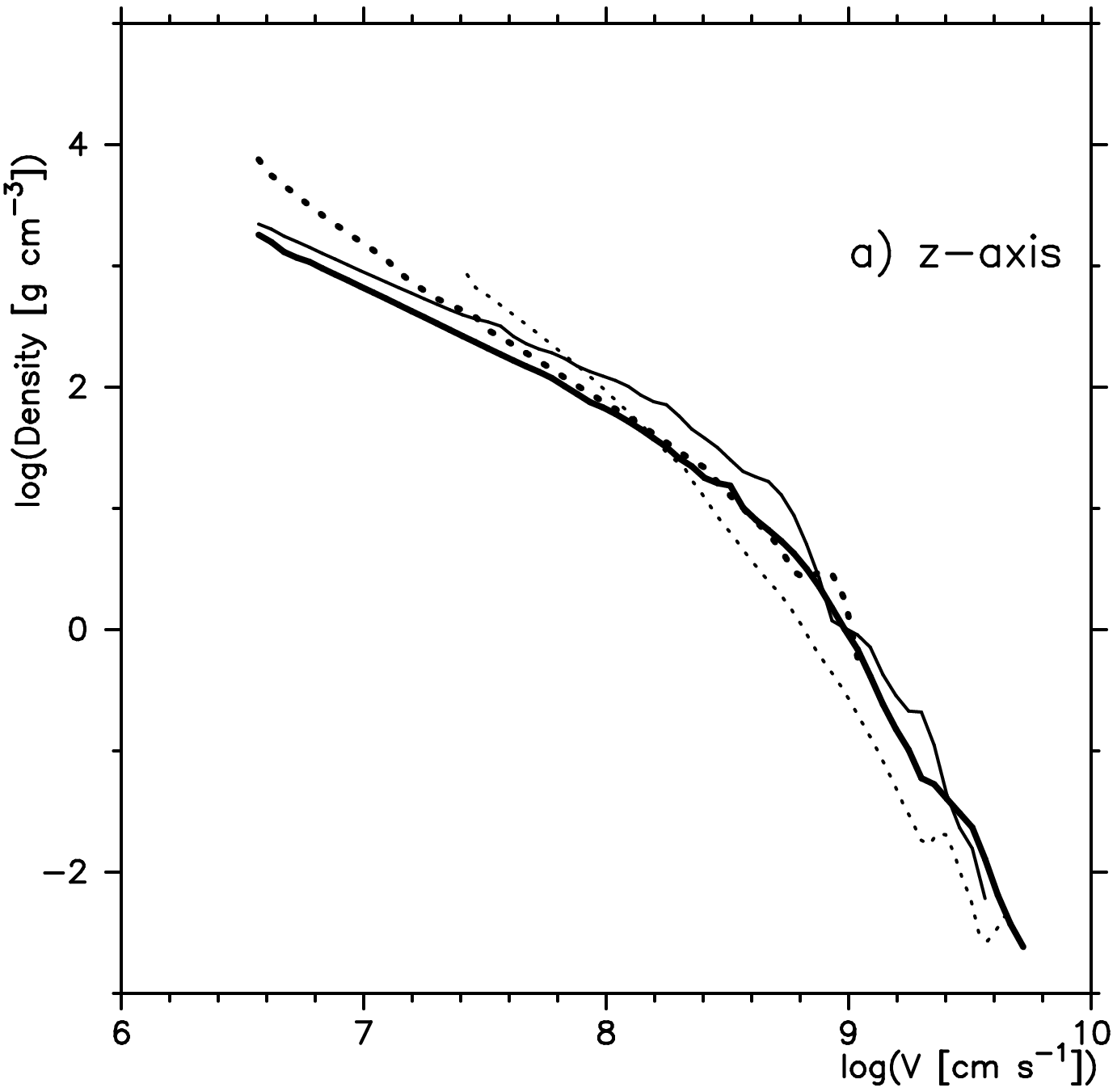}
	\end{minipage}
	\begin{minipage}[t]{0.4\textwidth}
		\epsscale{1.0}
		\plotone{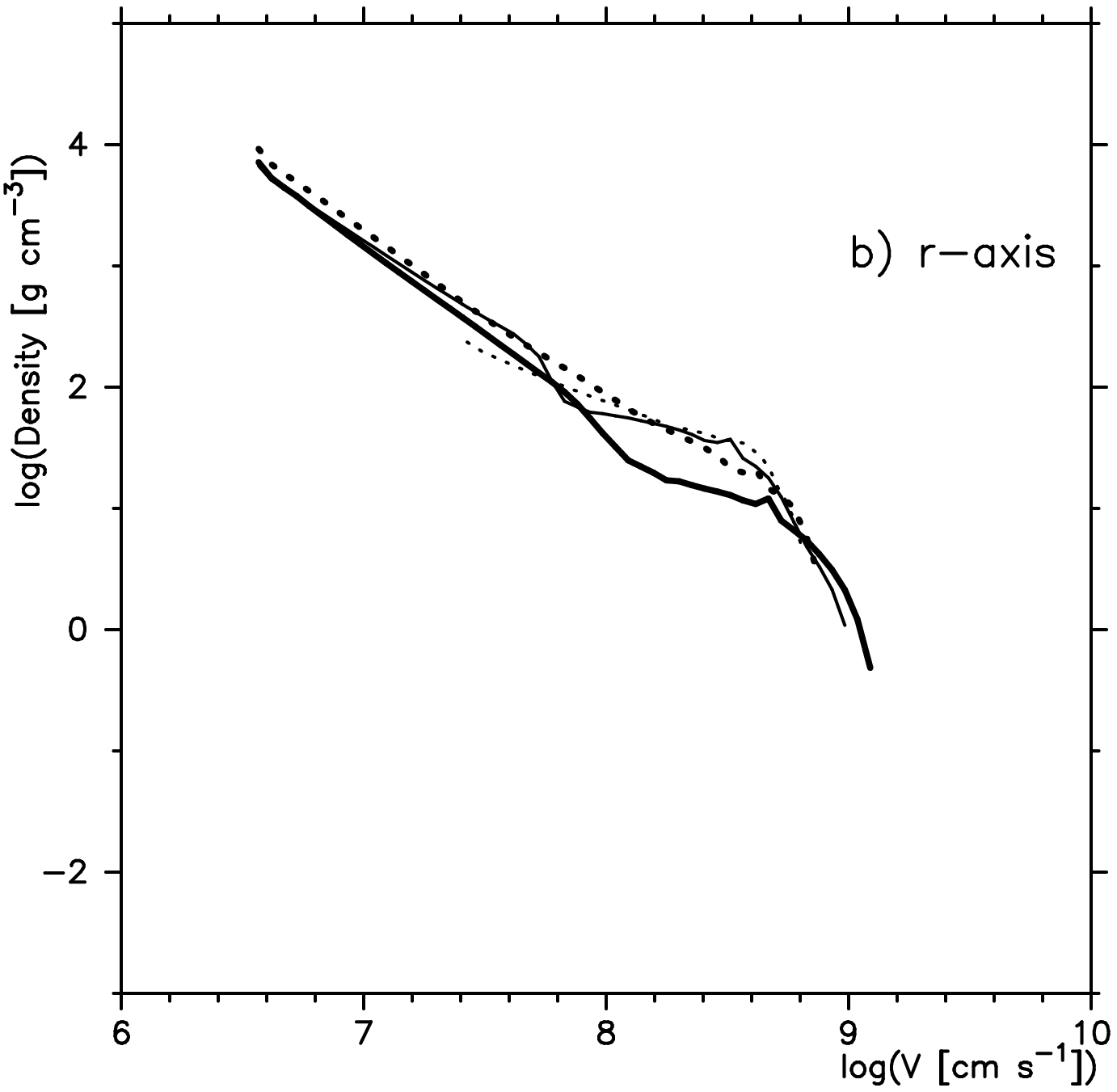}
	\end{minipage}
\end{center}
\caption{Density structure at 100 sec for different jet properties. 
(a) Along the $z$-axis of 40A (thick-solid), of 40B (thick-dotted),  
of 40C (thin-solid), and of 40D (thin-dotted). 
(b) Along the $r$-axis of 40A (thick-solid), of 40B (thick-dotted),  
of 40C (thin-solid), and of 40D (thin-dotted). 
\label{f5}}
\end{figure}

\begin{figure}
\begin{center}
	\begin{minipage}[t]{0.4\textwidth}
		\epsscale{1.0}
		\plotone{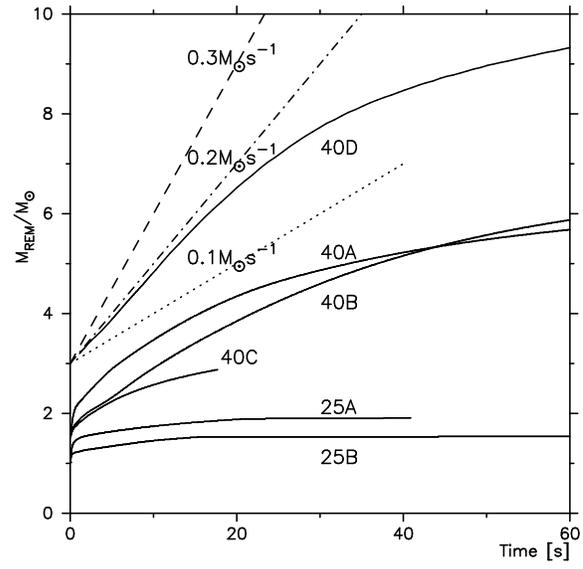}
	\end{minipage}
\end{center}
\caption{Growth of the mass of the central remnant ($M_{\rm REM}$). 
The three straight lines correspond to constant accretion rates. 
\label{f6}}
\end{figure}

\begin{figure}
\begin{center}
	\begin{minipage}[t]{0.4\textwidth}
		\epsscale{1.0}
		\plotone{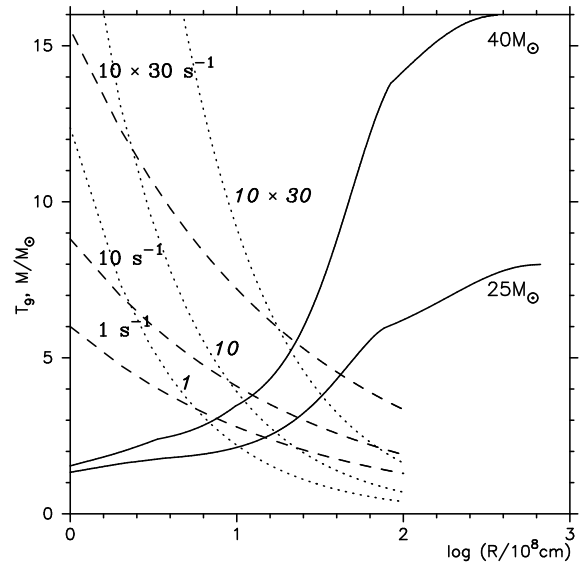}
	\end{minipage}
\end{center}
\caption{The post-shock temperature ($T_9$) 
as a function of radius predicted by equation (4), 
for $t < t_{\rm f}$ (dashed) and $t \ge t_{\rm f}$ (dotted). 
Also shown are the enclosed masses of the progenitor stars 
with main-sequence masses $M_{\rm ZAMS} = 40\msun$ and $25\msun$ 
(solid lines: Nomoto \& Hashimoto 1988). 
\label{f7}}
\end{figure}

We note that $M$($^{56}$Ni) in Figure~\ref{f1} has been obtained through 
analyzing light curves and spectra. 
Figure~\ref{f1} does not imply the absence of hypernovae/supernovae with 
little $M$($^{56}$Ni) such as model 40D, 
as the corresponding SNe~Ib/c must be too faint to be observed directly. 
Actually, the existence of such $^{56}$Ni-deficient supernovae is 
inferred from extremely metal-poor stars, which we discuss later. 

On the other hand, $M$($^{56}$Ni) obtained from modeling light curves 
and spectra shows that there exist hypernovae with ejection of 
large $M$($^{56}$Ni), highlighted by $M$($^{56}$Ni) $\sim 0.5\msun$ 
in SN~1998bw. 

To account for the latter case, i.e., hypernovae with $M$($^{56}$Ni) 
$\gsim 0.1\msun$, 
one of the following conditions must be satisfied. 
\begin{itemize}
\item[(1)]
To produce $^{56}$Ni in the shocked stellar material, $M_{\rm REM 0}$ 
must be sufficiently small ($M_{\rm REM 0} \lsim 2\msun$) and 
the energy injection rate sufficiently large 
($\dot E_{\rm jet} \gsim 10^{52}$ ergs s$^{-1}$). 
We note that the models presented here except for model 40D produce 
$\gsim 0.1\msun$ $M$($^{56}$Ni). 
\item[(2)]
The alternative way is to eject $^{56}$Ni from the inner most region, e.g., 
in a disk wind from the accretion disk surrounding the central object
(MacFadyen \& Woosley, 1999; MacFadyen 2003; see also Introduction). 
This would correspond to taking larger $\mu$ than in our models 
(and maybe larger $\theta_{\rm jet}$). 
In this case, the resultant flow from the inner region is more massive and slow. 
This possibility should be examined in more detail. 
In the present models, the mass carried in the jets
is small (with $E_{51} =10$, $\alpha = 0.01$, and $\mu = 0.1$, $M_{\rm jet} 
\sim 0.05\msun$), so that the newly 
synthesized elements are dominated by those in the shocked stellar material. 
\end{itemize}

\begin{figure}
\begin{center}
	\begin{minipage}[t]{0.4\textwidth}
		\epsscale{1.0}
		\plotone{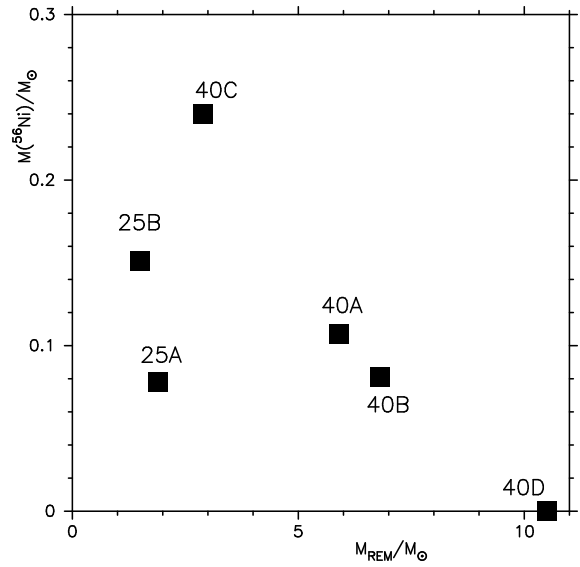}
	\end{minipage}
\end{center}
\caption{The mass of ejected $^{56}$Ni as 
a function of the mass of the central remnant. 
\label{f8}}
\end{figure}

Also, as shown in Figure~\ref{f8}, 
the present models predict anti-correlation between $M$($^{56}$Ni) 
and $M_{\rm REM}$ (thus correlation between $M$($^{56}$Ni) 
and $M_{\rm ej}$) for a given progenitor. 
For larger $M_{\rm REM}$, the mass available to 
$^{56}$Ni production is smaller. 
In addition, larger $M_{\rm REM}$ corresponds to longer time scale 
for the energy injection, i.e., longer $t_{\rm f}$. 
Both effects suppress the production of $^{56}$Ni. 

\begin{figure}
\begin{center}
	\begin{minipage}[t]{0.4\textwidth}
		\epsscale{1.0}
		\plotone{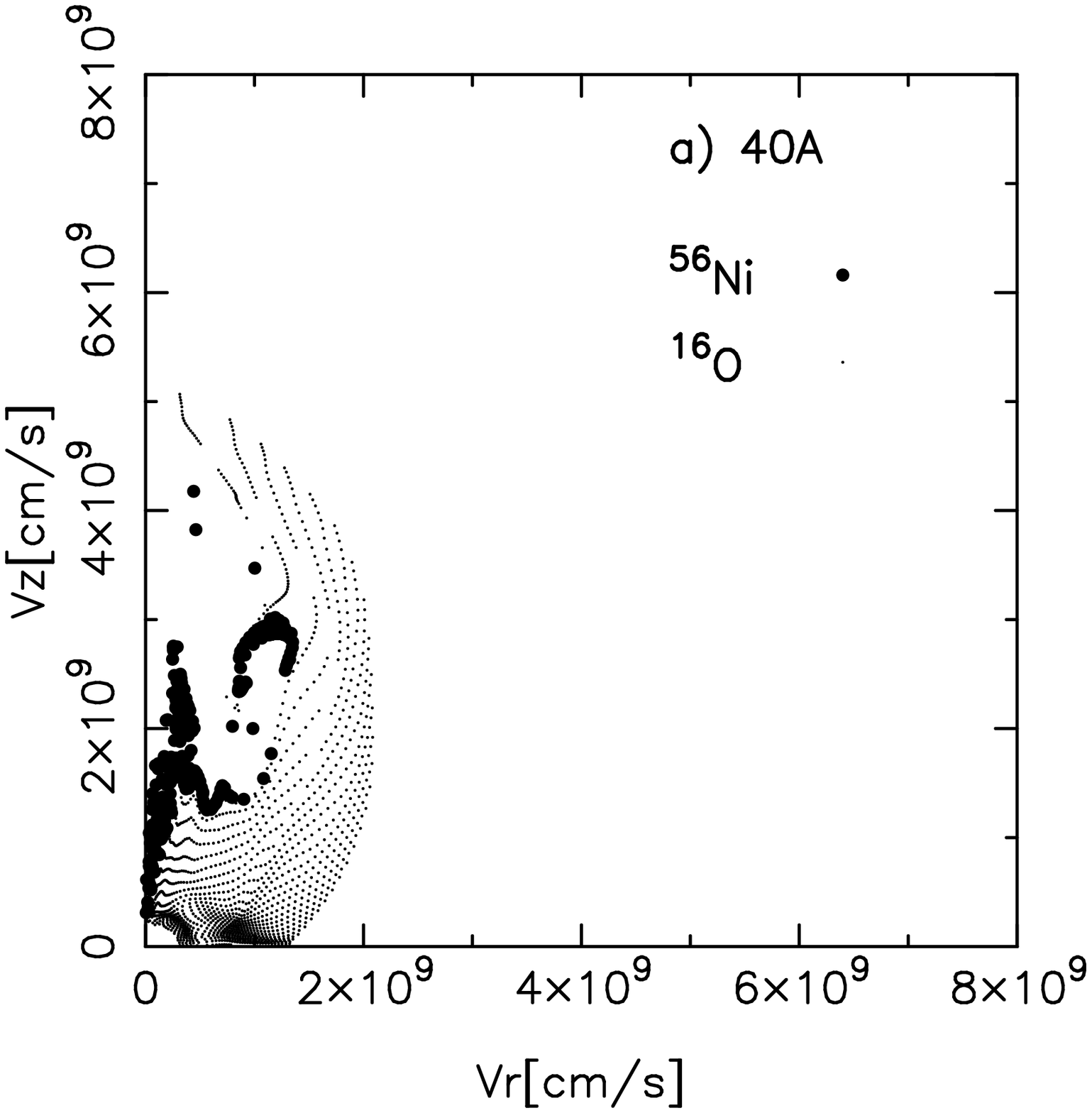}
	\end{minipage}
	\begin{minipage}[t]{0.4\textwidth}
		\epsscale{1.0}
		\plotone{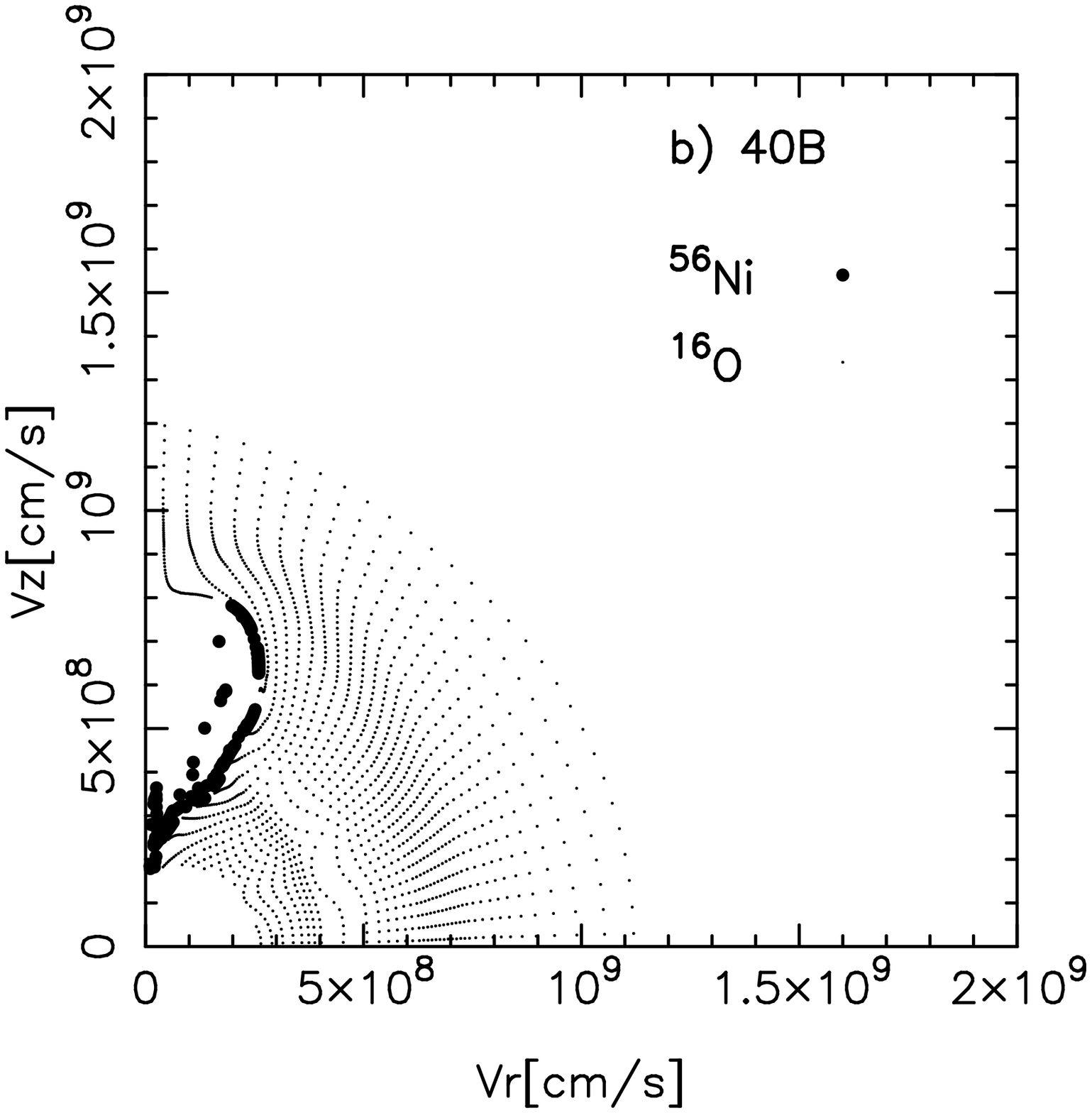}
	\end{minipage}
	\begin{minipage}[t]{0.4\textwidth}
		\epsscale{1.0}
		\plotone{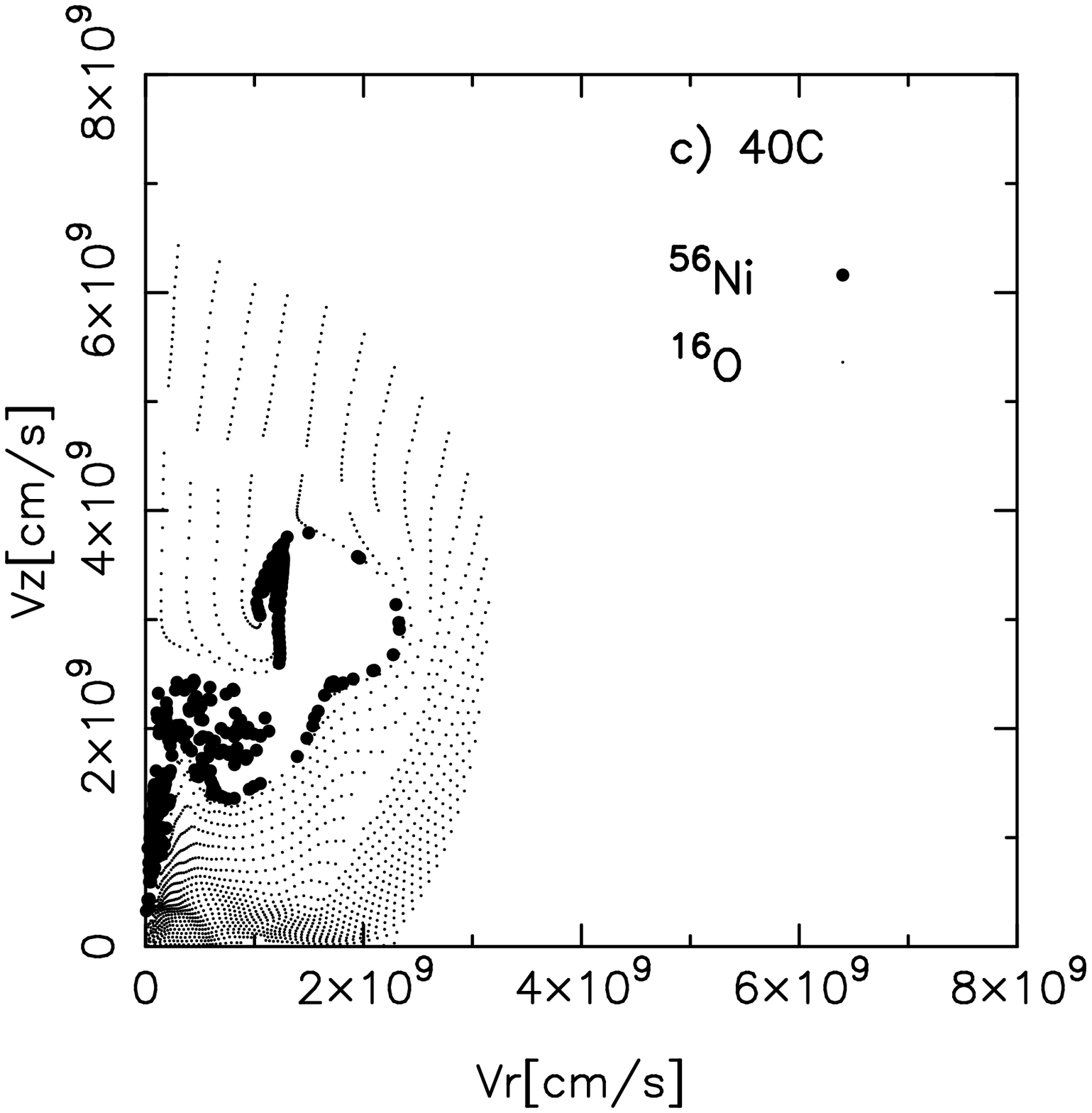}
	\end{minipage}
	\begin{minipage}[t]{0.4\textwidth}
		\epsscale{1.0}
		\plotone{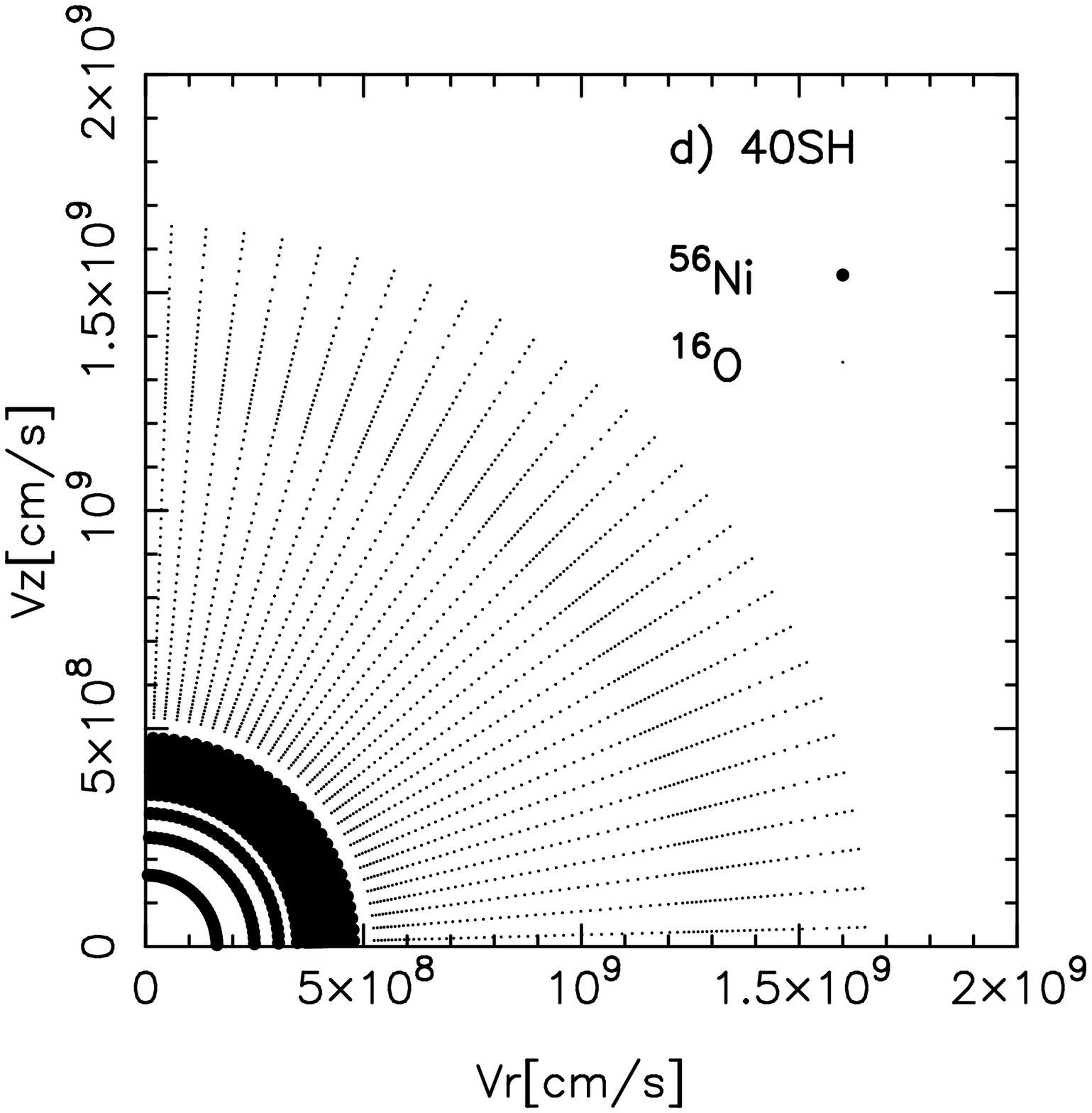}
	\end{minipage}
\end{center}
\vspace{2cm}
\caption{Distributions of $^{56}$Ni and $^{16}$O in the velocity space for 
(a) model 40A, (b) 40B, (c) 40C, and (d) 40SH. 
The filled circles and dots, respectively, show the mass elements in which the mass fraction of $^{56}$Ni and $^{16}$O exceeds 0.08. 
\label{f9}}
\end{figure}

Figure~\ref{f9} shows the distribution of $^{56}$Ni 
(which decays into $^{56}$Fe) 
and $^{16}$O at the homologous expansion phase 
for the models with $M_{\rm ZAMS} = 40\msun$ 
(we omit model 40D, because it produces very little amount of $^{56}$Ni). 
Along the jet axis, matter experiences a strong shock, thus being reprocessed 
through complete silicon burning. It produces a large amount of $^{56}$Ni. 
On the other hand, matter continues to accrete from the side. 
When the mass accretion rate becomes low enough, 
matter gets outward momentum large enough to compete with inflow. 
The peak temperatures of those materials are, however, too low to cause significant 
nuclear reactions. Thus their isotopic patterns reflect those of initial composition 
before the gravitational collapse. 
The result is characterized by the high velocity $^{56}$Ni-rich matter along the jet, 
and low velocity $^{16}$O-rich matter near the center. 
Maeda et al. (2002) and Mazzali et al. (2001) discussed that such configuration 
could naturally explain nebular spectra of SN~1998bw. 

The comparison among the models in Figure~\ref{f9} 
reveals how the overall asphericity 
is affected by the property of the jets. 
A larger opening angle ($\theta{\rm jet}$) leads to a less aspherical explosion 
as expected (40A vs. 40B). 
Larger efficiency ($\alpha$) leads to a more energetic and 
a less aspherical explosion (40A vs. 40D). 
The reason for the latter is as follows: 
a stronger shock (larger $\alpha$) 
deposits a larger amount of energy in the surrounding material 
at the working surface 
and causes a stronger lateral expansion. 
These are also obvious in Figure~\ref{f6}, which shows that models 40B and 40C 
keep the accretion rate smaller than 40A, indicating the stronger lateral expansion. 
In addition, final kinetic energies listed in Table 1 support this explanation. 
The final energies of models 40B and 40C 
are smaller than $E_{51} = 10.9 \times (\alpha/0.01)$, 
which is expected if the accretion rate is the same as in model 40A. 
This result is consistent with MacFadyen et al. (2001), 
who examined the interaction between bipolar jets and envelopes of failed supernovae 
by the similar formalism we adopt in this paper. 
We note that in model 40A, the jet material really 
drills the hole along its way up to the stellar surface.

\subsection{Nucleosnthesis}

\begin{figure}
\begin{center}
	\begin{minipage}[t]{0.4\textwidth}
		\epsscale{1.0}
		\plotone{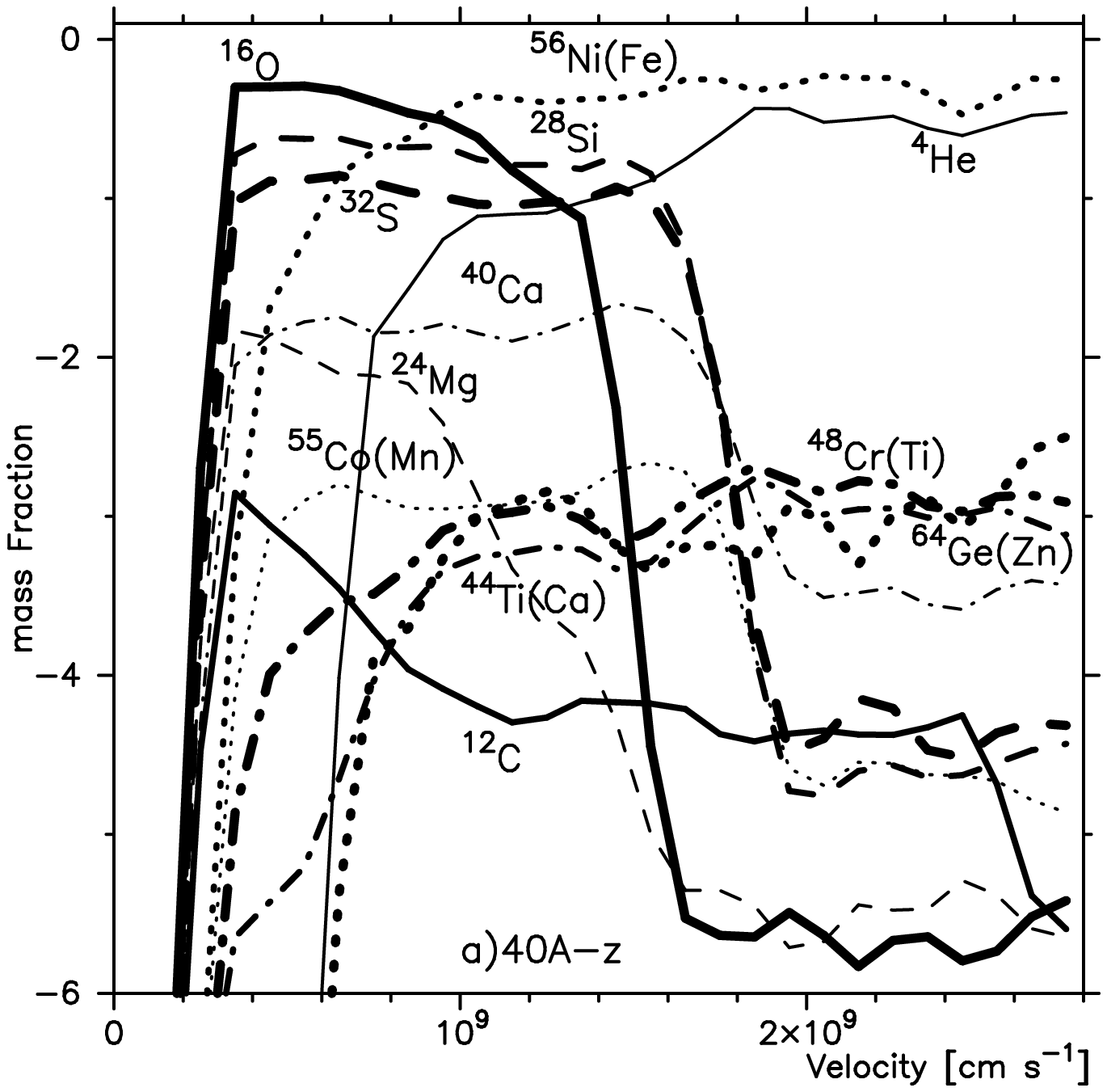}
	\end{minipage}
	\begin{minipage}[t]{0.4\textwidth}
		\epsscale{1.0}
		\plotone{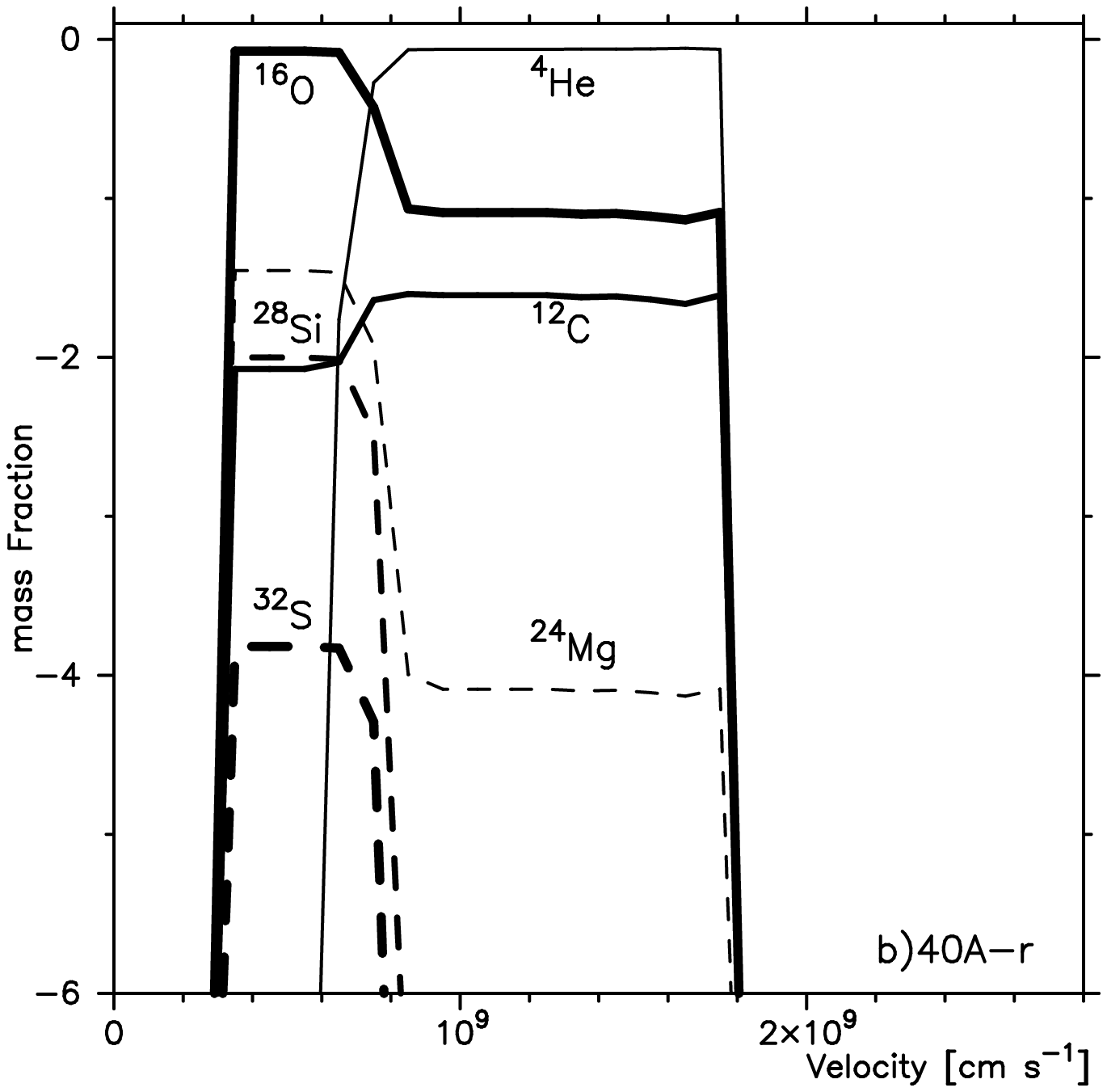}
	\end{minipage}
	\begin{minipage}[t]{0.4\textwidth}
		\epsscale{1.0}
		\plotone{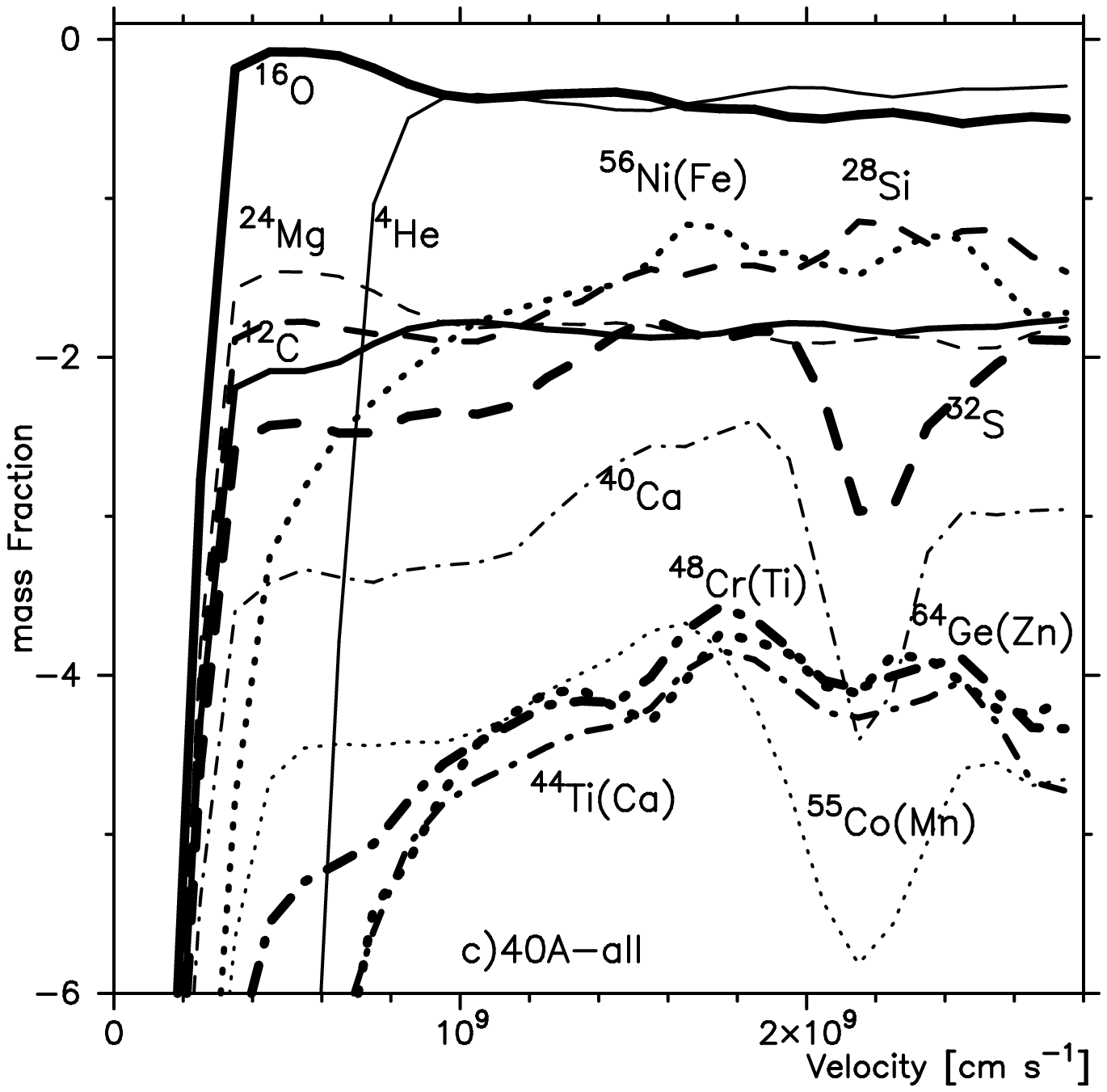}
	\end{minipage}
	\begin{minipage}[t]{0.4\textwidth}
		\epsscale{1.0}
		\plotone{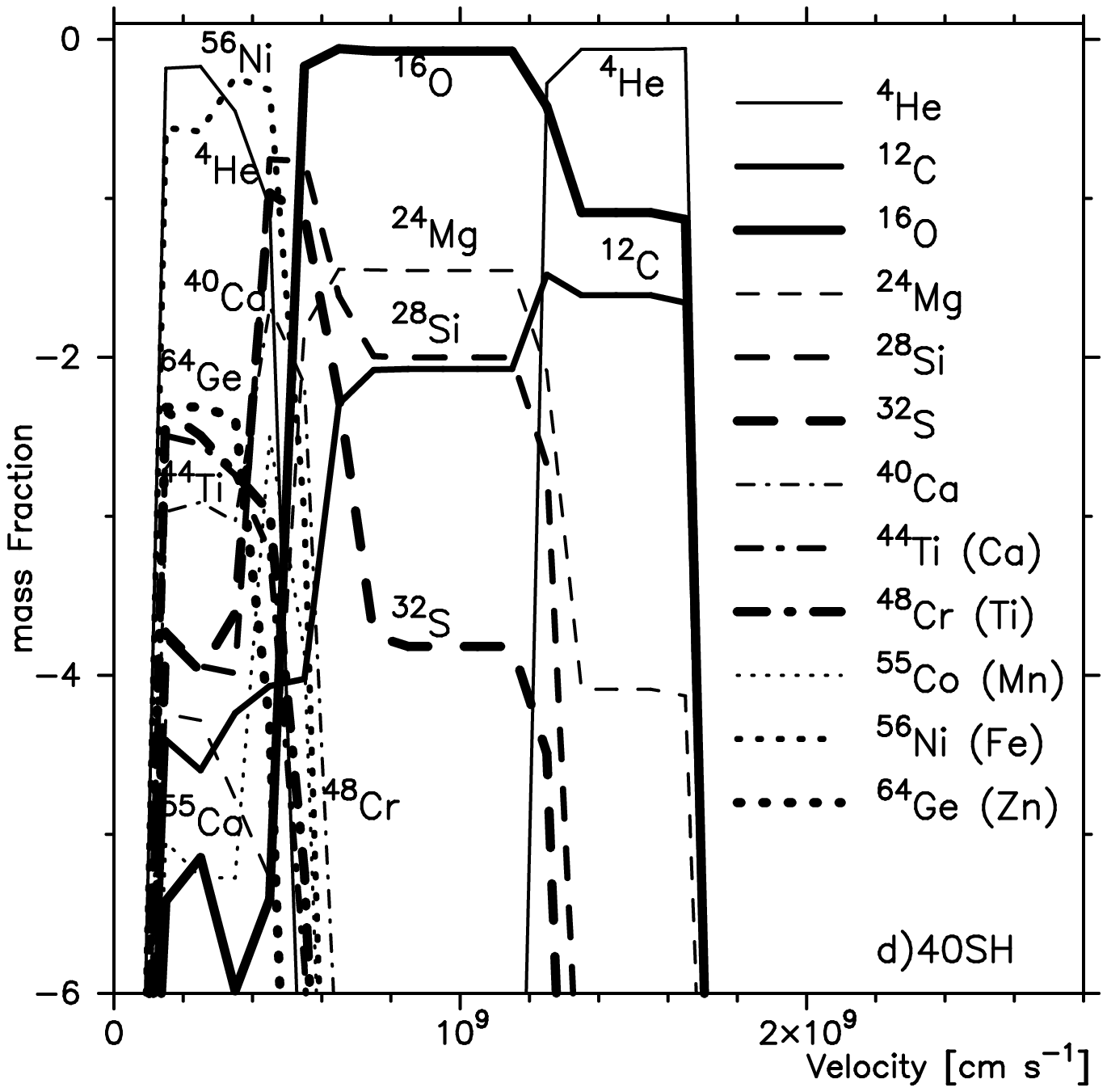}
	\end{minipage}
\end{center}
\caption{Mass fractions of selected isotopes in the velocity space for 
(a) model 40A along the $z$-axis (averaged over $0-15^{\circ}$), 
(b) 40A along the $r$-axis (averaged over $75-90^{\circ}$), 
(c) 40A averaged over all directions, and 
(d) a spherical model 40SH.
\label{f10}} 
\end{figure}

In this section, we discuss further details 
of the neucleosynthetic feature, 
i.e., the isotopic composition of the bipolar explosions. 
The isotopic abundance patterns of model 40A 
along the $z$-axis and the $r$-axis, 
as well as that averaged over all directions are shown in Figure~\ref{f10}. 
The abundance pattern of the spherical model 40SH is also shown. 

In explosive nucleosynthesis in a supernova explosion, 
the isotopic ratios depend on the peak temperature 
attained at the passing of the shock wave as follows 
(e.g., Thielemann, Nomoto, \& Hashimoto. 1996): 

\begin{itemize}
\item[(1)]
The regions with $T_9 > 5$ undergo 
complete silicon burning and $\alpha$-rich freezeout 
to produce $^{56}$Ni abundantly, as discussed in the previous section. 
Higher temperatures (or lower densities at given $T_9$) 
result in stronger $\alpha$-rich freezeout, 
thus producing $^{44}$Ti, $^{64}$Ge, and $^{59}$Cu more abundantly. 
\item[(2)]
The regions with $4 < T_9 < 5$ undergo incomplete silicon burning. 
Beside the dominant fuel nuclei $^{28}$Si and $^{32}$S, 
the burning produces $^{40}$Ca and some iron-peak isotopes 
including $^{55}$Co. 
\item[(3)]
At $3< T_9 < 4$, explosive oxygen burning takes place. 
$^{28}$Si, $^{32}$S, $^{36}$Ar, and $^{40}$Ca are abundantly produced 
at the consumption of $^{16}$O. 
\item[(4)]
At $T_9 < 3$, explosive nucleosynthesis does not proceed so much. 
Though a fraction of fuel nuclei, mostly $^{12}$C, 
are converted to heavier nuclei by carbon burning, 
this does not significantly change 
the isotopic composition. 
Thus, the isotopic composition of this region is almost the same as in 
the pre-explosion star, i.e., $^{16}$O, $^{12}$C, and $^{4}$He. 
\end{itemize}

\subsubsection{Spherical Models}

In spherical models, 
higher temperature is attained in inner regions, 
because the temperature after the passing of the shock wave 
decreases as the shock wave moves outward 
($T \propto R^{-3/4} E^{1/4}$ where $R$ is the radius 
of the position of the shock wave; see equations (3) and (4)). 
Thus, spherical models predict in general the following pattern 
in order of increasing radii: 
strong $\alpha$-rich freezeout 
(e.g., $^{44}$Ti (Ca), $^{59}$Cu (Co) , $^{64}$Ge (Zn)), 
complete silicon burning (e.g., $^{56}$Ni (Fe)), 
incomplete silicon burning 
(e.g., $^{40}$Ca, $^{55}$Co (Mn)), 
oxygen burning 
(e.g., $^{28}$Si, $^{32}$S, $^{36}$Ar), and 
no significant burning 
(e.g., $^{4}$He, $^{12}$C, $^{16}$O). 
(The stable isotopes after $\beta$-decays are 
indicated in parenthesis.) 
This feature in spherical models is evident in Figure 10d. 

We are interested in the isotopic yields in the whole ejecta, 
as these are responsible for the Galactic chemical evolution. 
In conventional spherical models, the yields 
depend on the 'mass cut', which divides 
the stellar material into those finally ejected 
and accreted onto a central remnant. 
We show how the resultant yields depend on $M_{\rm cut}$ 
(the mass contained below the mass cut) 
in spherical models, 
to clarify the deference between spherical models and the bipolar models 
in the later part. 

\begin{figure}
\begin{center}
	\begin{minipage}[t]{0.8\textwidth}
		\epsscale{1.0}
		\plotone{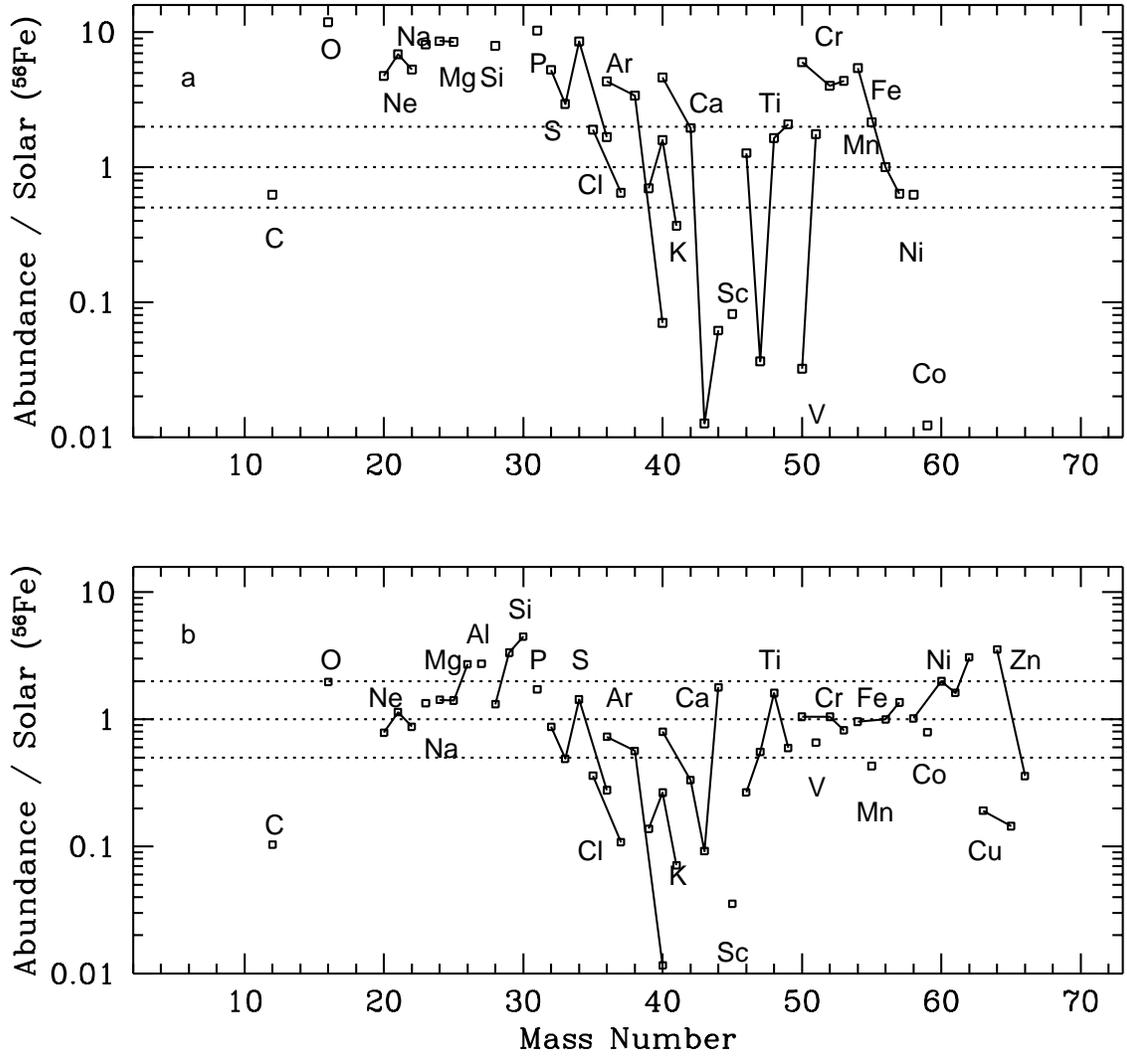}
	\end{minipage}
\end{center}
\caption{Isotopic yields of model the spherical model 40SH 
for different mass cuts $M_{\rm cut}$. 
Larger and smaller $M_{\rm cut}$ lead to 
(a) $M$($^{56}$Ni) $= 0.1\msun$ and (b) $0.54\msun$, respectively.
\label{f11}}
\end{figure}

Figure~\ref{f11} shows the isotopic patterns of spherical model 40SH 
for different mass cuts. 
The isotopic distribution in the ejecta of model 40SH 
shown in Figure 10d explains the behavior. 
For large $M_{\rm cut}$, $M$($^{56}$Ni) is small and the isotopes 
produced in the deepest regions such as $^{64}$Zn (produced as $^{64}$Ge) 
and $^{59}$Co (produced as $^{59}$Cu) are not ejected (Figure 11a). 
For small $M_{\rm cut}$, on the contrary, a large amount of $^{64}$Zn 
and $^{59}$Co are ejected (Figure 11b). 
The isotopes with mass numbers $A \lsim 40$ are 
produced mainly in the layers well above the mass cut, 
so that their abundances do not depend on the mass cut.
For these elements ($A \lsim 40$), therefore, 
the abundances normalized by iron, 
i.e., $\rm{[X/Fe]} = \log (X/Fe) - \log (X/Fe)_{\odot}$, 
are smaller for larger $M$ ($^{56}$Ni). 
The spherical models thus predict that the ejecta with smaller $M_{\rm cut}$ 
has larger [(Zn, Co)/Fe] and smaller [(O, Mg, Mn)/Fe] 
(the latter because of larger $M$($^{56}$Ni)), 
unless the mixing and fallback process takes place (Umeda \& Nomoto 2002). 
Such a correlation is not the case 
for the bipolar models because of the following reason. 

\subsubsection{Bipolar Models}
\begin{figure}
\begin{center}
	\begin{minipage}[t]{0.4\textwidth}
		\epsscale{1.0}
		\plotone{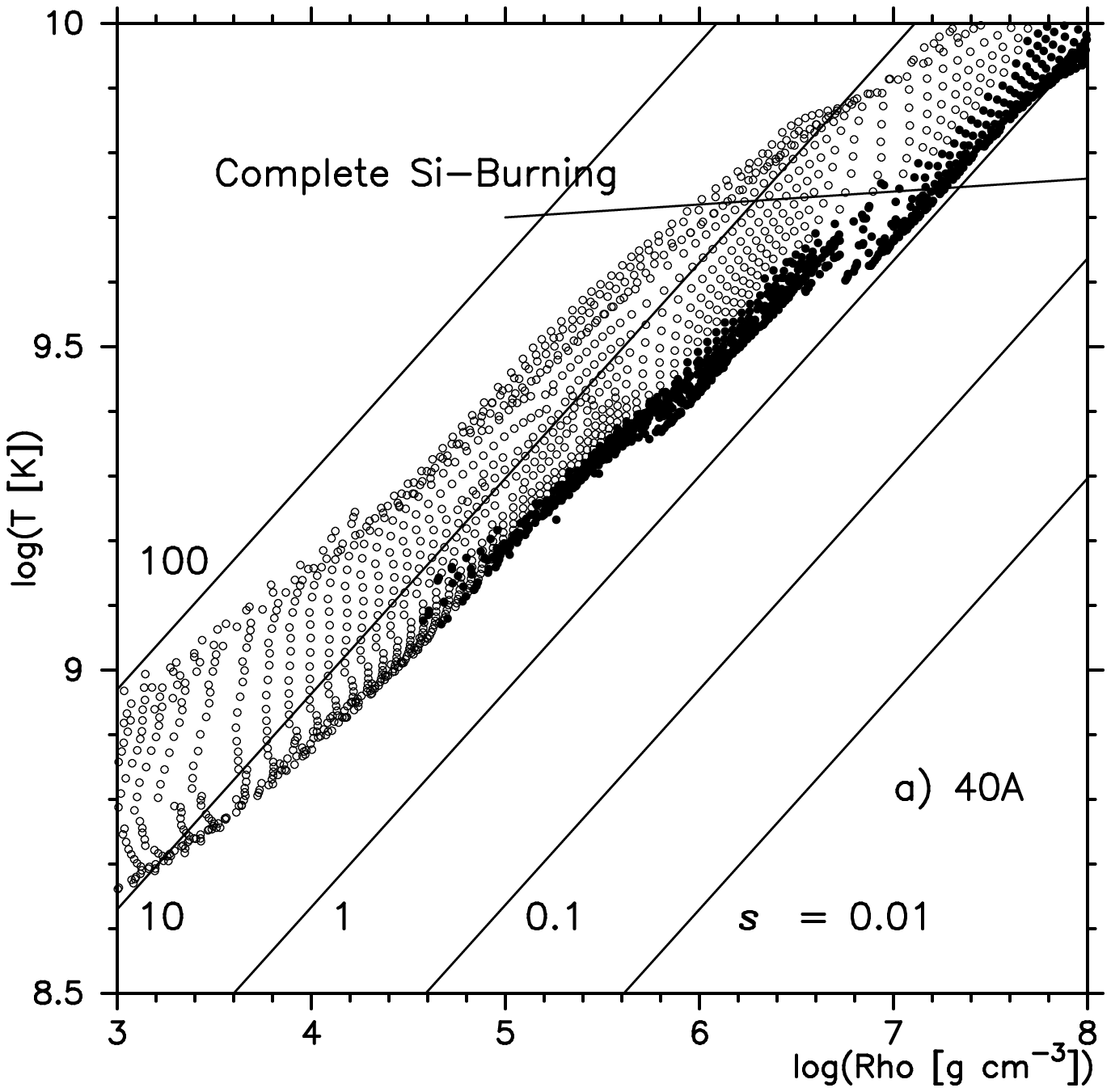}
	\end{minipage}
	\begin{minipage}[t]{0.4\textwidth}
		\epsscale{1.0}
		\plotone{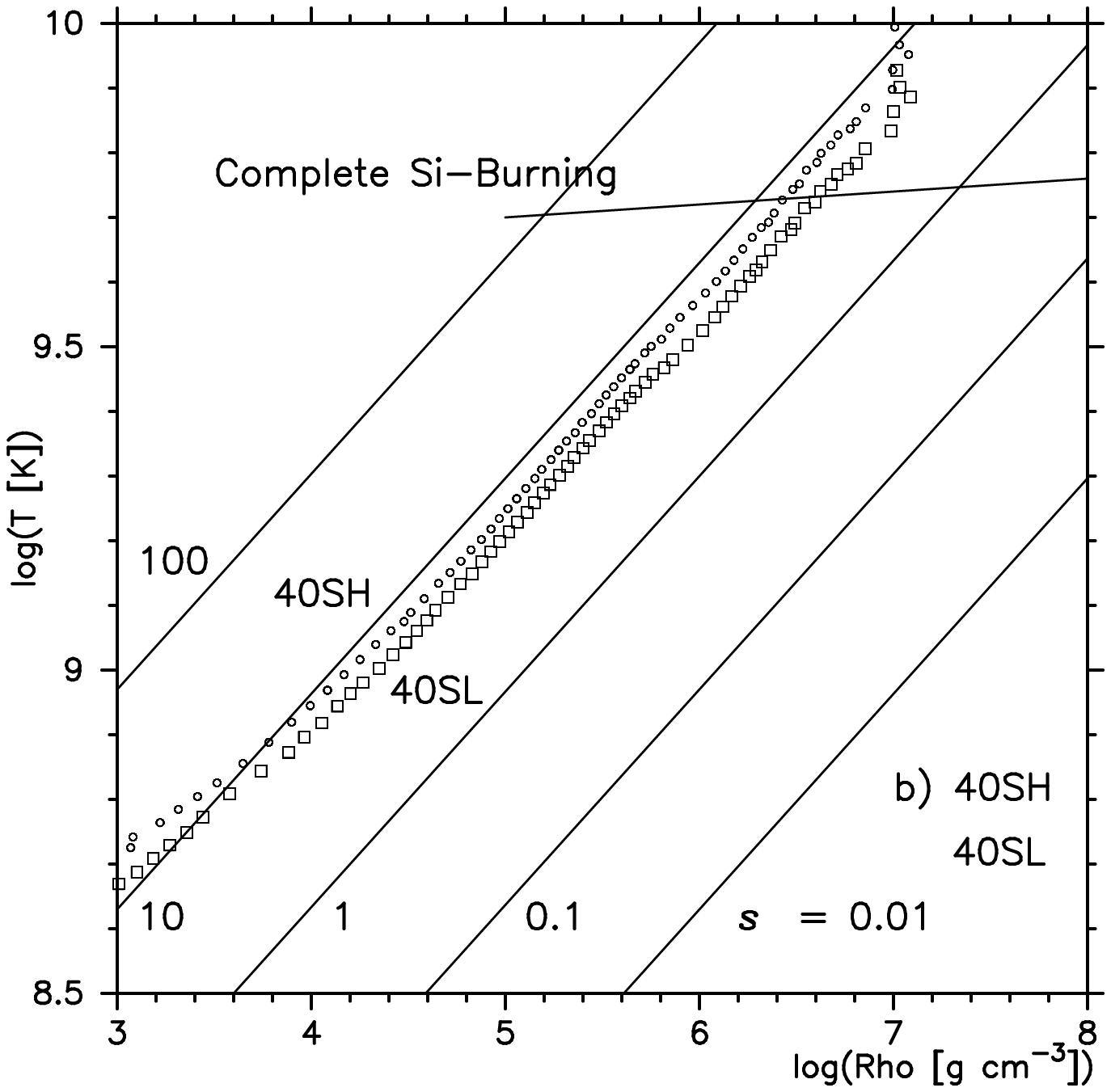}
	\end{minipage}
\end{center}
\caption{Peak temperatures of individual mass elements plotted 
against densities at the temperature maximum, 
for (a) model 40A and (b) the spherical models 40SH and 40SL. 
(a) The open circles and the filled circles denote the mass elements 
which are finally ejected and accreted, respectively. 
(b) The open circles and the squares denotes the mass elements for 
40SH ($E_{51} = 10$) and 40SL ($E_{51} = 1$), respectively. 
Entropies, $s \equiv S_{\gamma}/(k_B/m_u) 
= 4 a T^3 / 3 \rho /(k_B/m_u) 
\sim 0.12 T_9^3/\rho_6$, are shown (solid lines). 
Here $k_B$ and $m_u$ are Boltzmann constant and atomic unit mass, respectively. 
\label{f12}}
\end{figure}

Figure~\ref{f12} shows the peak temperatures of individual mass particles 
against the densities at the passing of the shock wave. 
At a given density, the peak temperatures spread over a wide range, 
in contrast to the single line in the $\rho - T$ plane 
in spherically symmetric models. 

Along the z-axis, the stronger shock heats up the stellar material 
to higher temperatures. 
Along the r-axis, temperatures are lower 
because of the weaker shock 
and densities are higher because of mass accretion. 
Therefore, the materials along the z-axis occupy the high entropy region 
in the $\rho - T$ plane, while those in the r-axis form the lowest bound 
of entropy. 
As a result, the paek temperatures distribute in a wide range 
for a given density. 
The variation in temperatures in model 40A is much larger 
than the deference between the spherical models 40SH ($E_{51}=10$) 
and 40SL ($E_{51}$=1) (Fig.~\ref{f12}).

Compared with the spherical model with the same $E$ (40SH), 
the entropy of material along the $z$-axis of model 40A is higher 
for the same density at $\rho \lsim 10^7$ g cm$^{-3}$. 
At $\rho \gsim 10^7$ g cm$^{-3}$, the entropy along the $z$-axis of model 40A 
is as low as that in the spherical model 40SH. 
This is because the energy injection 
by the jets in model 40A is not completed when the shock reaches 
the corresponding radius, so that the rise in the temperature is suppressed as 
discussed in the previous section.

In hydrodynamics, materials along the r-axis with lower temperature 
accrete onto the center, 
while materials undergoing higher temperatures 
are ejected along the z-axis. 
Thus, the bipolar explosions eject preferentially higher temperature matter 
as shown in Figure~\ref{f12}. 

This can also be seen in Figure~\ref{f10}. 
Along the $z$-axis, matter experiences strong $\alpha$-rich freezeout 
to synthesize $^{64}$Ge, $^{44}$Ti, $^{56}$Ni and is brought up to the surface. 
Along somewhat off-axis, matter experiences lower $T_9$, 
thus undergoing incomplete silicon burning to produce $^{55}$Co 
and oxygen burning to produce $^{28}$Si. 
The latter are heated by the bow shock expanding sideways and flows into the side.  
As the energy deposited is smaller than along the $z$-axis, 
some of them experience circulation and flow into behind 
the $\alpha$-rich freezeout isotopes. 

The matter aside the jets moves toward the center, which forms the oxygen-rich 
high density core (section 3.1). This O-rich core is seen in both plots 
along the $z$-axis and along the $r$-axis. 
Thus, the result is the inversion of the velocities among various isotopes along 
the $z$-axis, as compared with spherical models: 
$^{64}$Zn, $^{56}$Ni, and $^{44}$Ti at the highest velocity, 
$^{55}$Co and $^{28}$Si at the intermediate velocity, 
and $^{16}$O at the lowest velocity. 
This is the case not only along the $z$-axis, 
but also for the average over all the direction (Fig. 10c). 
This feature is due to the action between the jets and gravity modeled 
in this study, and was not examined to this extent in the simpler models 
in the previous studies (Nagataki 2000; Maeda et al. 2002).

\begin{figure}
\begin{center}
	\begin{minipage}[t]{0.4\textwidth}
		\epsscale{1.0}
		\plotone{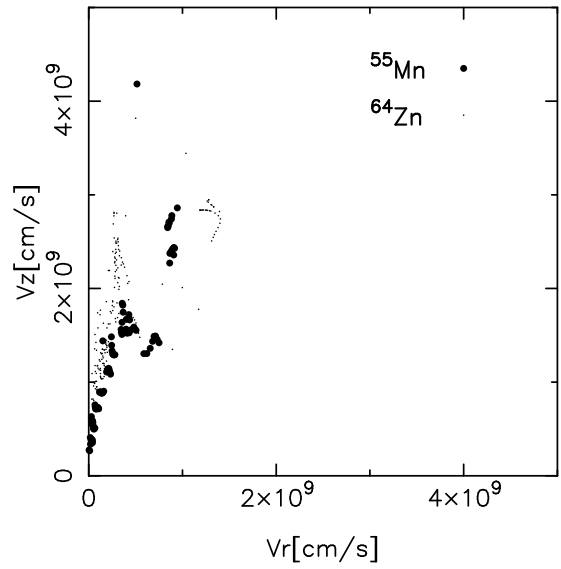}
	\end{minipage}
\end{center}
\vspace{3cm}
\caption{Distribution of $^{55}$Mn and $^{64}$Zn in the velocity space 
for model 40A. 
The filled circles and dots, respectively, show the mass elements 
in which the mass fraction of $^{55}$Mn and $^{64}$Zn exceeds $10^{-4}$. 
\label{f13}}
\end{figure}

Figure~\ref{f13} demonstrates that 
$^{64}$Zn is ejected with higher velocities than $^{55}$Mn. 
Here Zn represents the products of 
strong $\alpha$-rich freezeout in complete Si burning, 
while Mn represents the products of incomplete Si burning. 
The velocity inversion between these elements is obvious. 
The bipolar model predicts that Zn is ejected with higher velocities 
than Mn, in contrast to spherical models.

\begin{figure}
\begin{center}
	\begin{minipage}[t]{0.8\textwidth}
		\epsscale{1.0}
		\plotone{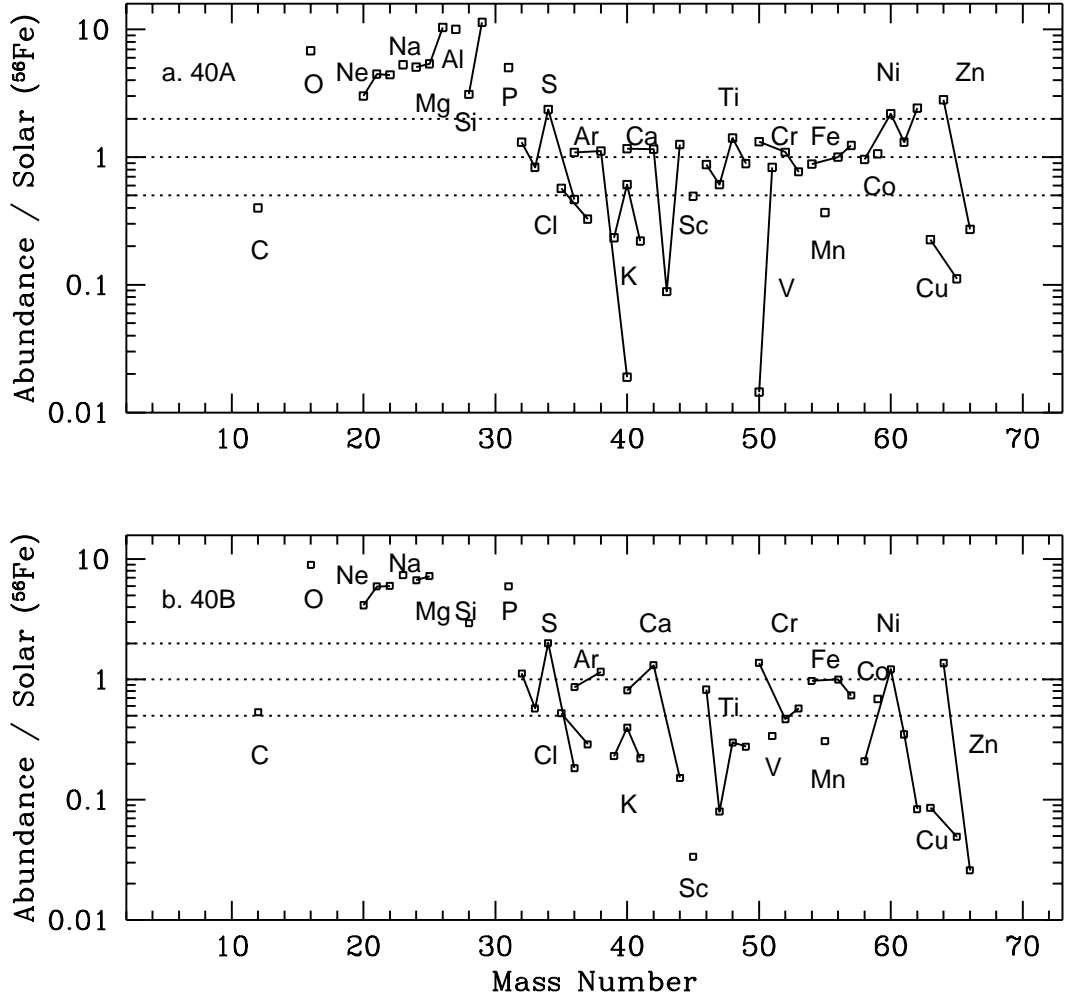}
	\end{minipage}
\end{center}
\caption{Isotopic yields of the bipolar models (a) 40A, (b) 40B, 
(c) 40C, (d) 40D, (e) 25A, and (f) 25B.
\label{f14}} 
\end{figure}

The above features are seen in the isotopic yields of the bipolar models 
shown in Figure~\ref{f14}, Tables 2-4 
(after $\beta$-decays), 
Tables 5-7 (before $\beta$-decays), 
and Table 8 (Radioactive isotopes). 
Comparison between 40A and 40SH (spherical) reveals 
the following characteristics in the bipolar models: 
\begin{itemize}
\item[(1)]
Production of heavy isotopes with $A \gsim 60$, 
especially $^{64}$Zn and $^{59}$Co, is enhanced relative to $^{56}$Fe. 
This is attributed to the high entropy of matter ejected along the z-axis. 
Comparison between Figures~\ref{f11} and~\ref{f14} shows that the yields 
of these isotopes relative to $^{56}$Fe ($^{56}$Ni) in 40A are similar 
to those of 40SH with large $M$ ($^{56}$Ni) (i.e., small $M_{\rm cut}$). 
Both models are dominated by the material 
which experiences strong $\alpha$-rich freezeout. 
For the same reason, $^{55}$Mn/$^{56}$Fe ratio is suppressed. 
\item[(2)]
On the other hand, model 40A produces a relatively small 
amount of $^{56}$Ni compared to 40SH with small $M_{\rm cut}$. 
As a result, 
the ratios [($^{16}$O, $^{24}$Mg)/$^{56}$Fe] are much larger in 40A. 
These ratios in model 40A are close to those in 
40SH with small $^{56}$Ni (large $M_{\rm cut}$). 
\end{itemize}

\subsubsection{$\alpha$-rich freezeout and 
complete silicon burning products} 

Regarding the ejection of isotopes produced in strong $\alpha$-rich 
freezeout, the masses of $^{44}$Ti and $^{64}$Ge 
(which decay into $^{44}$Ca and $^{64}$Zn, respectively) 
are shown against $M$($^{56}$Ni) in Figure~\ref{f15}. 
Because the bipolar models preferentially eject 
the materials experiencing higher temperatures (higher entropies) 
in complete silicon burning, the ratios 
$^{44}$Ti/$^{56}$Ni and $^{64}$Ge/$^{56}$Ni are significantly 
larger in the bipolar models than spherical models 
(except for model 25B, as this is a very weak explosion 
with $E_{51} =0.6$). 

The previous study on aspherical explosions by 
Maeda et al (2002) attained $M$($^{44}$Ti) $\sim 1.6 \times 10^{-3}\msun$ 
and $M$($^{64}$Ge) $\sim 1.6 \times 10^{-4}\msun$ for 
$M$($^{56}$Ni) $\sim 0.4\msun$ (their model C), 
thus $^{44}$Ti/$^{56}$Ni and $^{64}$Ge/$^{56}$Ni relative to the solar values 
are $\sim 3.4$ and $\sim 0.5$, respectively. 

In the present models, $M$($^{44}$Ti), $M$($^{56}$Ni), 
and the ratio $^{44}$Ti/$^{56}$Ni are smaller 
than in Maeda et al. (2002). 
The smaller $^{44}$Ti/$^{56}$Ni is mainly attributed to 
the time scale of the energy injection as discussed 
in the previous section: 
the present models are driven by jets which are injected over a few seconds 
(as they are powered by accretion), while 
Maeda et al. (2002) assumed prompt aspherical explosions. 
Because the continuous energy injection leads to lowering the post-shock 
temperature (section 3.2), the production of $^{44}$Ti 
is suppressed as compared to the prompt explosions.

For $^{64}$Ge (which decays into $^{64}$Zn), the present models 
predict much larger $^{64}$Ge/$^{56}$Ni ratio than Maeda et al (2002). 
This mainly stems from the difference in $Y_{e}$ in the deepest region: 
In the present study, we assumed $Y_{e} = 0.5$ 
down to the mass cut. 
Because the production of $^{64}$Ge is very sensitive to $Y_{e}$ 
and most abundantly produced with $Y_{e} = 0.5$ as this isotope has 
the same number of protons and neutrons, 
this choice results in a large amount of $^{64}$Ge 
(Umeda \& Nomoto 2002). 

In spherical models without any mixing, ejecta with large ratios of 
($^{44}$Ti, $^{64}$Ge)/$^{56}$Ni relative to the solar ratios should 
contain such a large amount of $^{56}$Ni as 
$M$($^{56}$Ni) $\gsim 0.4\msun$.
In contrast, bipolar models predict large ratios of 
$M$($\alpha$-rich freezeout)/$M$($^{56}$Ni) relative to the solar values 
in ejecting relatively small amount of $^{56}$Ni ($\sim 0.1 - 0.2\msun$). 
This feature may have significant implication for the abundances in extremely 
metal-poor stars, as discussed in the coming section. 

\subsubsection{Incomplete silicon burning products}

Incomplete silicon burning in 
the bipolar models shows different characteristics 
as compared to spherical models.  
Figure~\ref{f16} exhibits the ratios, [Ti/Ca], [Ca/Si], 
[S/Si], [Si/O], and [C/O] (where $[{\rm X/Y}] \equiv 
\log({\rm X/Y})) - \log({\rm X/Y})_{\odot}$) 
as a function of the remnant's mass ($M_{\rm REM}$). 

Ti and Ca are dominated by $^{48}$Ti and $^{40}$Ca, respectively, 
which are produced by incomplete silicon burning. 
$^{48}$Ti is a decay product of $^{48}$Cr whose synthesis 
requires higher temperature than $^{40}$Ca. 
This is analogous to the relation between 
strong $\alpha$-rich freezeout products and $^{56}$Ni. 
Thus, the trend of Ti/Ca ratio is expected to be the same 
as the trends of ($^{44}$Ti, $^{64}$Ge)/$^{56}$Ni (\S 3.3.3).

Figure~\ref{f16} shows that 
the Ti/Ca ratio is basically larger for smaller $M_{\rm REM}$, 
and thus for larger $^{56}$Ni (Figure~\ref{f8}), 
with the exception of the very weak explosion model 25B. 
This relation and the variation of [Ti/Ca] as a function of $M_{\rm REM}$ 
are qualitatively similar to those seen in spherical models. 
However, we should note that 
$M_{\rm REM}$ varies in a much wider range ($1.5 - 11\msun$) 
in the bipolar models than in the spherical models ($<3\msun$). 

The trend in [Ca/Si] is similar to [Ti/Ca] for the same reason. 
For large $M_{\rm REM}$, however, 
the difference between [Ti/Ca] and [Ca/Si] becomes evident 
As ejected Si is produced both before and during the explosion, 
the behavior of the element ratios with respect to Si is complicated 
as will be discussed in the next subsection. 

\subsubsection{Oxygen burning and hydrostatic burning products}

S, Si, O, Mg, and C are produced in hydrostatic burning 
before the explosion. 
During the explosion 
S and Si are also produced by explosive oxygen burning. 
In spherical models, all these elements are produced 
well above the mass cut, so that their abundance ratios  
are basically independent of $M_{\rm cut}$ 
as is evident in Figure~\ref{f16}. 
However, this is not the case for the bipolar models, 
as shown by the large variations of these ratios in Figure~\ref{f16}.

First, the abundance ratios depend on the progenitor's mass 
as follows:
\begin{itemize}
\item[(1)]
For $M_{\rm ZAMS} = 40\msun$, these ratios show large variations. 
The difference from the spherical models is larger 
for larger $M_{\rm REM}$. 
\item[(2)]
For $M_{\rm ZAMS} = 25\msun$, the bipolar models 25A and 25B 
produce similar ratios to the spherical model 25S. 
\end{itemize}
The accretion of heavy elements onto the central remnant 
is responsible for the variation of these abundance ratios. 
Since the final remnant mass, $M_{\rm REM}$, 
is a measure of the amount of accreted material, 
accretion yields the correlation between these ratios and $M_{\rm REM}$. 
Such a correlation is more pronounced in more massive stars 
because $M_{\rm REM}$ takes a larger range of values.

\begin{itemize}
\item[(3)]
Si, S: In spherical models, Si and S in presupernova stars 
are mostly burned into Fe-peak elements during the explosion. 
However, these elements are newly synthesized by explosive O-burning 
and ejected with [Si/S] $\sim 0$. 
In the bipolar models, the silicon layer aside the jets are 
ejected without undergoing significant explosive burning. 
The ejected masses of Si and S produced during the explosion 
are comparable to those produced before the explosion. 

In the presupernova star, 
S is more centrally concentrated than Si, 
thus accreting more easily onto the central remnant. 
This yields smaller [S/Si] for larger $M_{\rm REM}$. 
This might explain the observed low S/Si ratio in M87 
(Matsushita, Finoguenov, \& B\"ohringer, 2003). 
As Si and S are produced in the deeper region than O, 
larger $M_{\rm REM}$ means a larger amount of Si and S 
in the accreted matter, thus leading to smaller [(S, Si)/O]. 

In model 40D, however, [Si/O] is larger than that in models 40A and 
40B with smaller $M_{\rm REM}$. 
For very large $M_{\rm REM}$ like in 40D, 
Si in the presupernova star is totally 
accreted onto the center, thus Si ejected are dominated by 
the explosive burning products. 
The amount of Si, therefore, does not decrease with larger $M_{\rm REM}$ anymore. 
The amount of O, however, even decreases with larger $M_{\rm REM}$ as 
O is distributed in the progenitor star in the wide range. 
Thus, for model 40D, [Si/O] is large.

\item[(4)]
O, C: 
The carbon-rich layer is located above the oxygen-rich layer, 
so that O more easily accretes onto the central remnant. 
Therefore, larger [C/O] results from larger $M_{\rm REM}$. 
\end{itemize}

\section{Implications on Abundances in Extremely Metal-Poor Stars}

Figure~\ref{f17} shows [(Si, S, Sc, Ti, Cr, Mn, Co, Zn)/Fe] 
against [Mg/Fe] for the bipolar models and 
the spherical models (with different $M_{\rm cut}$). 
Figure~\ref{f17} clearly shows that the yields of the 
bipolar models are different from spherical models, irrespective 
of the choice of $M_{\rm cut}$. 
The features of the bipolar explosions are summarized as follows. 
\begin{itemize}
\item[(1)]
Si, S: 
[(Si, S)/Fe] are smaller than in spherical models, 
because of the substantial loss of presupernova Si and S by accretion. 
This is evident in more massive progenitors with 
$M_{\rm ZAMS} = 40\msun$, since a more massive star undergoes 
a larger amount of mass accretion. 
The difference between the bipolar models and the spherical models 
Is larger for [S/Fe] than [Si/Fe], 
especially for larger $M_{\rm REM}$ 
(i.e., larger [Mg/Fe]), 
because of more centrally concentrated distribution of S than Si. 
\item[(2)]
Sc, Ti: 
For explosions with larger energies and asphericity (models 40A, 
40C, and 25A), [(Sc, Ti)/Fe] are larger 
compared to the less energetic and spherical models, because 
the production of Sc and Ti is enhanced at strong $\alpha$-rich freezeout, 
i.e., high entropy condition, in complete Si burning. 
The enhancement is particularly strong for Sc, as 
its production of is very sensitive to the entropy and 
needs very low density (i.e., high entropy). 
For 40A, 40C, and 25A [Sc/Fe] $\> 0$ is reached $\gsim 0.0$, 
while [Sc/Fe] $\lsim -1.0$ for spherical models irrespective 
of $M_{\rm cut}$. 
The caveat is that the production of Sc is very sensitive to the 
numerical resolution employed (see Appendix), though qualitatively 
enhancement of [Sc/Fe] is found irrespective of the numerical resolution. 
Future higher resolution studies are needed to quantitatively 
confirm the above result.

\item[(3)]
Cr, Mn: 
For a given [Mg/Fe], [(Cr, Mn)/Fe] in the bipolr models 
are much smaller than in the spherical models. 
Cr and Mn are mainly produced in the incomplete silicon burning region, i.e., 
at lower peak-temperatures than complete silicon burning which produces Fe. 
Since the bipolar explosions 
preferentially eject higher temperature matter, 
[(Cr, Mn)/Fe] are smaller than in spherical models. 
\item[(4)]
Co, Zn: 
For a given [Mg/Fe], [(Co, Zn)/Fe] in the bipolar models 
are much larger than in the spherical models. 
Because these ratios are larger at higher entropy (Umeda \& Nomoto, 2002), and 
and the ejection of higher entropy matter is enhanced in 
the bipolar models, [(Co, Zn)/Fe] are significantly enhanced 
in the bipolar models. 
\end{itemize}

The bipolar supernova/hypernova explosions may have significantly contributed 
to the early Galactic chemical evolution. 
Abundances in extremely metal-poor stars provide us with important clues to know 
what were the first supernova explosions and 
what were the first stars in Galaxy.  

At early phases, the Galaxy was not chemically well-mixed yet. 
Extremely metal-poor stars store the chemical information 
of an individual (core-collapse-induced) 
supernova, because those stars are likely to have been contaminated 
by only a single supernova 
(Audouze \& Silk 1995; Shigeyama \& Tsujimoto 1998). 
This expectation has led to many attempts to 
compare abundances in extremely metal-poor stars with theoretical 
supernova yields (e.g., Umeda \& Nomoto 2002). 

In this section, we compare the yields of the bipolar models and 
the abundances in extremely metal-poor stars. 
It should be taken in mind that the models presented here are 
solar metalicity models (see section 2). 
Therefore, we are mainly concerned in this section with 
explosive burning products and major isotopes of intermediate elements, 
which are little affected by the metalicity of the progenitor models.

\subsection{Iron-Peak Elements}

Observational studies on metal-poor halo stars have shown 
that there exist interesting trends in the abundances of iron peak elements 
for [Fe/H] $\lsim -2.5$ 
(McWilliam et al. 1995; Ryan, Norris, \& Beers 1996; 
Primas et al. 2000; Blake et al. 2001). 
Both [Cr/Fe] and [Mn/Fe] decrease toward smaller [Fe/H], 
while [Co/Fe] and [Zn/Fe] increase to reach $\sim 0.3 -0.5$ at [Fe/H] $\sim -3$ 
(Figure~\ref{f18}).  
Also [(O, Mg)/Fe] are as large as $\sim 0.5$ for [Fe/H] $\lsim -2.5$, indicating 
that the amount of Fe was not excessively large in the ejecta of supernovae 
which were responsible for forming those stars 
(see also section 3.2 on the production of $^{56}$Ni which decays into $^{56}$Fe).

The large ratios of [Co/Fe] and [Zn/Fe] have not been predicted in previous 
Nucleosyntheic yields of spherical thermal bomb models. 
As pointed out by Umeda \& Nomoto (2002), 
it is difficult to reproduce these ratios together with [(O, Mg)/Fe] 
$\sim 0.4 - 0.5$. 
Co and Zn are produced in the deepest layer 
of the ejecta through strong $\alpha$-rich freezeout.  
Mg and O, on the other hand, are mostly produced by hydrostatic burning, 
so that these yields are insensitive to the explosion. 
The ejection of a large amount of Zn and Co inevitably leads to the ejection 
of a large amount of Fe (produced as $^{56}$Ni), because Fe 
is synthesized not only in the deep layer with Zn, 
but also in the outer regions with Si. 
This decreases [Mg/Fe] and [O/Fe] (Figure 11b), 
contrary to the observations. 
With [Mg/Fe] $\sim 0.4 - 0.5$, [(Co, Zn)/Fe] $\lsim 0.0$ and even much smaller than 
this for most of the case (Figure 11a).

To reproduce the observed trends of 
increasing [(Zn, Co)/Fe] and decreasing [(Mn, Cr)/Fe] toward smaller [Fe/H], 
nucleosynthesis in the SNe which are responsible for the chemical enrichment of 
small [Fe/H] stars should have the following characteristics. 
\begin{itemize}
\item[(1)]
Matter which undergoes complete Si-burning should be ejected abundantly 
without ejecting much incomplete Si-burning products. 
\item[(2)]
The mass ratio between the complete Si-burning products 
to incomplete Si-burning products 
should be larger at lower metalicities. 
\end{itemize}

One possible model to realize the above nucleosynthesis is the deep mass cut 
(small $M_{\rm cut}$). 
As shown in Figure~\ref{f11}, 
smaller $M_{\rm cut}$ makes [(Zn,Co)/Fe] larger and [(Mn,Cr)/Fe] smaller 
simultaneously. 
Figures 17e-17h also clearly show these trends with varying $M_{\rm cut}$, 
where smaller [Mg/Fe] results from smaller $M_{\rm cut}$. 
In this case, however, [(O, Mg)/Fe] are too small 
because of the presence of a large amount of Fe above the mass cut (Figure 11b). 
Figures 17e-17h show that in the spherical models, 
[Cr/Fe] $\lsim 0.0$, [Mn/Fe] $\lsim -0.5$, and 
[(Co, Zn)/Fe] $\gsim 0.0$ as observed in the very metal-poor stars require 
so deep a mass cut which results in [Mg/Fe] $\lsim 0.2$, contrary 
to the observations. 

We point out that the bipolar models naturally account for these features. 
Large [(Co, Zn)/Fe], small [(Cr, Mn)/Fe], and large [(O, Mg)/Fe] are simultaneously 
realized as seen in Figures~\ref{f14} and~\ref{f17}. 
Compared with the spherical models, 
the bipolar models yield large [(Co, Zn)/Fe] and small [(Cr, Mn)/Fe] 
for given [(Mg, O)/Fe]. 
This is due to the characteristic in hydrodynamics of the bipolar models, 
i.e., the ejection of higher temperature (and higher entropy) matter along the 
$z$-axis and the accretion of lower temperature matter along the $r$-axis. 

If the formation of metal-poor stars was mainly driven by a supernova shock wave, 
[Fe/H] in those stars is approximately determined by the ratio of 
the ejected mass of Fe to 
the amount of interstellar hydrogen swept up by the shock wave 
(Ryan et al.1996). 
Then [Fe/H] of the newly born metal-poor star is expressed as 
\begin{equation}
	{\rm [Fe/H]}  =  
\log_{10}(M({\rm Fe})_{\rm SN}/M({\rm H)}_{\rm SW}) 
-  \log_{10}(X({\rm Fe})/X({\rm H}))_\odot \ .
\end{equation}
Here $M({\rm Fe})_{\rm SN}$ is the mass of ejected iron 
and $M({\rm H})_{\rm SW}$ is the mass of interstellar 
hydrogen swept up by the shock wave. 

According to Shigeyama \& Tsujimoto (1998), $M({\rm H})_{\rm SW}$ can be 
estimated as 
\begin{equation} 
M({\rm H})_{\rm SW} = 5.1 \times 10^4 M_{\odot} (E_{51})^{0.97} 
n_1^{-0.062} (C_{10})^{-9/7} \ , 
\end{equation}
where $n_1$ denotes the number density of hydrogen in 
circumstellar/interstellar medium 
(cm$^{-3}$) and $C_{10}$ is the sound speed in units of 10 km s$^{-1}$. 
The values of [Fe/H] estimated by these formulae are listed in Table 9 
along with the values of [X/Fe].

Assume that the spherical models 40SL and 25S with $M$($^{56}$Ni) $= 0.1\msun$ 
represent normal supernovae, then the bipolar models 40A-C and 25A 
can well account for the abundance trends observed in extremely metal-poor stars. 
These models have larger [(Zn,Co)/Fe], smaller [(Mn,Cr)/Fe], and smaller [Fe/H] 
than those in 40SL and 25S. 
In addition, [Mg/Fe] and [O/Fe] are large enough to be consistent with observed ones.  
The smaller [Fe/H] is due to larger explosion energies and 
thus large $M({\rm H})_{\rm SW}$ (Equation (8)) in these models. 
We emphasize that these good coincidence can not be achieved by simply 
varying a mass cut in spherical models. 
An example is the model 40SH with large $M$($^{56}$Ni), which has the problems of 
too small [(Mg, O)/Fe] as already noted, as well as too large [Fe/H]. 
Figure~\ref{f18} demonstrates that the models 40A and 25A 
are in good agreement with the abundances in extremely metal-poor stars, 
well-reproducing the overabundance of [(Zn, Co)/Fe] 
as well as the trends seen in metal-poor stars. 

\subsection{Intermediate Mass Elements}

Enhancement of [(Sc, Ti)/Fe] is also characteristic of the bipolar models. 
These ratios are sensitive to the asphericity in the explosion. 
For larger asphericity, these ratios are larger,  
which is especially evident for [Sc/Fe]. 
We note that the production of Sc is very sensitive to how the 
detailed jet phenomenology can be resolved in the numerical calculation 
(see Appendix). The enhancement of [Sc/Fe] is probably the case, though 
the detailed value of [Sc/Fe] is to be examined by higher resolution 
simulations.

The enhancement of [Sc/Fe] was not clearly shown in the previous studies on 
bipolar explosions (Nagataki 2000; Maeda et al. 2002). 
For example, Maeda et al. (2002) yields [Si/Fe] $= -0.74$. 
This is not surprising, because these previous works only 
addressed the limited case, i.e., prompt explosions. 

[(Sc, Ti)/Fe] have been problematic in the previous supernova yields 
based on conventional spherical models. 
[(Sc, Ti)/Fe] $\sim 0 - 0.5$ are observed in the very metal-poor stars, 
while those predicted by the models were less than the solar values 
(even much less for [Sc/Fe]). 
The present bipolar models predict [(Sc, Ti)/Fe] $\gsim 0$ 
(for highly aspherical explosions), 
thus being in good agreement with observations 
(though these values, especially of Sc,  
have to be examined by higher resolution calculations. See Appendix).
 
\subsection{Carbon-rich Metal-Poor Stars}

Model 40D produces very small amount of $^{56}$Ni $\sim 10^{-7}\msun$, 
while the explosion is very energetic with $E_{51} \sim 10$. 
Such Fe-poor explosions might have 
caused the formation of a certain class of carbon-rich metal-poor stars, 
e.g., CS22949-037 
(Norris et al. 2001; Aoki et al. 2002; Depagne et al. 2002) 
as postulated by Umeda \& Nomoto (2003). 

Model 40D ejects very little amount of explosive burning products, 
while a large amount of C, O, and Mg are ejected. 
This results in the enhancement of [(C, O, Mg)/Fe] as 
shown in Figure~\ref{f14} and Table 9. 
The enhancement of [(Si, S)/Fe] is smaller because of 
accretion of a large fraction of Si and S. 
Heavier elements, such as iron-peak elements, relative to the solar values 
are not enhanced. 
These features are consistent with the abundance feature of CS22949-037. 
It would also be possible for larger amount of Mg to accrete to become 
[Mg/Fe] $\sim 0$ as postulated in Umeda \& Nomoto (2003) to reproduce 
the observed feature of the most Fe-poor star HE0107-5249 
(Clriestlieb et al. 2002). 

The bipolar models predict other characteristic abundance ratios, 
which can be tested with observations. 
For example, the bipolar models predict 
smaller [S/Si] and [O/C] ratios for Fe-poor stars 
as discussed in section 3.3. 

Also, in the context of the bipolar explosion, 
Fe-poor explosions are accompanied with the formation of 
very massive central remnants, 
i.e., $M_{\rm REM} \gsim 10\msun$. 
Model 40D results from large $M_{\rm REM0}$, 
the mass of the remnant at the initiation of the jets. 
Though we treated $M_{\rm REM0}$ as a parameter, 
we expect $M_{\rm REM0}$is related to 
the angular momentum in the presupernova star. 
If the Fe core is rapidly rotating before the explosion, 
$M_{\rm REM0}$ would be small. 
On the other hand, if the core rotates relatively slowly, 
the core would not generate the jets immediately. 
In this case, continuous accretion would spin up the core, 
thus the core would be large at the formation of the jets, 
resulting in larger $M_{\rm REM0}$.

\section{Conclusions \& Discussion}

In this paper, we have shown the hydrodynamical and nucleosynthetic features 
of bipolar supernova/hypernova explosions. 
In our 2D hydrodynamical models, 
the properties of jets (such as $\dot M_{\rm jet}$ and $\dot E_{\rm jet}$) are 
determined by the mass accretion rate $\dot M_{\rm accretion}$ 
as described in equation (1), while 
the interaction between the jets and the stellar material affects 
$\dot M_{\rm accretion}$ and thus the properties of the jets. 
In our models, therefore, the explosion and the fallback are self-regulated. 

Two possible sites have been suggested to be responsible for nucleosynthesis 
in jet-driven supernova/hypernova explosions: 
stellar materials heated by the jets 
and jets (winds) themselves from the central region. 
Which is more prominent will be determined by how much fraction of 
accreting materials is ejected from the accretion disk 
rather than is accreted onto the central remnant. 
In the present study, we have only addressed the case where 
a small amount of the accreting gas is ejected in the jets, i.e., 
fast and light jets. 
The other possibility, i.e., slow and massive jets (disk wind), 
is also interesting, though we postpone the detailed study to future 
works as it should require a different approach (both in physically and numerically) 
from the present study (see sections 1 \& 2). 

With the caveat mentioned above, 
our important and interesting findings are summarized as follows. 
\begin{enumerate}
\item[(1)]
{\bf Explosion Energies}:
In our models, the resultant explosion energy ranges from 
$E_{51} \sim 0.5$ to $\gsim 30$, 
thus covering the observed hypernovae whose kinetic energies 
are estimated to be $E_{51} \sim 6-10$ (SN~2002ap: Mazzali et al. 2002), 
$E_{51} \sim 10-20$ (SN~1997ef: Iwamoto et al. 2000; Mazzali et al. 2000), 
and $E_{51} \gsim 30$ (SN~1998bw; Iwamoto et al. 1998). 
Reasonable parameters of the jets' properties (e.g., $\alpha=0.01$) yield 
explosions with $E_{51} \sim 10$,  
which implies that these are promising models for hypernovae. 
Moreover, we have found that 
a more massive star produces a more energetic explosion, 
which reproduces the observed trend in the 'hypernova branch' 
(Figure~\ref{f1}; see also Nomoto et al. 2003). 
\item[(2)] 
{\bf Remnant's Mass}: 
The mass of the central remnant shows a large variation, 
$M_{\rm REM} \sim 2 - 10\msun$, 
being larger for more massive progenitors.  
$M_{\rm REM}$ reaches $\gsim 5\msun$ 
for the progenitor star with $M_{\rm ZAMS}=40\msun$. 
This can be tested with observations of supernova remnants and 
X-ray binaries, such as X-ray Nova Sco (Israelian et al. 1999). 
In fact Nova Sco shows 
the evidence of nucleosynthesis in a hypernova explosion 
accompanied by the formation of 
a black hole with $M_{\rm BH} \sim 5\msun$ (Podsiadlowski et al. 2002). 
\item[(3)] 
{\bf Density Structure}:
The density structures of the ejecta of the bipolar models 
do not resemble those of conventional spherical models. 
Though the detailed structure becomes more evident with higher 
numerical resolutions, the overall slope is rather insensitive to 
the resolution. 
The slope depends on the assumed jet properties 
and probably on time variability of the central source, while 
the bipolar models in general have much flatter density distributions 
in the outer layers and a higher density central core 
than the spherical models. 
These features agree with the those inferred from modeling 
spectra and light curves of the observed 
hypernovae (Branch 2001 and Nakamura et al. 2001a for SN~1998bw; 
Iwamoto et al. 2000 and Mazzali et al. 2000 for SN~1997ef; 
Mazzali et al. 2002 and Yoshii et al. 2003 for SN~2002ap; 
Maeda et al. 2003 for SNe~1998bw, 1997ef, 2002ap).  
\item[(4)] 
{\bf Production of $^{56}$Ni}:
The amount of synthesized $^{56}$Ni, $M$($^{56}$Ni), depends 
on the time scale of energy injection. 
For slow energy injection, $M$($^{56}$Ni) is small because of 
pre-expansion of the ejecta. 
For $\dot E_{51} \lsim 1$ s$^{-1}$ or $M_{\rm REM 0} \gsim 3\msun$, especially, 
$M$($^{56}$Ni) $\lsim 0.1\msun$. 
The present models predict that there is rough 
anti-correlation between $^{56}$Ni and $M_{\rm REM}$. 
To explain the large masses of $^{56}$Ni in the observed hypernovae 
($M$($^{56}$Ni) $\gsim 0.1\msun$), 
the explosion mechanism must satisfy one of the following conditions: 
\begin{enumerate}
\item 
$\dot E_{51}$ is as large as $\gsim 10$ s$^{-1}$ and $M_{\rm REM 0} \lsim 3\msun$, 
i.e., the compact remnant begins to generate jets before it evolves too much. 
In this case, the amount of $^{56}$Ni 
produced at the shocked stellar material could be as large as $\sim 0.5\msun$ 
as observed in SN1998bw (Iwamoto et al. 1998). 
\item 
The jets (outflow from the central region) are more massive and 
have smaller velocities 
than those in the present models. 
Such jets could contain abundant $^{56}$Ni in the jet material itself. 

\end{enumerate}
In both cases we expect that the distribution of $^{56}$Ni is highly 
aspherical and elongated along the $z$-axis. 
Such configuration can account for the unusual feature in late time 
spectra of SN1998bw (Maeda et al. 2002). 
\item[(5)]
{\bf Distribution of isotopes}: 
The distribution of various isotopes in the bipolar models 
is completely different from conventional spherical models. 
Nucleosynthetic products are distributed as follows 
in the order of decreasing velocities (radii):
\begin{enumerate}
\item
products of strong $\alpha$-rich freezeout 
(e.g., $^{64}$Zn, $^{59}$Co, $^{44}$Ca) 
at the highest velocity, 
\item
products of complete silicon burning (e.g., $^{56}$Fe), 
\item
products of incomplete silicon burning 
(e.g., $^{55}$Mn, $^{52}$Cr, $^{40}$Ca, $^{32}$S, $^{28}$S), and 
\item
products of hydrostatic burning 
(e.g., $^{32}$S, $^{28}$Si, $^{16}$O, $^{12}$C, $^{4}$He) 
at the lowest velocity. 
\end{enumerate}
\item[(6)] 
{\bf Abundance patterns}:
\begin{enumerate}
\item
$\alpha$-rich freezeout and radioactive isotopes: 
The ratio between $\alpha$-rich freezeout products and $^{56}$Ni(Fe) is 
enhanced for more aspherical explosions. 
In particular, 
$^{44}$Ti/$^{56}$Ni is enhanced significantly. 
\item
Iron-peak elements: 
The bipolar models provide unique yields for iron-peak elements, 
where [Zn/Fe] and [Co/Fe] are enhanced while [Cr/Fe] and [Mn/Fe] 
are suppressed for a given [(Mg, O)/Fe]. 
The reason is that the bipolar explosions blow up the material 
from the innermost region 
which undergoes strong $\alpha$-rich freezeout. 
\item
$^{45}$Sc: 
Another important feature is the production of $^{45}$Sc. It has been difficult to 
produce $^{45}$Sc in spherical models, even 
for very large explosion energies. 
Spherical models always predict [Sc/Fe] $< -1$ because of 
too high densities (too low entropies) in the regions 
where $\alpha$-rich freezeout takes place to produce 
a large amount of $^{45}$Sc . 
The bipolar models, on the other hand, provide [Sc/Fe] $\sim 0$, 
because $\alpha$-rich freezeout proceeds at lower densities 
along the $z$-axis than in spherical models. 
Inversely, $^{45}$Sc could be an indicator of 
the degree of asphericity in supernovae/hypernovae. 
At the same time, we have found that the production of 
Sc is very sensitive to how the detailed jet phenomenology can 
be resolved. Therefore, future higher resolution studies are needed 
to confirm this result.

\item
Intermediate mass elements:
The bipolar models have different characteristics 
in the ratios among S, Si, Mg, O, C  
as compared to spherical models. 
In contralt to spherical models, which predict that these 
ratio are insensitive to the explosion process, 
the bipolar models result in large variations in these ratios. 
The accretion decreases the amount of those elements in a different 
way (e.g., S is more easily accreted onto the central remnant than Si), 
thus these ratios depends on the amount of the matter accreted onto the center. 
Large $M_{\rm REM}$ results in the reduction of heavier elements 
(e.g., larger $M_{\rm REM}$ leads to smaller [S/Si], which might 
explain the low S/Si in M87 (Matsushita et al. 2003)). 
\end{enumerate}
\item[(7)]
{\bf Extremely metal-poor stars}:
\begin{enumerate}
\item
Trends in iron-peak elements: 
The nucleosynthetic features of the bipolar models (described in (6b) above) 
are compatible with the abundance 
patterns and trends in extremely metal-poor stars. 
This is the result of the combined effects of the jets and the fallback. 
In spherical explosion models, 'mixing and fallback' provides a 
good agreement between the ejecta yields and the observed abundances
(Umeda \& Nomoto 2003). 
The bipolar models provide naturally (without fine tuning) almost the same 
effect as 'mixing and fallback' by ejecting 
preferentially higher entropy matter. 
Moreover, it is possible that the jets induce a global mixing, 
through share instability and so on. 
\item
Fe-poor explosions and black hole formation: 
Some bipolar models also explain the existence of Fe-poor explosions 
with very little amount of Fe. 
Such explosion would be responsible for the formation of 
the carbon-rich metal-poor stars including the most Fe-poor star. 
The Fe-deficient explosion results from large $M_{\rm REM0}$, 
which may be interpreted as the small angular momentum 
in the pre-explosion core. 
%Such an explosion is accompanied by the formation of a very massive compact 
%remnant, $M_{\rm REM} \gsim 10\msun$. 
%As the accretion is so significant in such an explosion, 
%the ratios [S/Si] and [O/C] are predicted to be small. 
\end{enumerate}

\end{enumerate}

We have shown that the bipolar models have potential to explain 
a wide range of observations. 
We conclude that 
(1) hypernovae explosions are driven by bipolar jets, and 
(2) their contribution on the early Galactic chemical evolution may be significant.

\acknowledgments
We would like to thank Hideyuki Umeda for useful discussion.  
The computation was partly carried out on Fujitsu VPP-700E at the Institute of 
Physical and Chemical Research (RIKEN). 
This work has been supported in part by the grant-in-Aid for 
COE Scientific Research (14047206, 14540223) of 
the Ministry of Education, Science, Culture, Sports, and Technology in Japan.


%% command and will number appendix equations (A1), (A2), etc. \appendix

%\appendix 

\section*{A. Sensitivity to the numerical resolution}

In this section, we show what extent the results depend on 
the numerical grid resolution. 
We have run model 40A with the numerical grids 
$200 \times 60$ (higher resolution), 
$100 \times 30$ (standard run), 
and $70 \times 21$ (lower resolution) 
to see if there are significant differences. 

Figure~\ref{f19} shows time evolution of $M_{\rm REM}$ for different 
numerical resolutions. 
The evolution of $M_{\rm REM}$ is determined by the interaction 
between the jets and stellar materials, thus is a good indicator 
for a convergence test. Also, the final $M_{\rm REM}$ is of essential 
importance to the resultant isotopic composition in the ejected materials, 
because it determines which regions are finally ejected.

Except for the lowest resolution ($70 \times 21$), 
the curves are almost identical. We, therefore, are confident 
that the numerical resolution of $100 \times 30$ used throughout the paper 
is sufficient to catch the overall structure of the interaction. 
In the lowest resolution, it is most likely that the bow shock expands 
sideways incorrectly faster than in the higher resolutions, making the 
accretion rate smaller.

Another test is to see the detailed structure of the ejecta. 
Figure~\ref{f20} shows the density structures along the $z$- and $r$-axis 
for different resolutions. 
As implied by the first test ($M_{\rm REM}$), the overall structure 
from the lowest to the highest velocities is very similar to one another. 
The curves of the higher resolution runs, especially, match with each other 
almost identically, except at the highest velocities 
($\gsim 2.5 \times 10^{9}$ cm s$^{-1}$) along the $z$-axis. 

With higher numerical resolution appears richer jet phenomenology 
(Aloy et al. 2000; Zhang, Woosley, \& MacFadyen 2003), making the deviation of the detailed 
structure at the highest velocities along the $z$-axis among the 
different numerical resolutions. 
We do not see the numerical convergence in this sense, thus need 
a higher resolution (more than $200 \times 60$) to resolve the exact 
jet phenomenology. We postpone it to future works. 
We note, however, that the averaged behavior 
(e.g., the averaged slope at $> 10^{8}$ cm s$^{-1}$) is rather insensitive 
to the detailed density structure, as low resolutions just smear the details.

The final and the most important test is to see the effects of the numerical 
resolution on the isotopic composition of the ejected materials. 
Figure~\ref{f21} compares the isotopic yield of model 40A with different 
resolutions. 
The isotopic compositions, especially for major species, are strikingly 
similar to one another. 
This is because the main feature of the jet induced explosion 
is the ejection of higher entropy materials associated with the jets 
and the accretion of lower entropy materials initially at outer envelopes, 
and the process is appropriately resolved with the resolution of $100 \times 30$ 
(Fig.~\ref{f19}). 

A few isotopes, however, show quantitative difference for different resolutions. 
They are $^{44}$Ca, $^{48}$Ti (enhanced with higher resolution), 
and $^{45}$Sc (suppressed). 
The difference of ($^{44}$Ca, $^{48}$Ti)$/^{56}$Fe 
between different resolutions is still not large and the results with 
the resolution of $100 \times 30$ are almost unchanged, 
while $^{45}$Sc$/^{56}$Fe shows the difference by a factor of $2 - 3$.

The difference most likely comes from the formation of richer jet phenomenology 
with higher resolution (Fig.~\ref{f20}). 
Because these isotopes, especially $^{45}$Sc, are minor isotopes and 
produced by $\alpha$-rich freezeout, production of these isotopes is sensitive 
to the detailed structure (e.g., recollimation shocks; see Aloy et al. 2000 and 
Zhang et al. 2003) of the interaction between the jets and stellar materials. 

In sum, we evaluate sensitivity of the results to numerical resolution as follows. 
\begin{itemize}
\item[(1)]
The overall structure of the hydrodynamic interaction can be resolved 
with the resolution of $100 \times 30$ used throughout the paper. 
It guarantees the results on the global quantities in the paper, e.g., 
$M_{\rm REM}$, $E_{\rm total}$, and $M$($^{56}$Ni). 
For most isotopes, we see the convergence at the resolution $100 \times 30$. 
\item[(2)]
To resolve the detailed jet phenomenology, we need higher numerical resolutions. 
This affects the detailed density structure at the highest velocities along 
the $z$-axis, and production of a few isotopes. The isotope which is most sensitive 
to the resolution is $^{45}$Sc. 
The detailed ejecta structure is of course important, though we note that 
the averaged behavior does not sensitively depend on the numerical resolution. 
For model 40A, the density structure above $10^{9}$ cm s$^{-1}$ shows significant 
flattening compared to spherical model 40SH, and the average slope is similar 
irrespective of the resolution.  For $^{45}$Sc, the result of 
the present study, i.e., the enhancement of $^{45}$Sc$/^{56}$Fe is probably 
correct qualitatively, while we need future works with higher resolutions 
to conform the result quantitatively. 
\end{itemize}

\clearpage



\clearpage

\clearpage

\clearpage

\clearpage

\clearpage

\clearpage

\clearpage

\clearpage

\clearpage

\clearpage

\clearpage

\clearpage

\clearpage

\clearpage

\begin{figure}
\begin{center}
	\begin{minipage}[t]{0.8\textwidth}
		\epsscale{1.0}
		\plotone{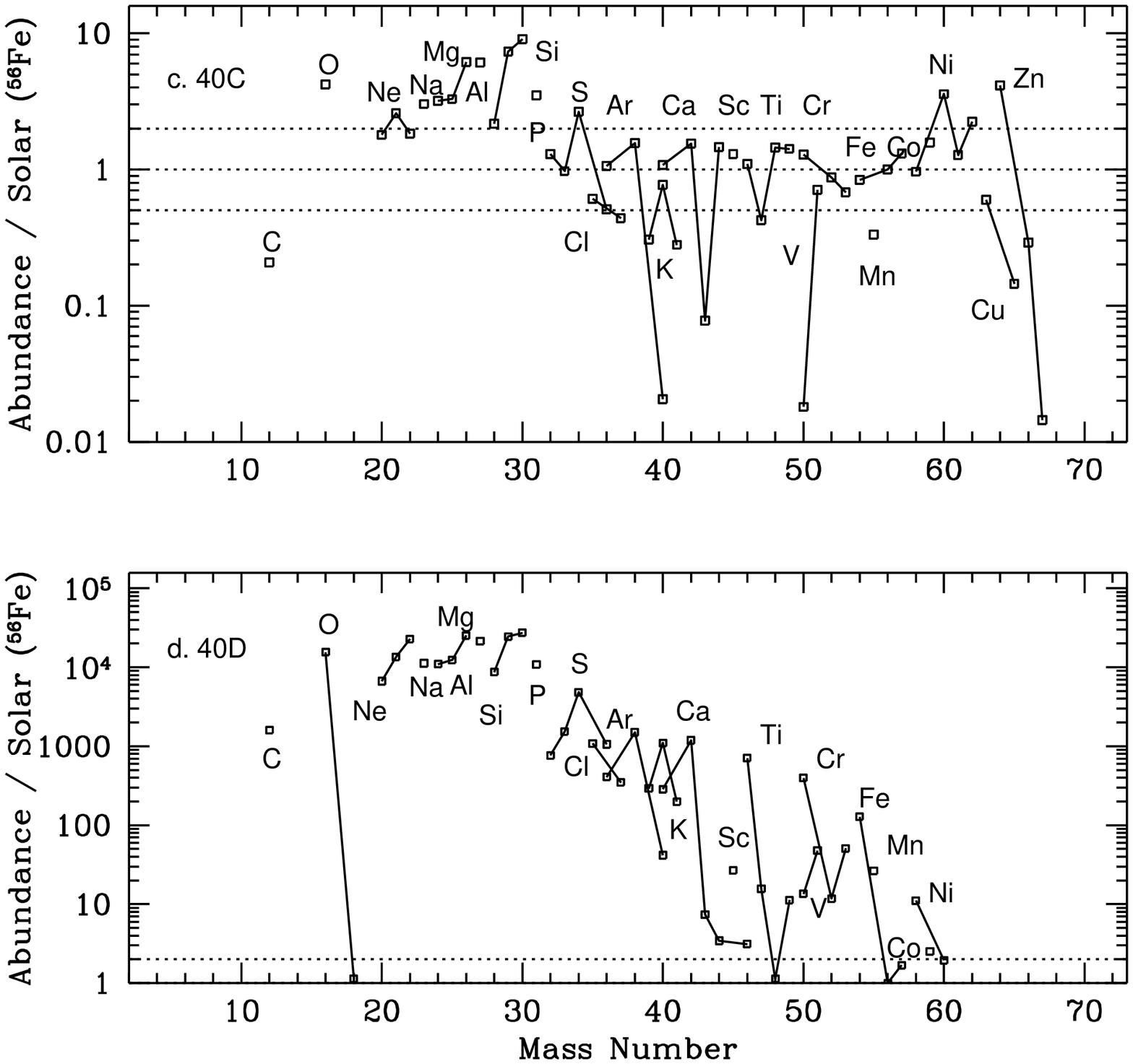}
	\end{minipage}
\end{center} 
\end{figure}

\clearpage

\begin{figure}
\begin{center}
	\begin{minipage}[t]{0.8\textwidth}
		\epsscale{1.0}
		\plotone{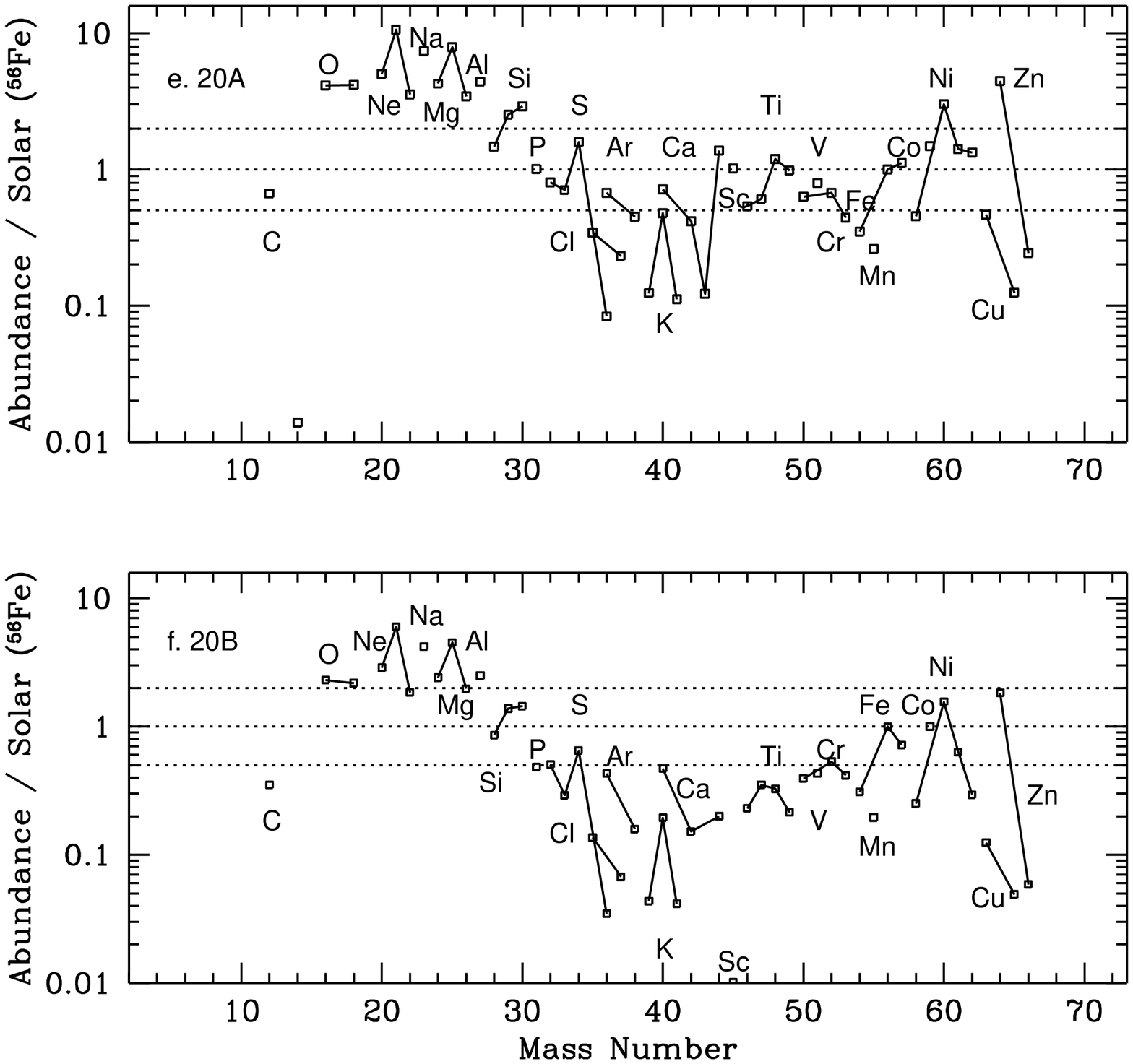}
	\end{minipage}
\end{center} 
\end{figure}

\clearpage

\begin{figure}
\begin{center}
	\begin{minipage}[t]{0.4\textwidth}
		\epsscale{1.0}
		\plotone{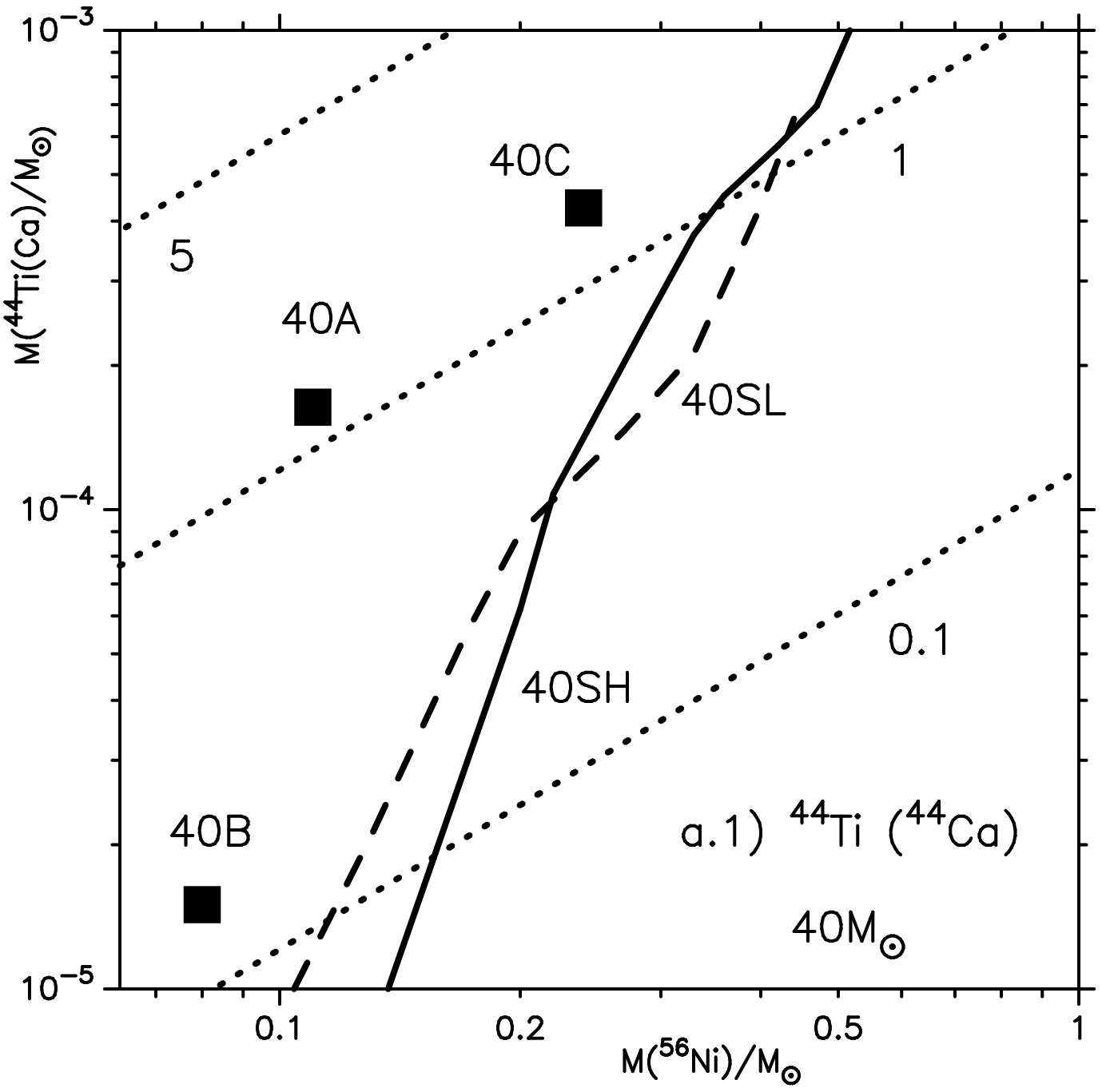}
	\end{minipage}
	\begin{minipage}[t]{0.4\textwidth}
		\epsscale{1.0}
		\plotone{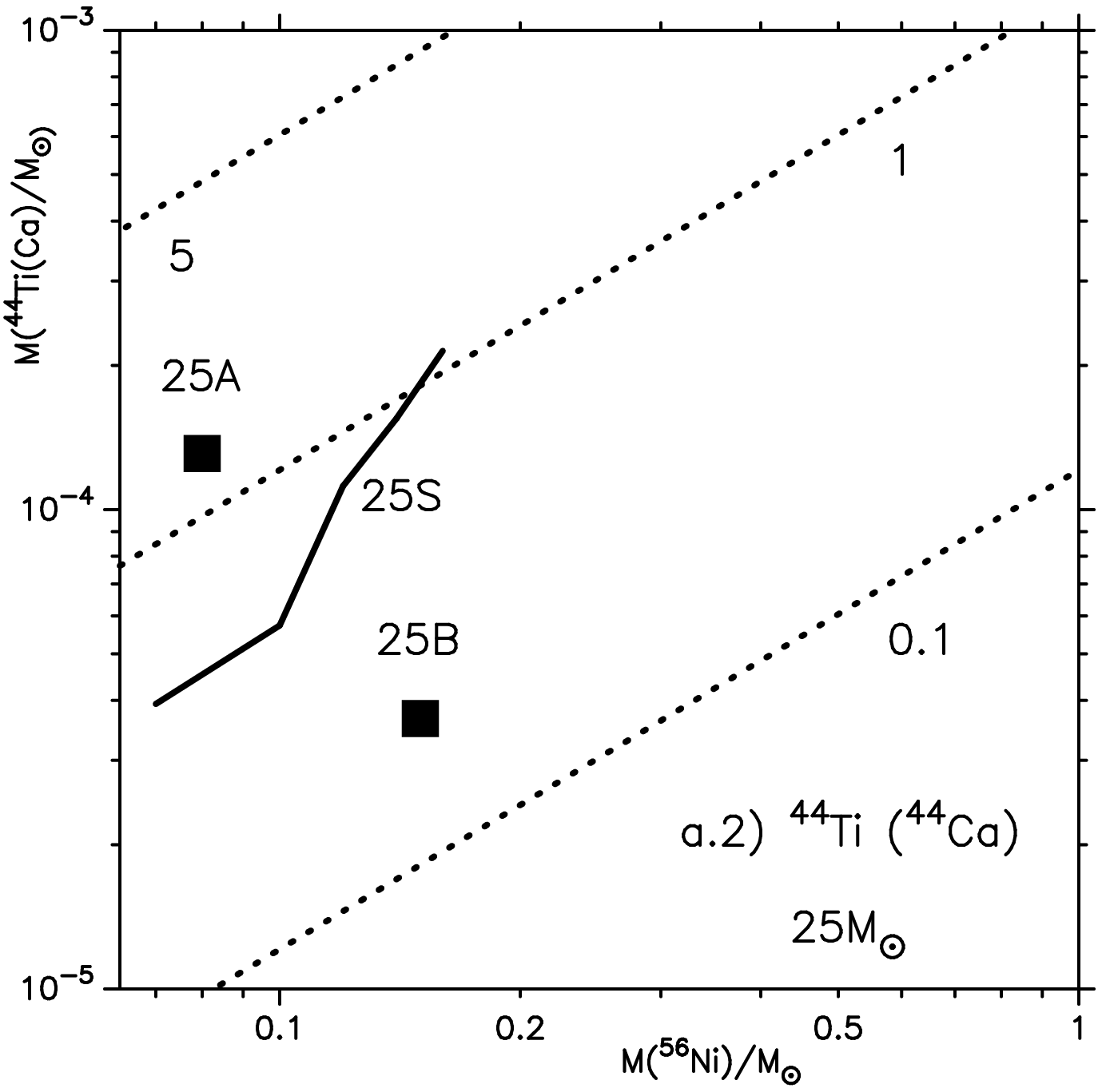}
	\end{minipage}
	\begin{minipage}[t]{0.4\textwidth}
		\epsscale{1.0}
		\plotone{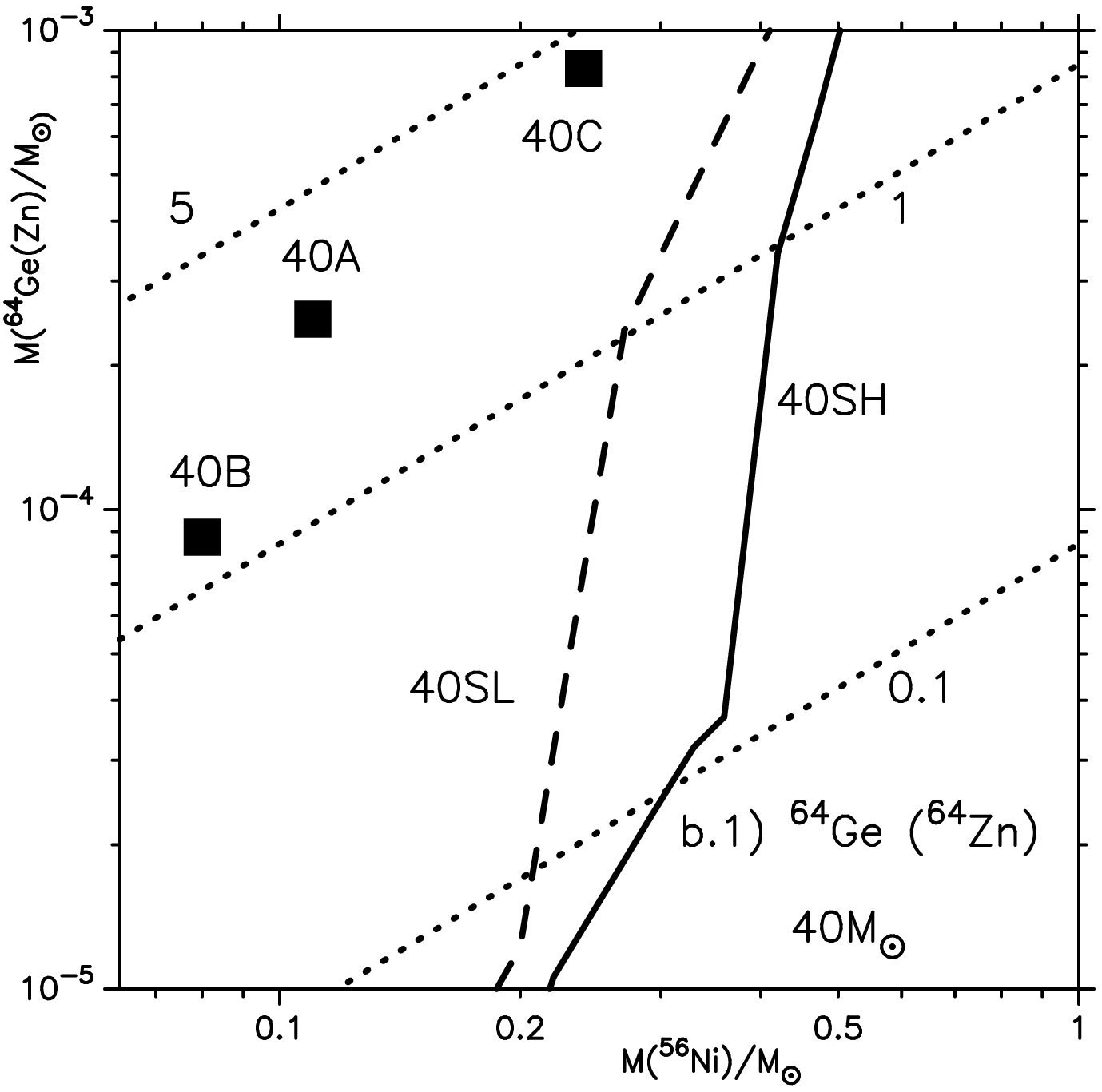}
	\end{minipage}
	\begin{minipage}[t]{0.4\textwidth}
		\epsscale{1.0}
		\plotone{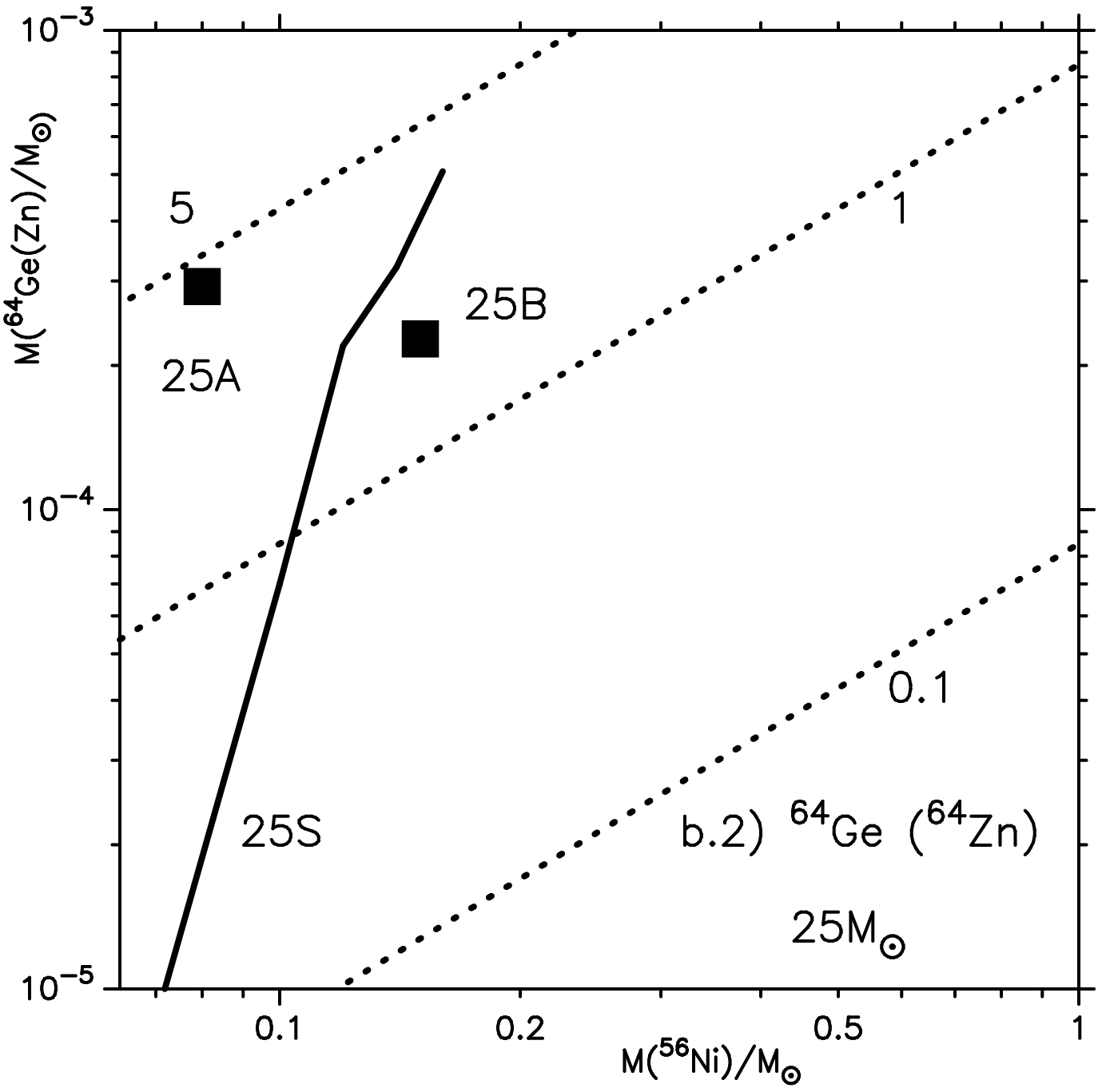}
	\end{minipage}
\end{center}
\caption{Ejected masses of (a) $^{44}$Ti and (b) $^{64}$Ge 
against ejected masses of $^{56}$Ni for the bipolar models 
(filled-squares), and for the spherical models 
40SH (solid), 40SL (dashed), and 25S (solid). 
The dashed lines show the ratio between 
($^{44}$Ca,$^{64}$Zn)/$^{56}$Ni relative to the solar values.
\label{f15}}
\end{figure}

\clearpage

\begin{figure}
\begin{center}
	\begin{minipage}[t]{0.4\textwidth}
		\epsscale{1.0}
		\plotone{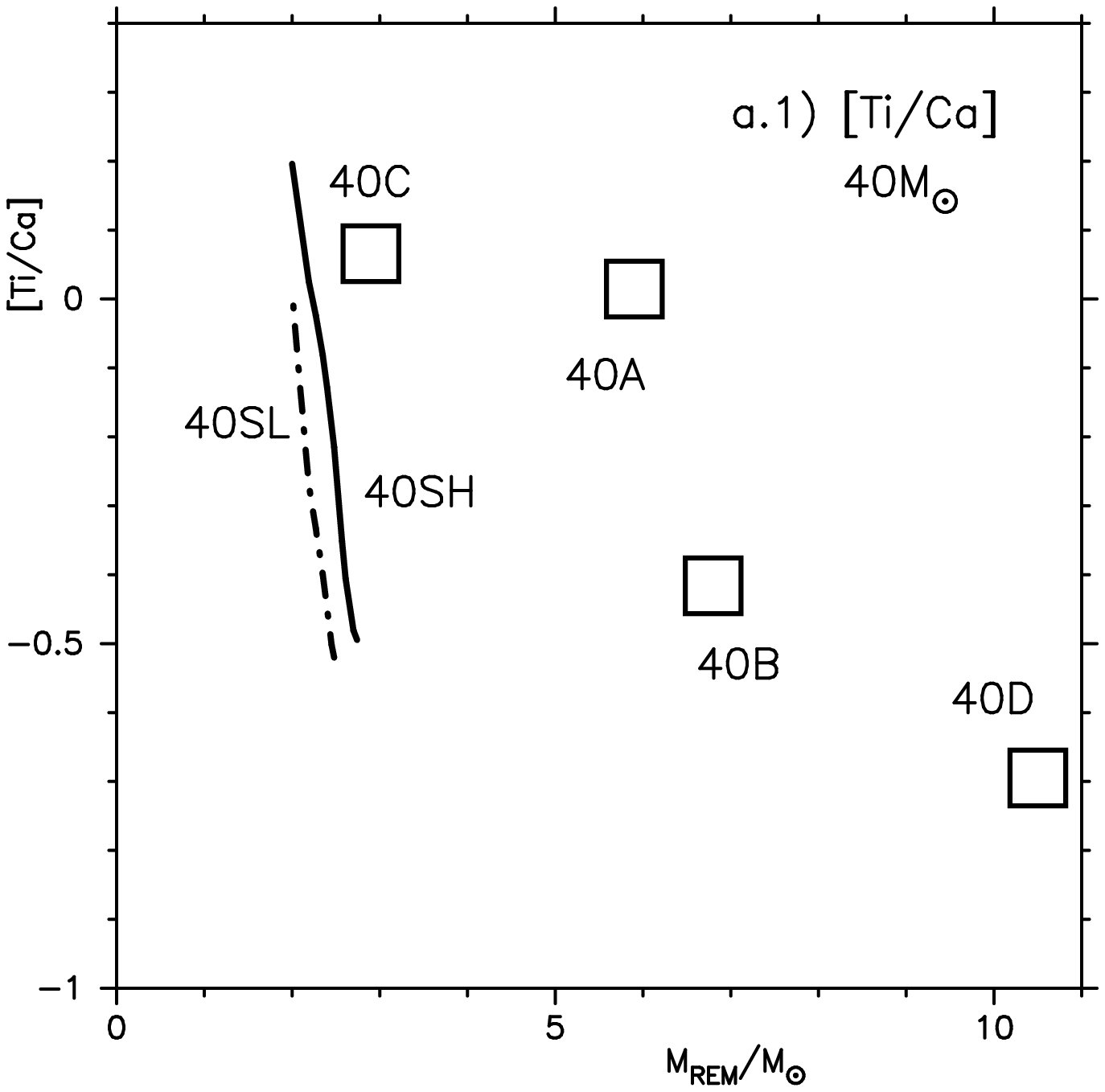}
	\end{minipage}
	\begin{minipage}[t]{0.4\textwidth}
		\epsscale{1.0}
		\plotone{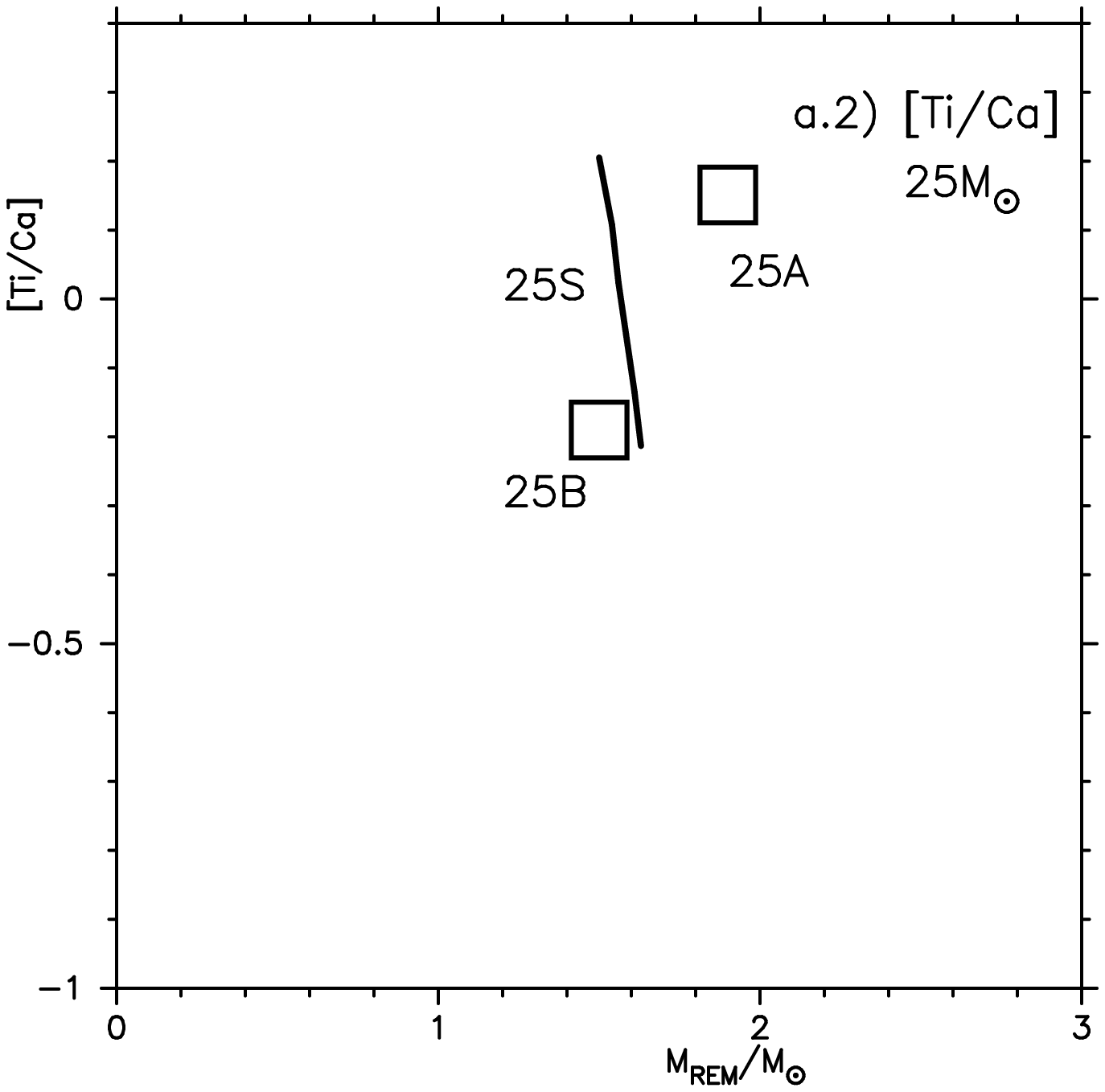}
	\end{minipage}
	\begin{minipage}[t]{0.4\textwidth}
		\epsscale{1.0}
		\plotone{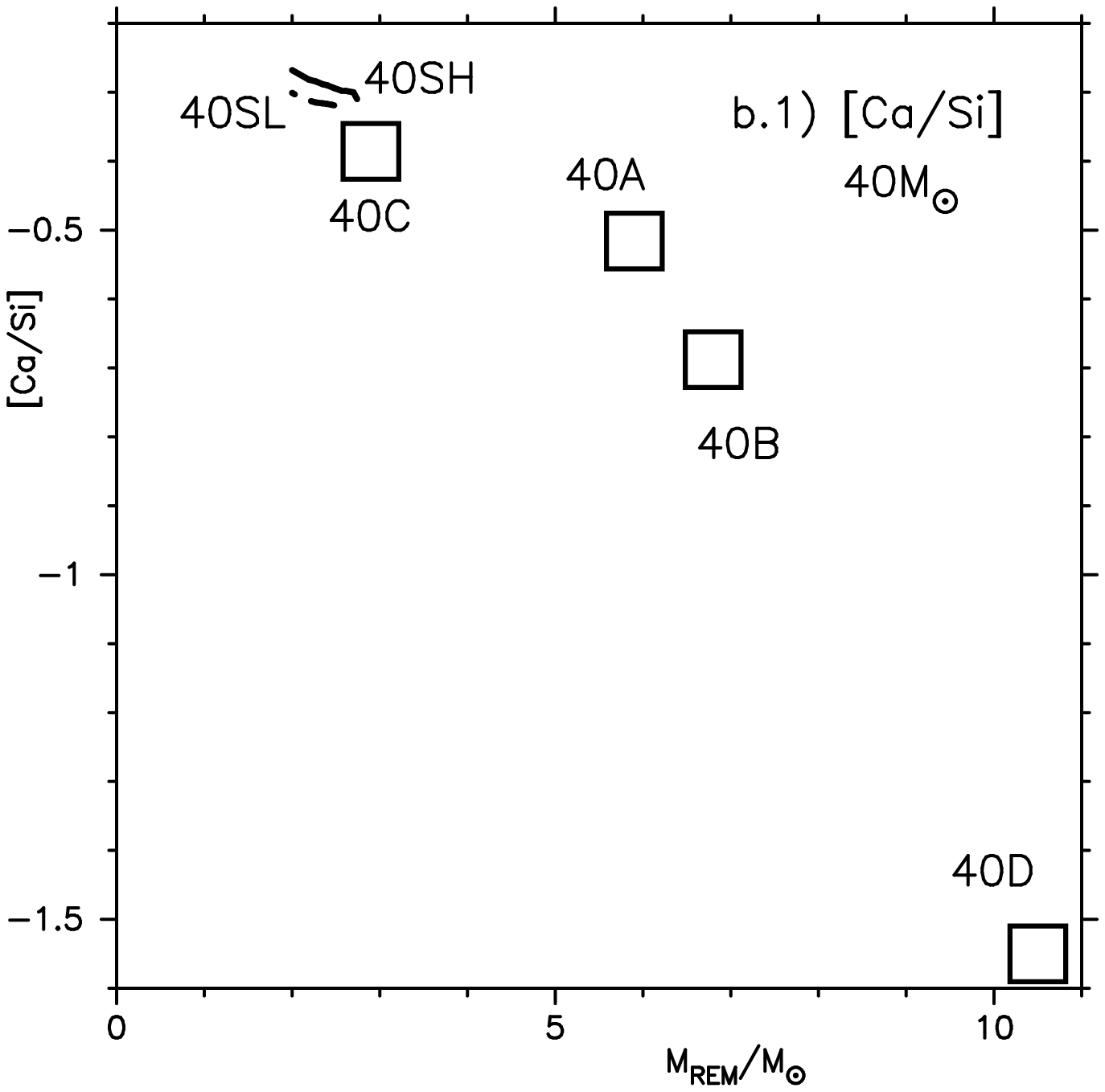}
	\end{minipage}
	\begin{minipage}[t]{0.4\textwidth}
		\epsscale{1.0}
		\plotone{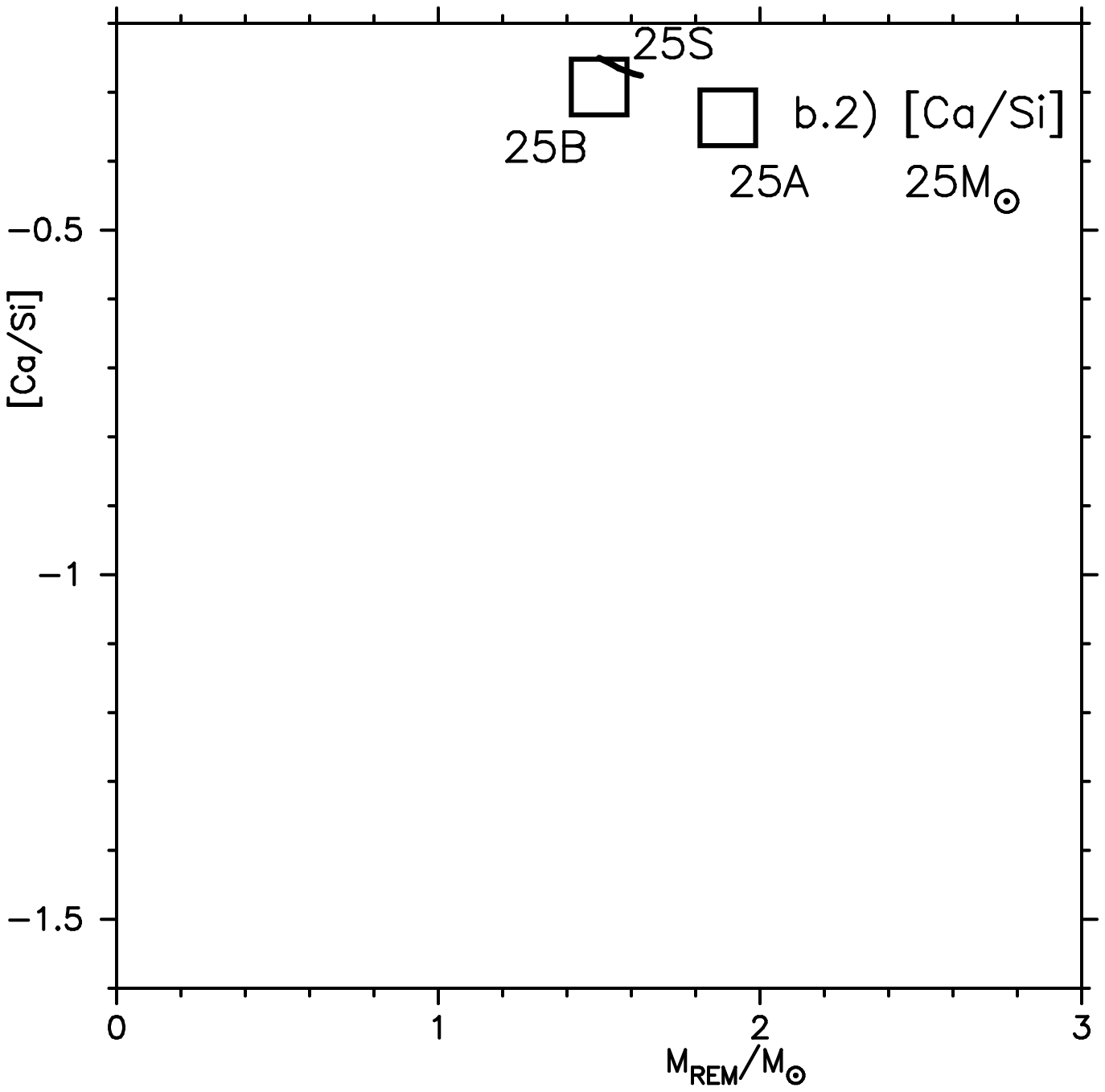}
	\end{minipage}
\end{center}
\caption{Abundance ratios, 
(a) [Ti/Ca], (b) [Ca/Si], 
(c) [S/Si], (d) [Si/O], and (e) [C/O], 
as a function of the final remnant mass ($M_{\rm REM}$) 
for the bipolar models (open squares). 
The lines shows those for the spherical models 
40SH (solid), 40SL (dashed), and 25S (solid) with 
different $M_{\rm cut}$. 
\label{f16}}
\end{figure}

\clearpage

\begin{figure}
\begin{center}
	\begin{minipage}[t]{0.4\textwidth}
		\epsscale{1.0}
		\plotone{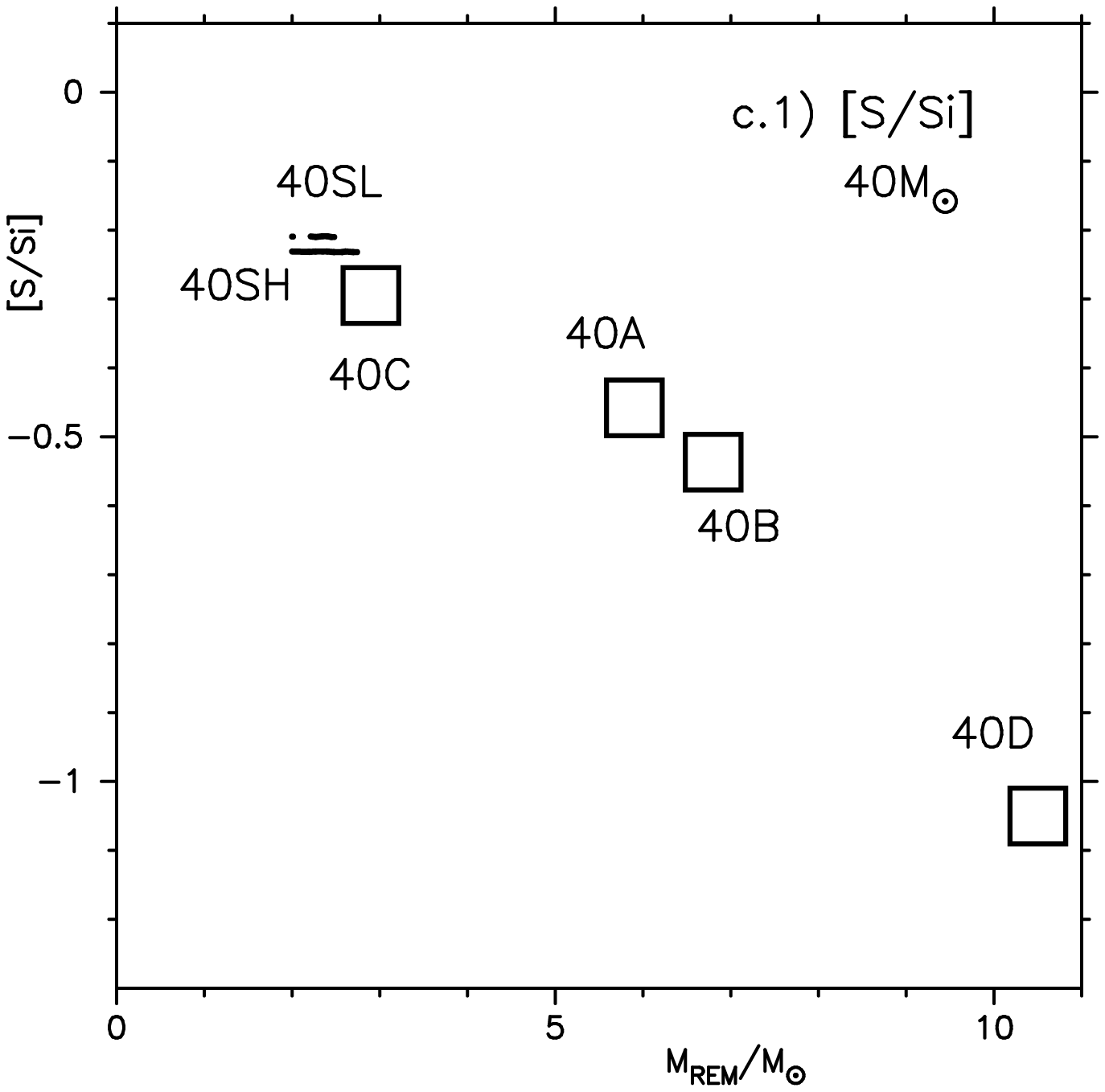}
	\end{minipage}
	\begin{minipage}[t]{0.4\textwidth}
		\epsscale{1.0}
		\plotone{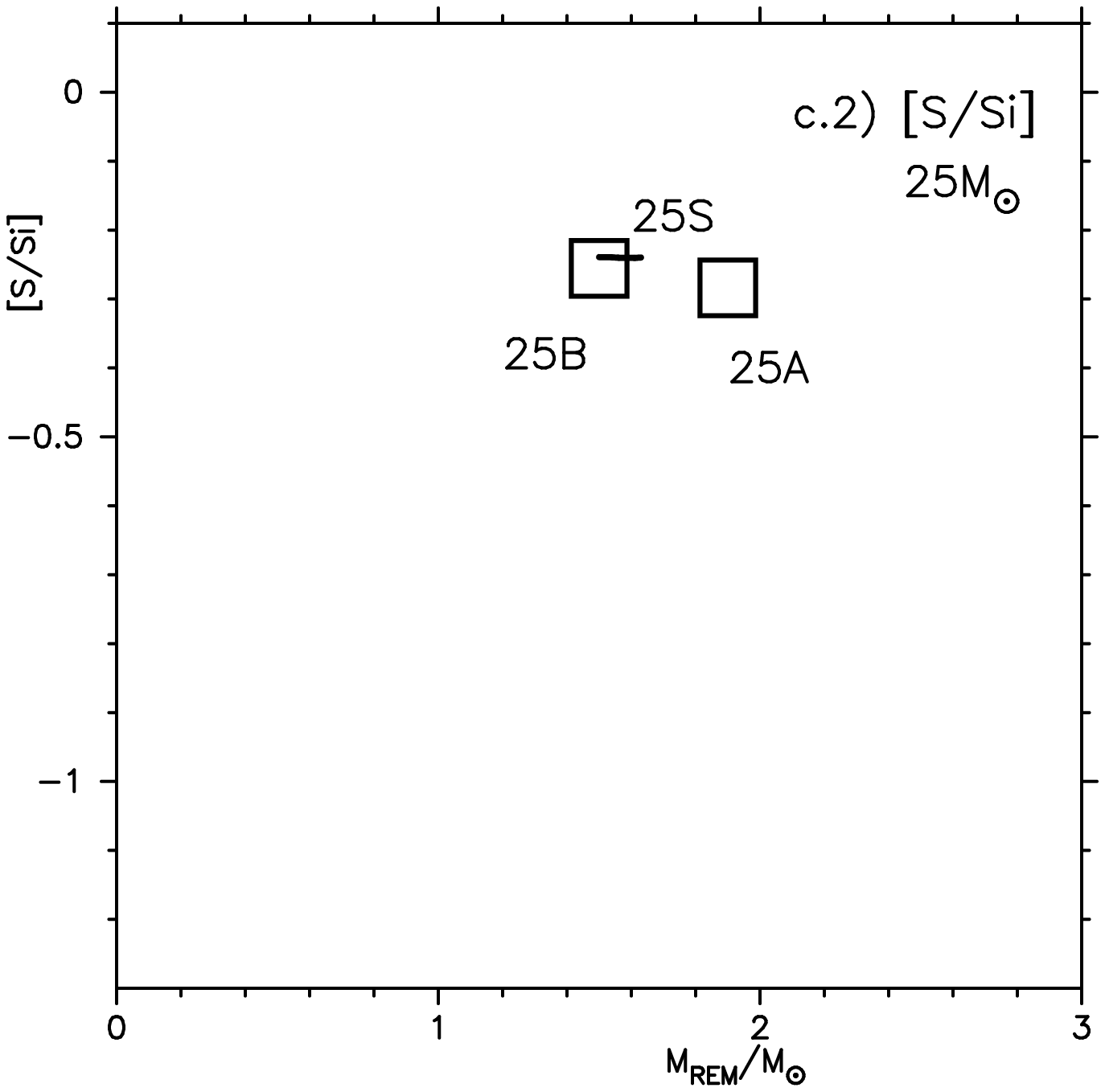}
	\end{minipage}
	\begin{minipage}[t]{0.4\textwidth}
		\epsscale{1.0}
		\plotone{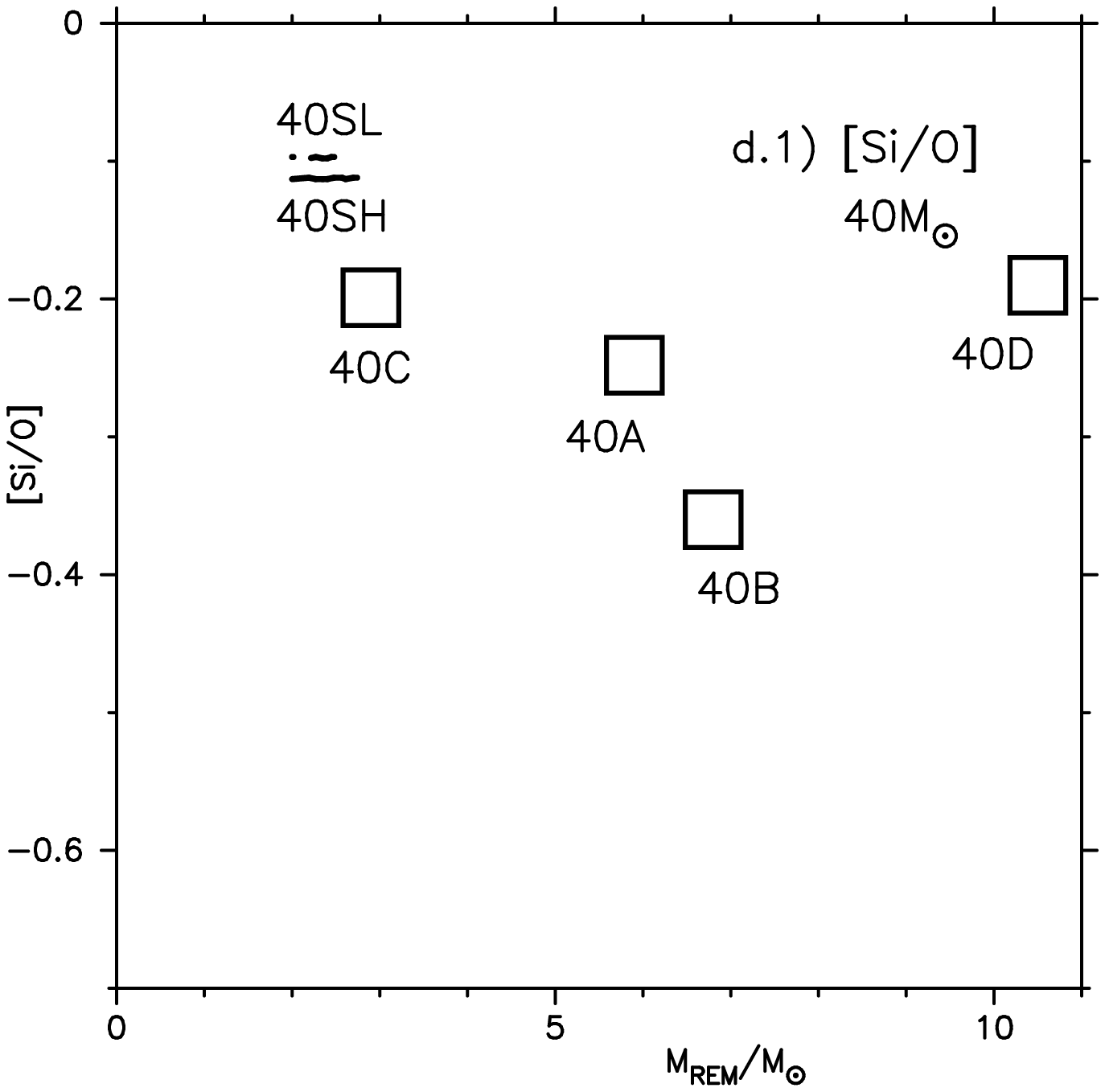}
	\end{minipage}
	\begin{minipage}[t]{0.4\textwidth}
		\epsscale{1.0}
		\plotone{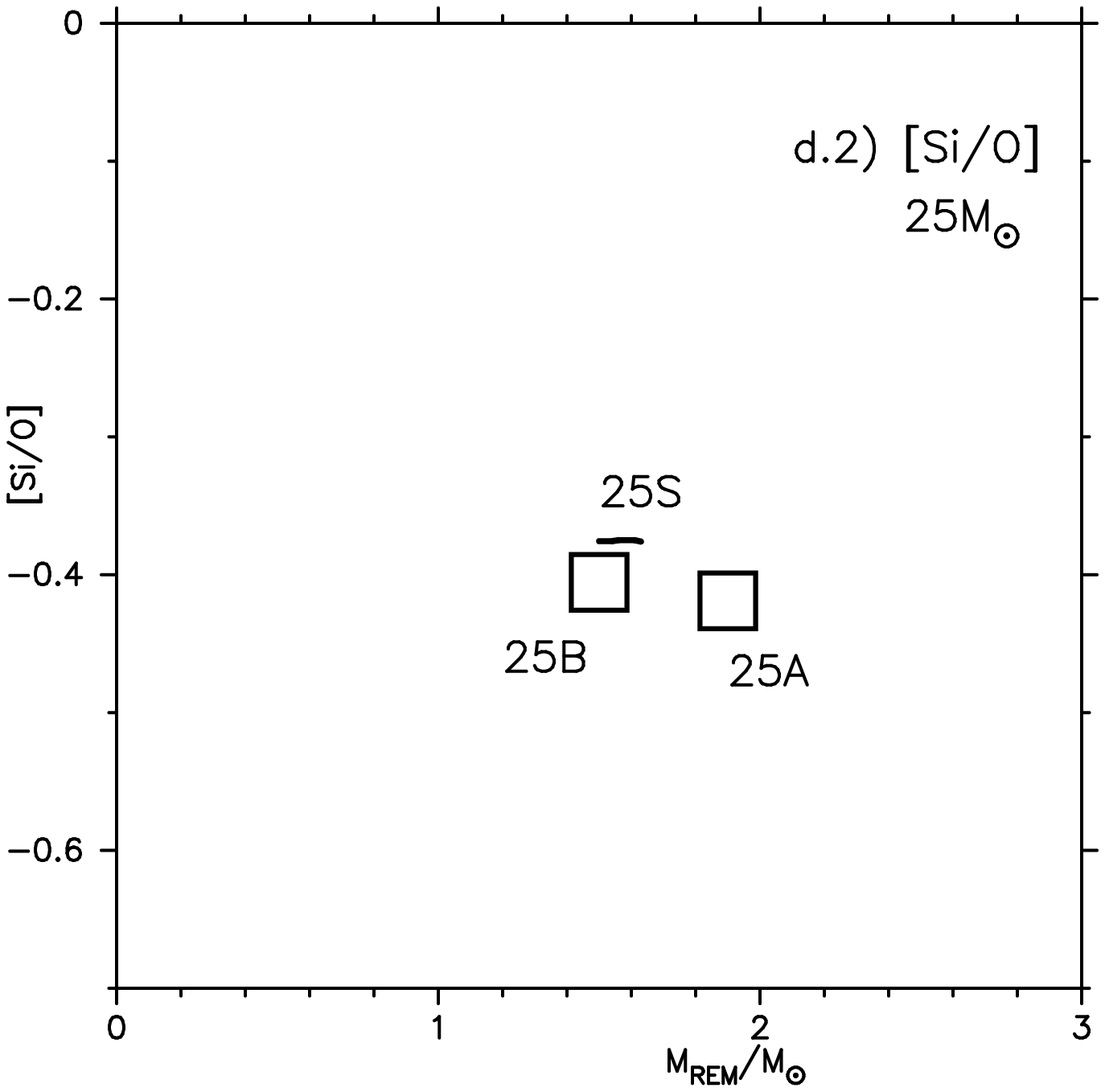}
	\end{minipage}
\end{center}
\end{figure}

\clearpage

\begin{figure}
\begin{center}

	\begin{minipage}[t]{0.4\textwidth}
		\epsscale{1.0}
		\plotone{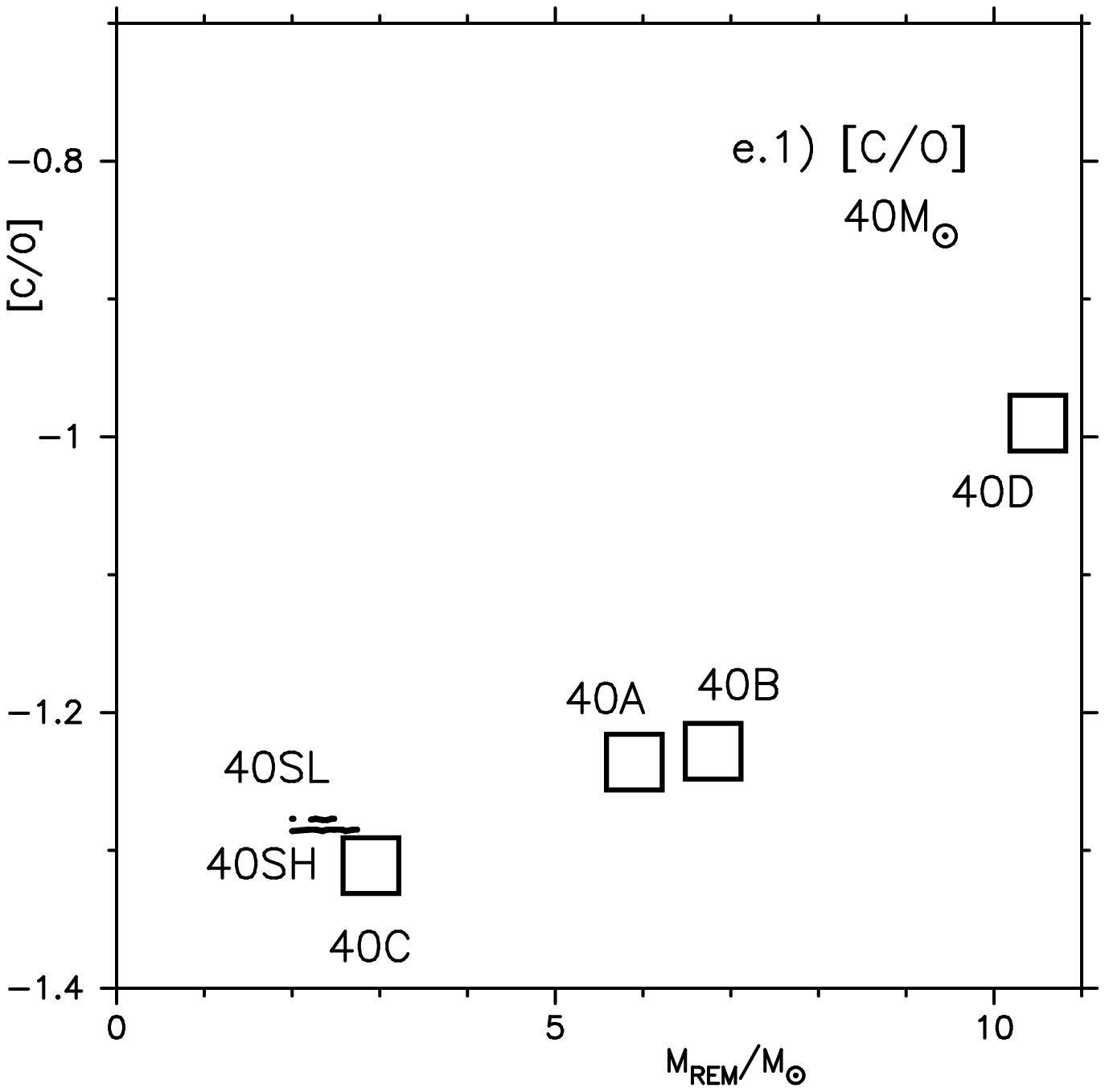}
	\end{minipage}
	\begin{minipage}[t]{0.4\textwidth}
		\epsscale{1.0}
		\plotone{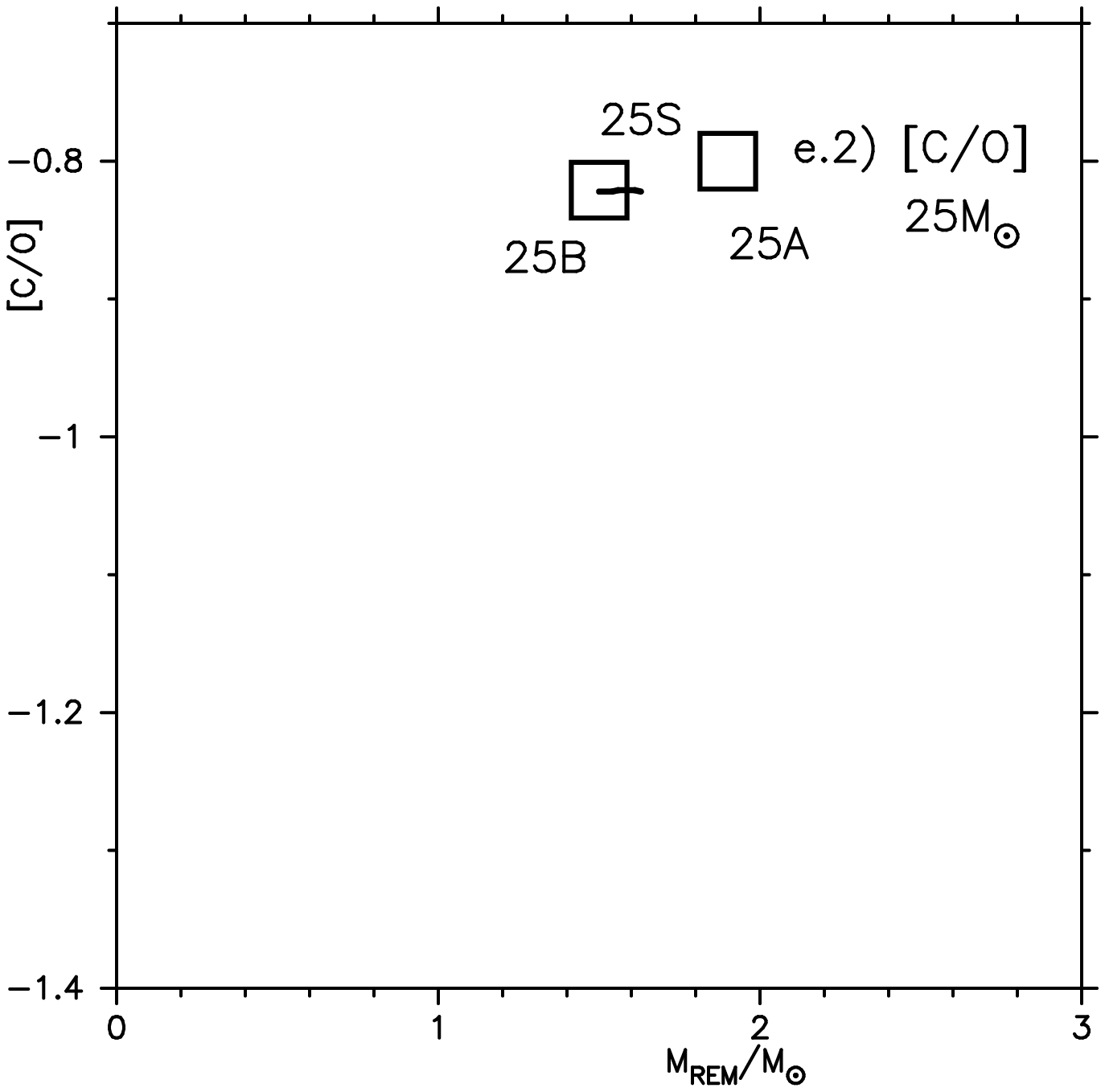}
	\end{minipage}
\end{center}
\end{figure}

\clearpage

\begin{figure}
\begin{center}
	\begin{minipage}[t]{0.4\textwidth}
		\epsscale{1.0}
		\plotone{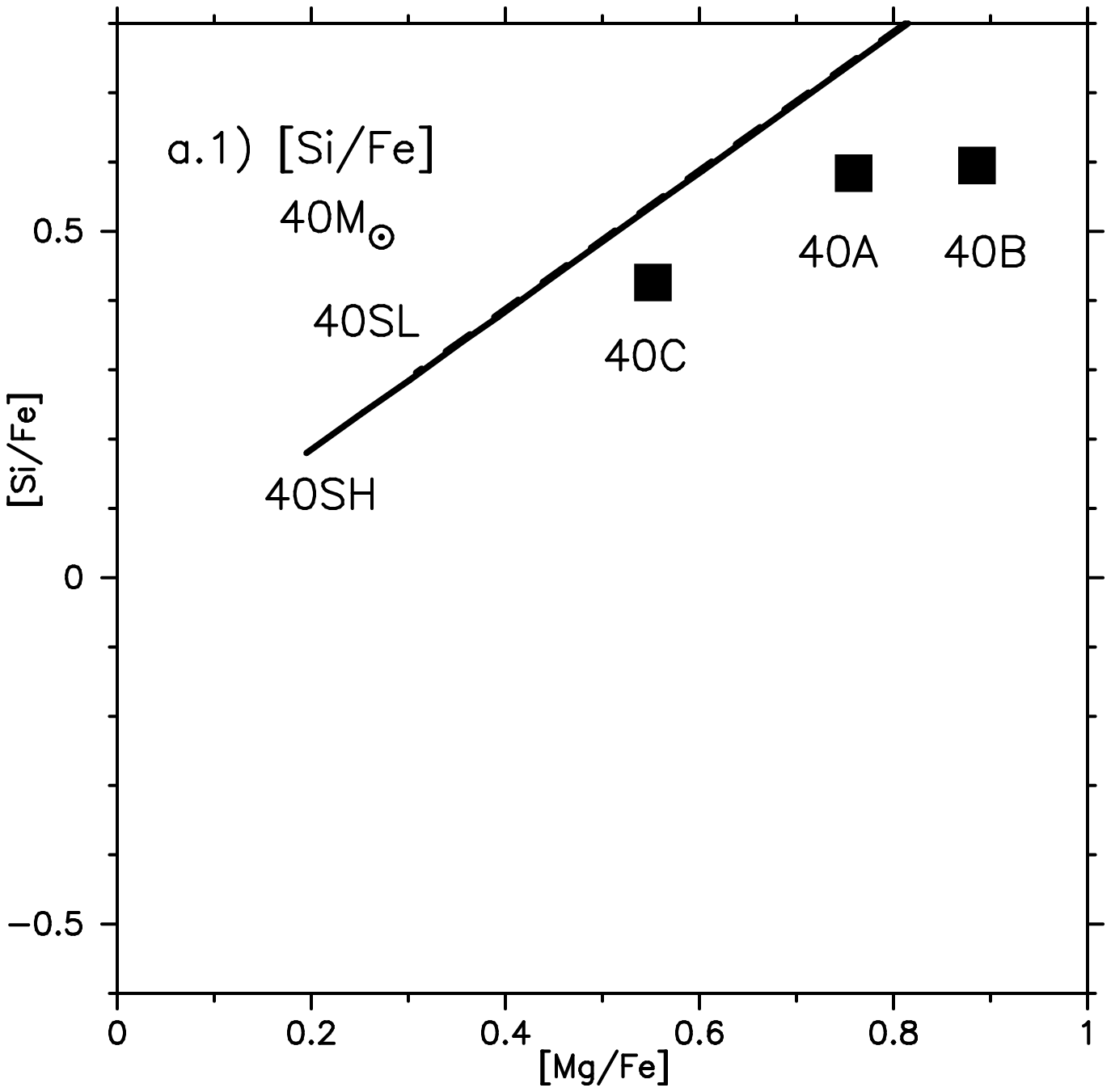}
	\end{minipage}
	\begin{minipage}[t]{0.4\textwidth}
		\epsscale{1.0}
		\plotone{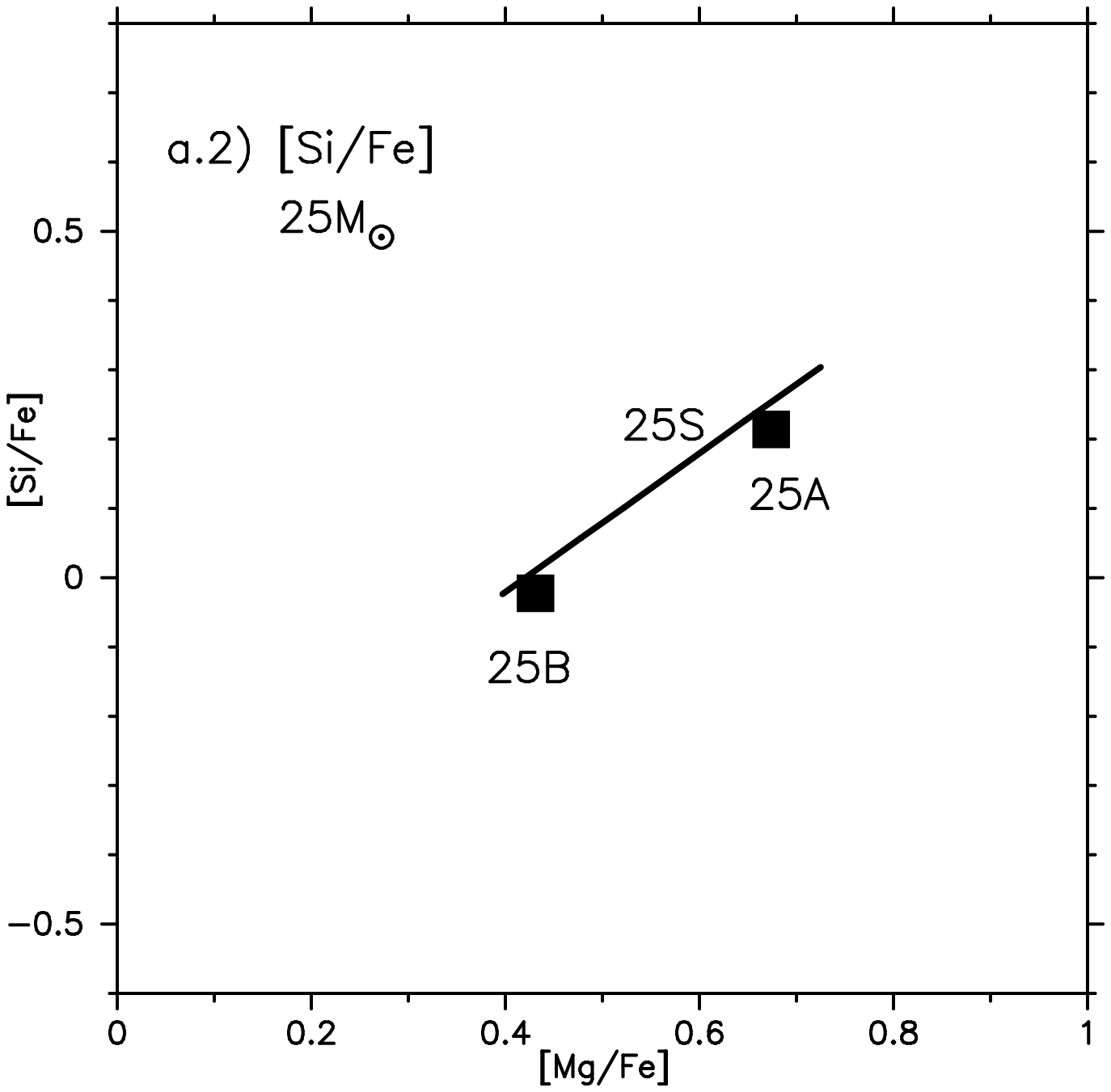}
	\end{minipage}
	\begin{minipage}[t]{0.4\textwidth}
		\epsscale{1.0}
		\plotone{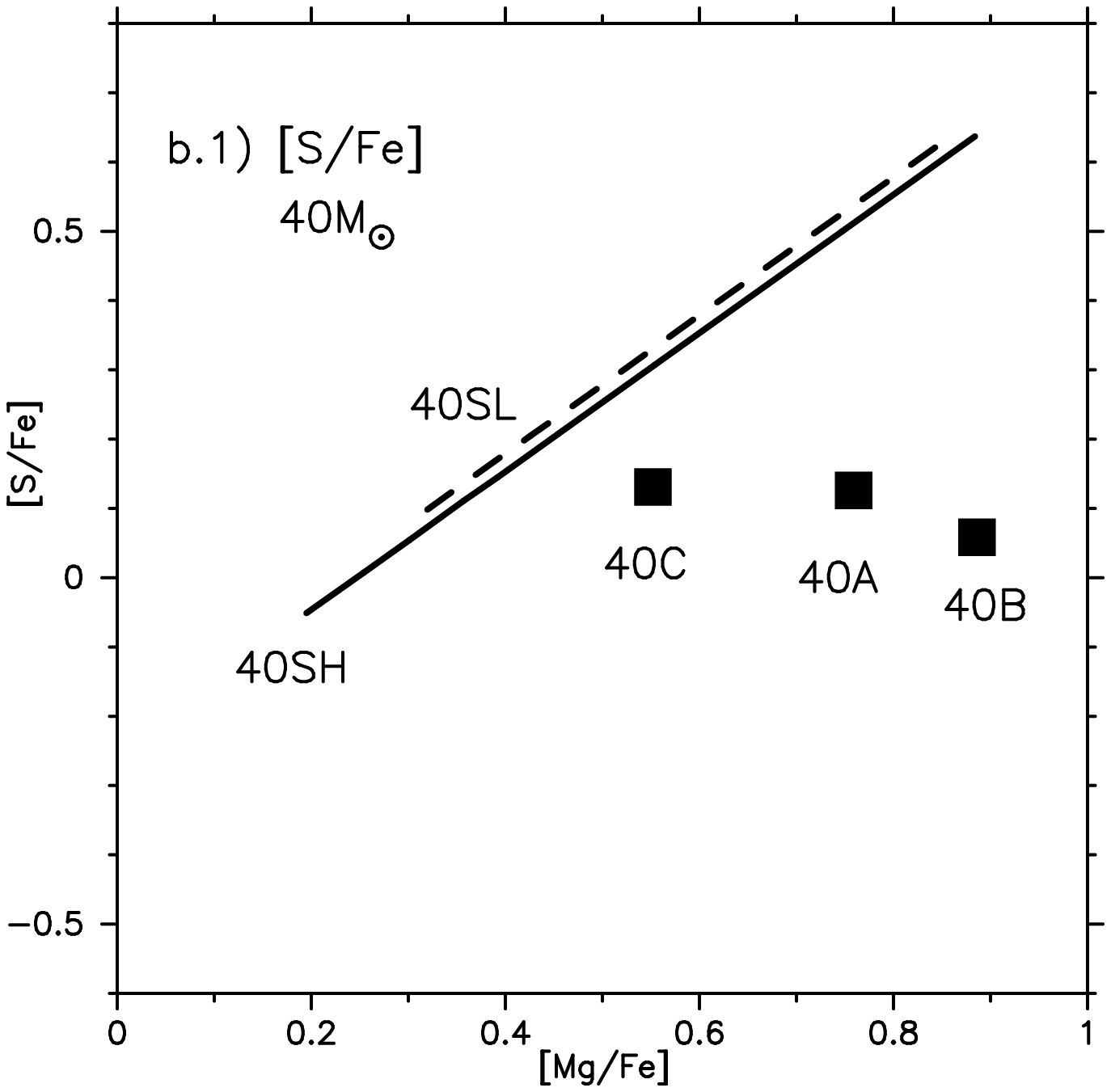}
	\end{minipage}
	\begin{minipage}[t]{0.4\textwidth}
		\epsscale{1.0}
		\plotone{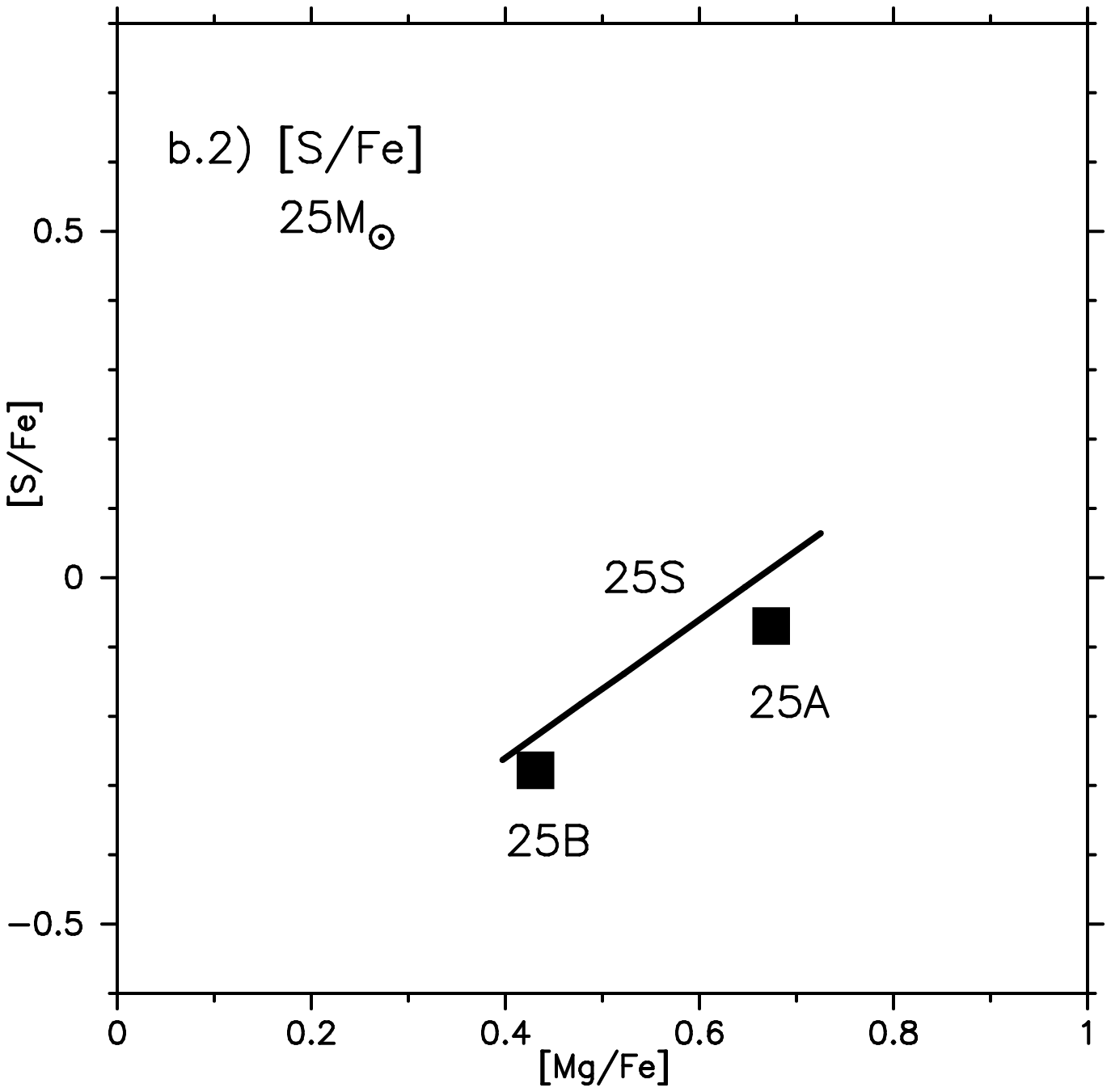}
	\end{minipage}
\end{center}
\caption{Abundances of stable isotopes relative to the solar values, 
for the bipolar models 40A, 40B, 40C, 25A, 25B (filled squares). 
The abundances for the spherical models with various $M_{\rm cut}$ are 
shown for 40SH (solid), 40SL (dashed), and 25S (solid). 
The isotopes plotted are (a) Si, (b) S, (c) Sc, (d) Ti, 
(e) Cr, (f) Mn, (g) Co, and (h) Zn. 
\label{f17}}
\end{figure}

\clearpage

\begin{figure}
\begin{center}
	\begin{minipage}[t]{0.4\textwidth}
		\epsscale{1.0}
		\plotone{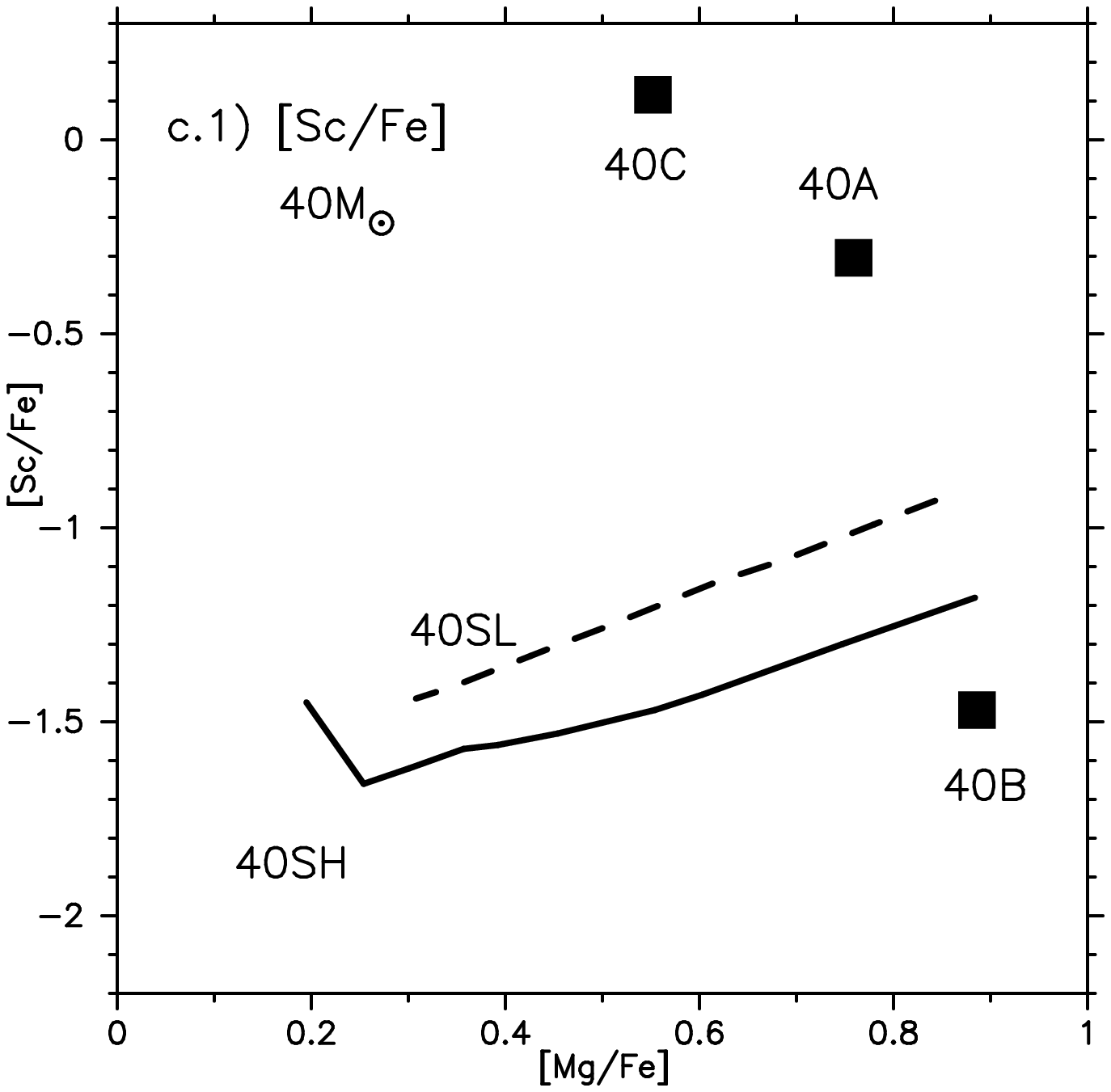}
	\end{minipage}
	\begin{minipage}[t]{0.4\textwidth}
		\epsscale{1.0}
		\plotone{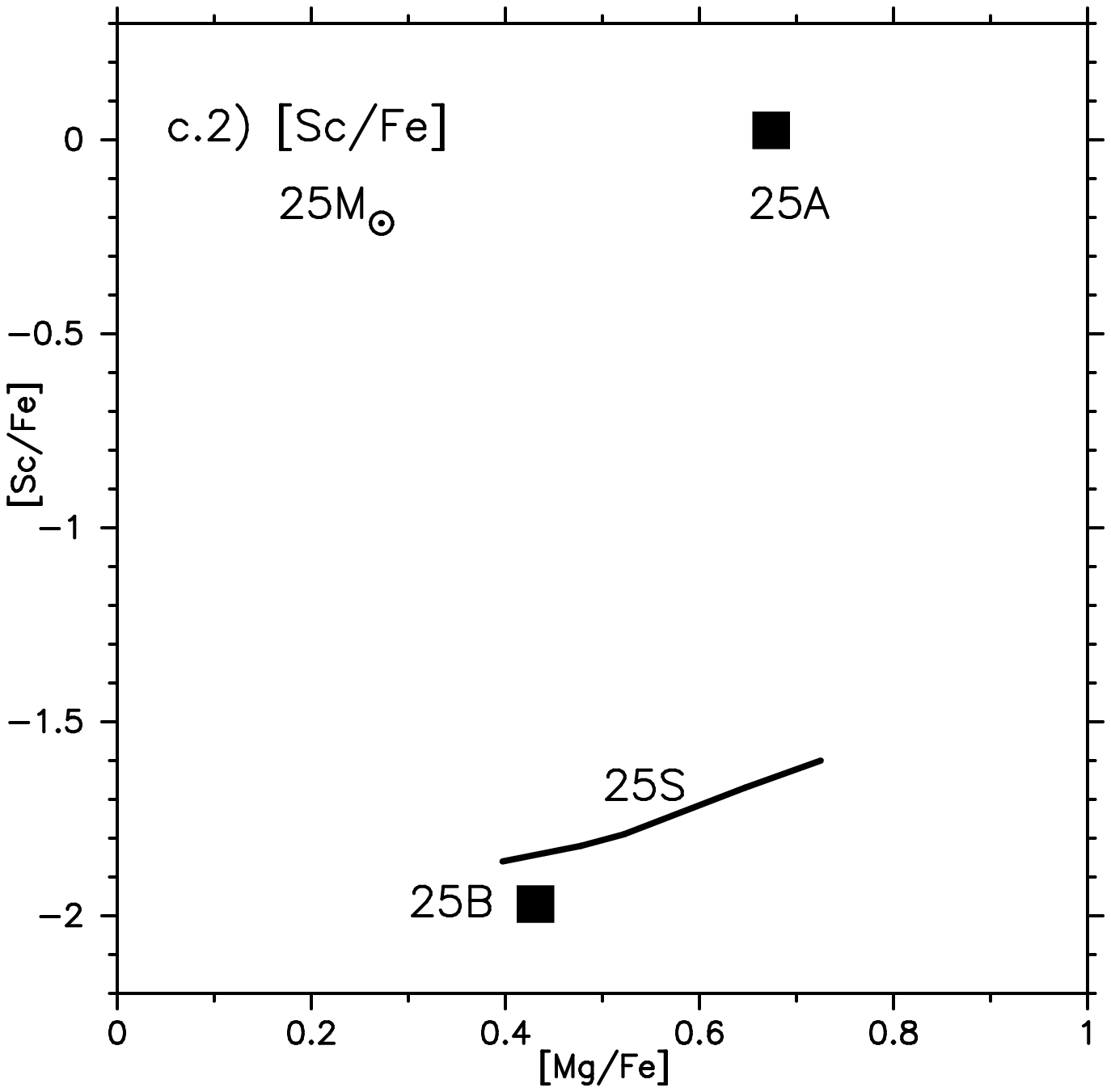}
	\end{minipage}
	\begin{minipage}[t]{0.4\textwidth}
		\epsscale{1.0}
		\plotone{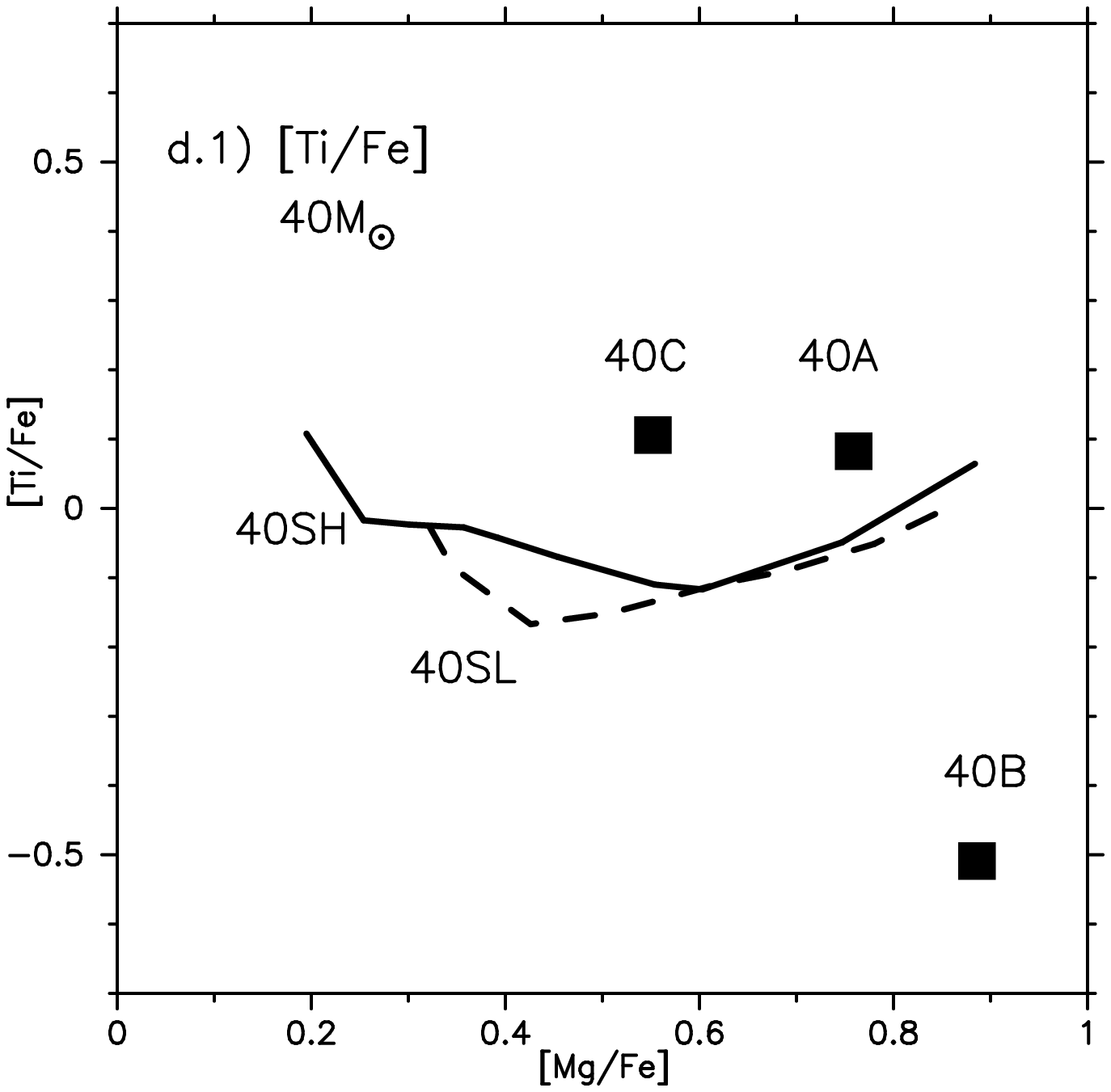}
	\end{minipage}
	\begin{minipage}[t]{0.4\textwidth}
		\epsscale{1.0}
		\plotone{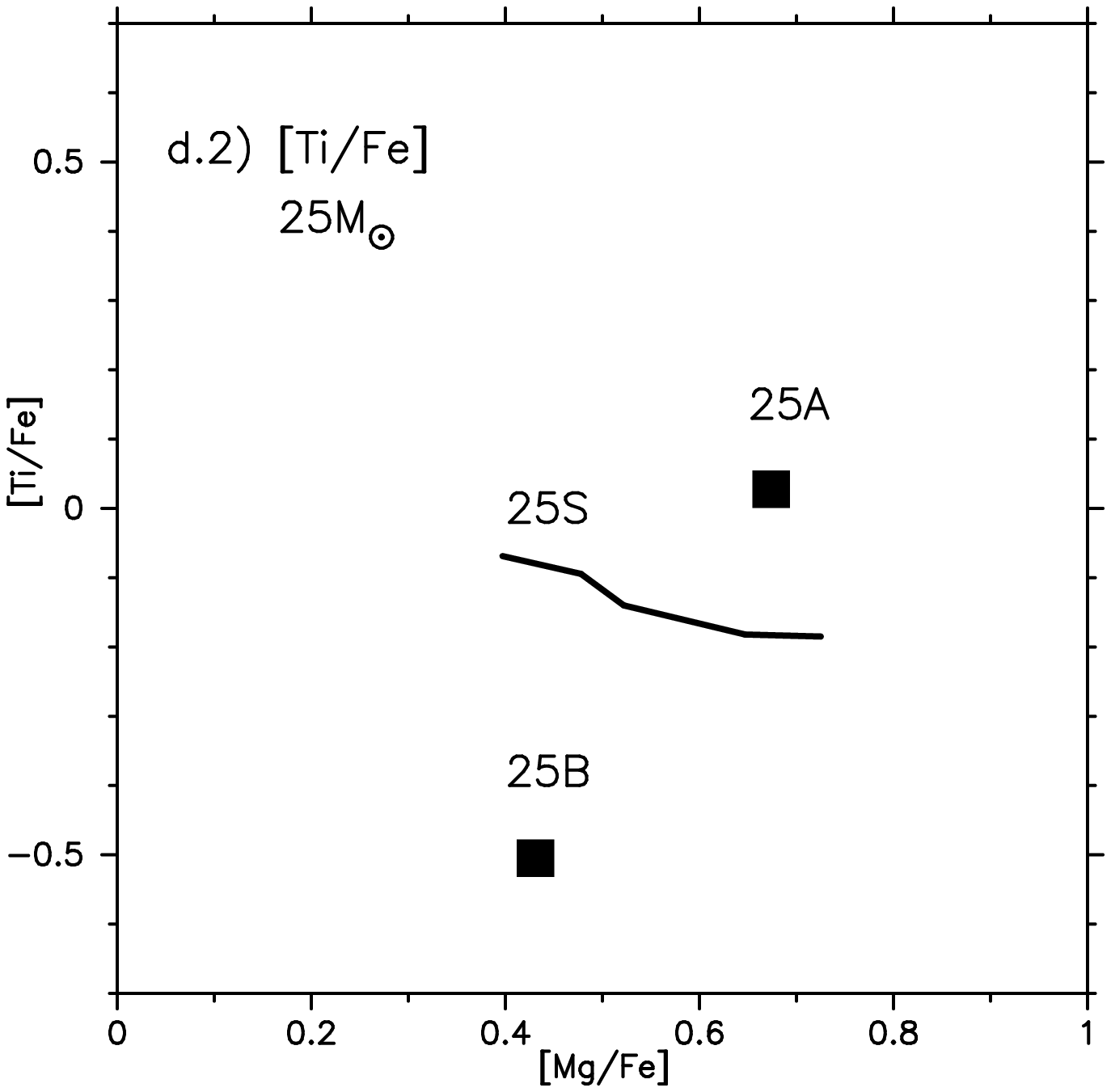}
	\end{minipage}
\end{center}
\end{figure}

\clearpage

\begin{figure}
\begin{center}
	\begin{minipage}[t]{0.4\textwidth}
		\epsscale{1.0}
		\plotone{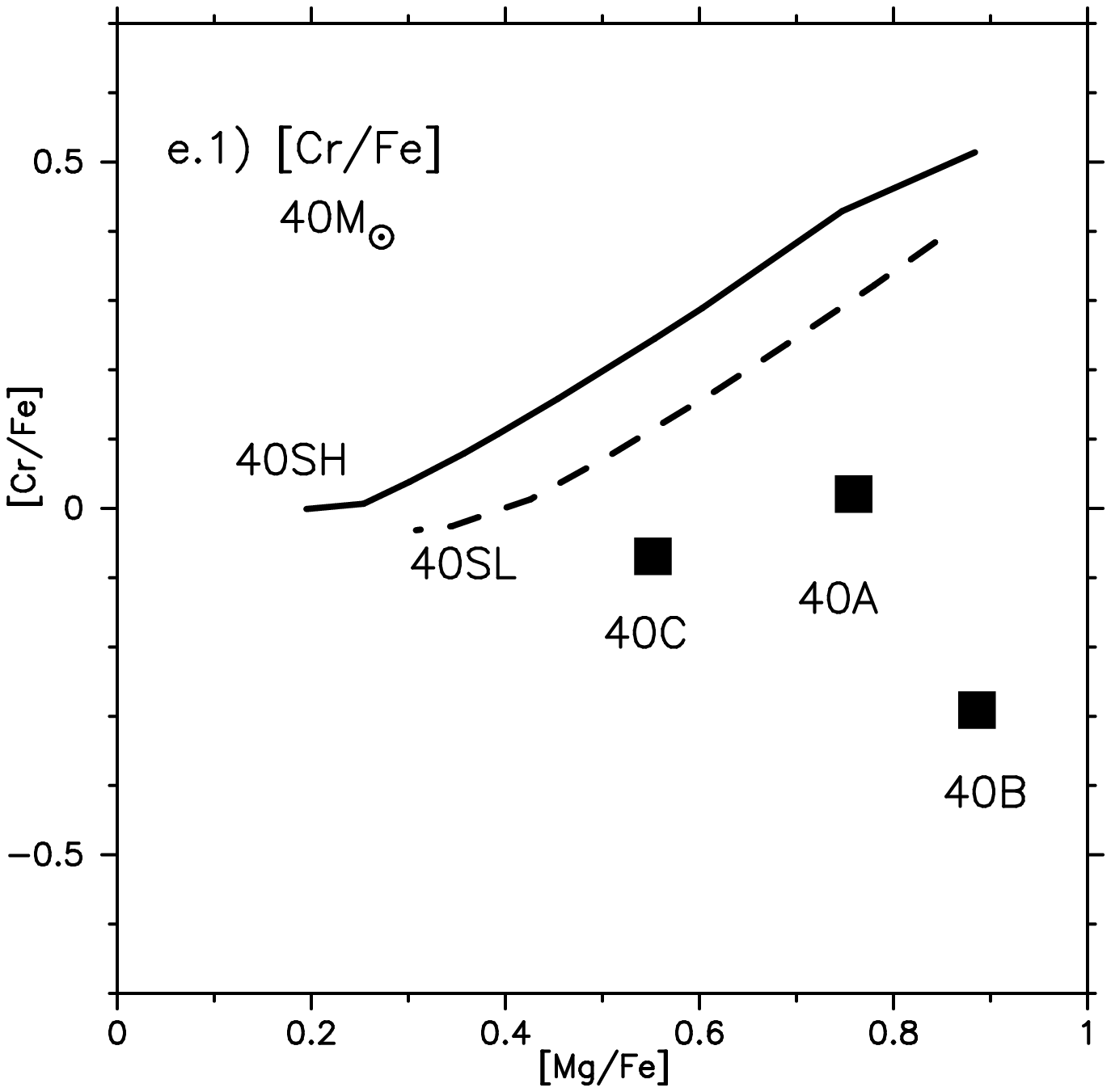}
	\end{minipage}
	\begin{minipage}[t]{0.4\textwidth}
		\epsscale{1.0}
		\plotone{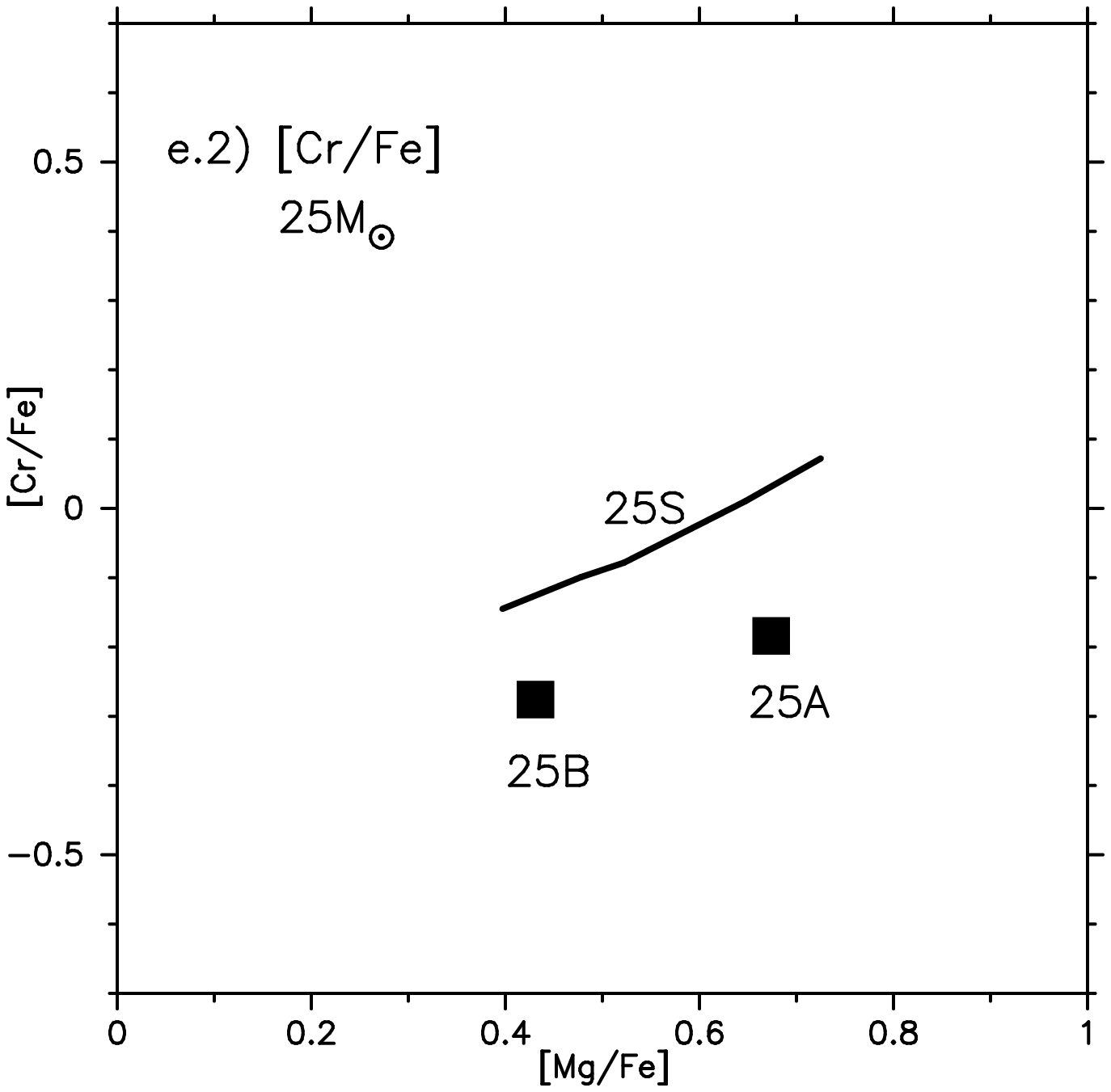}
	\end{minipage}
	\begin{minipage}[t]{0.4\textwidth}
		\epsscale{1.0}
		\plotone{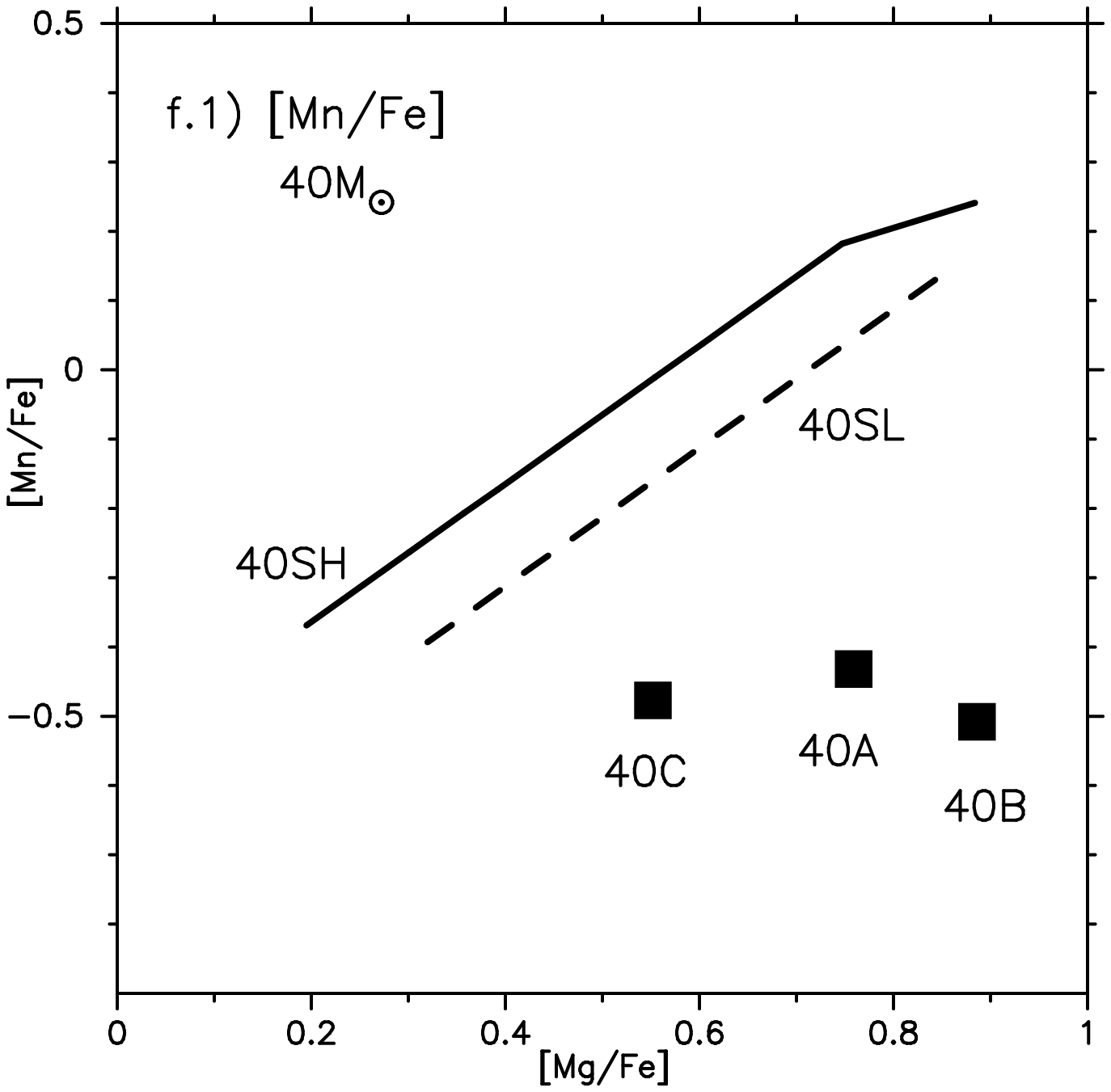}
	\end{minipage}
	\begin{minipage}[t]{0.4\textwidth}
		\epsscale{1.0}
		\plotone{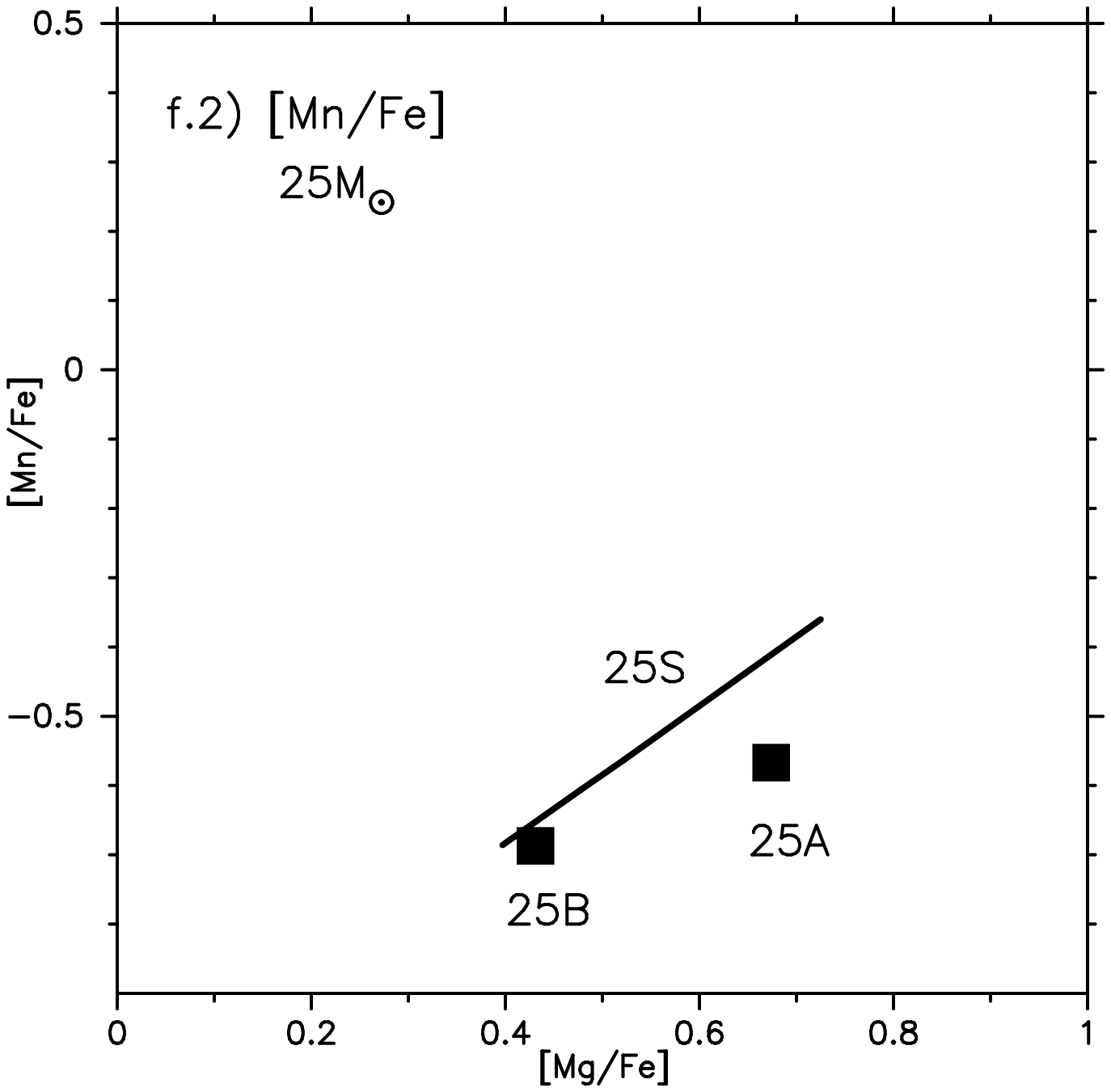}
	\end{minipage}
\end{center}
\end{figure}

\clearpage

\begin{figure}
\begin{center}
	\begin{minipage}[t]{0.4\textwidth}
		\epsscale{1.0}
		\plotone{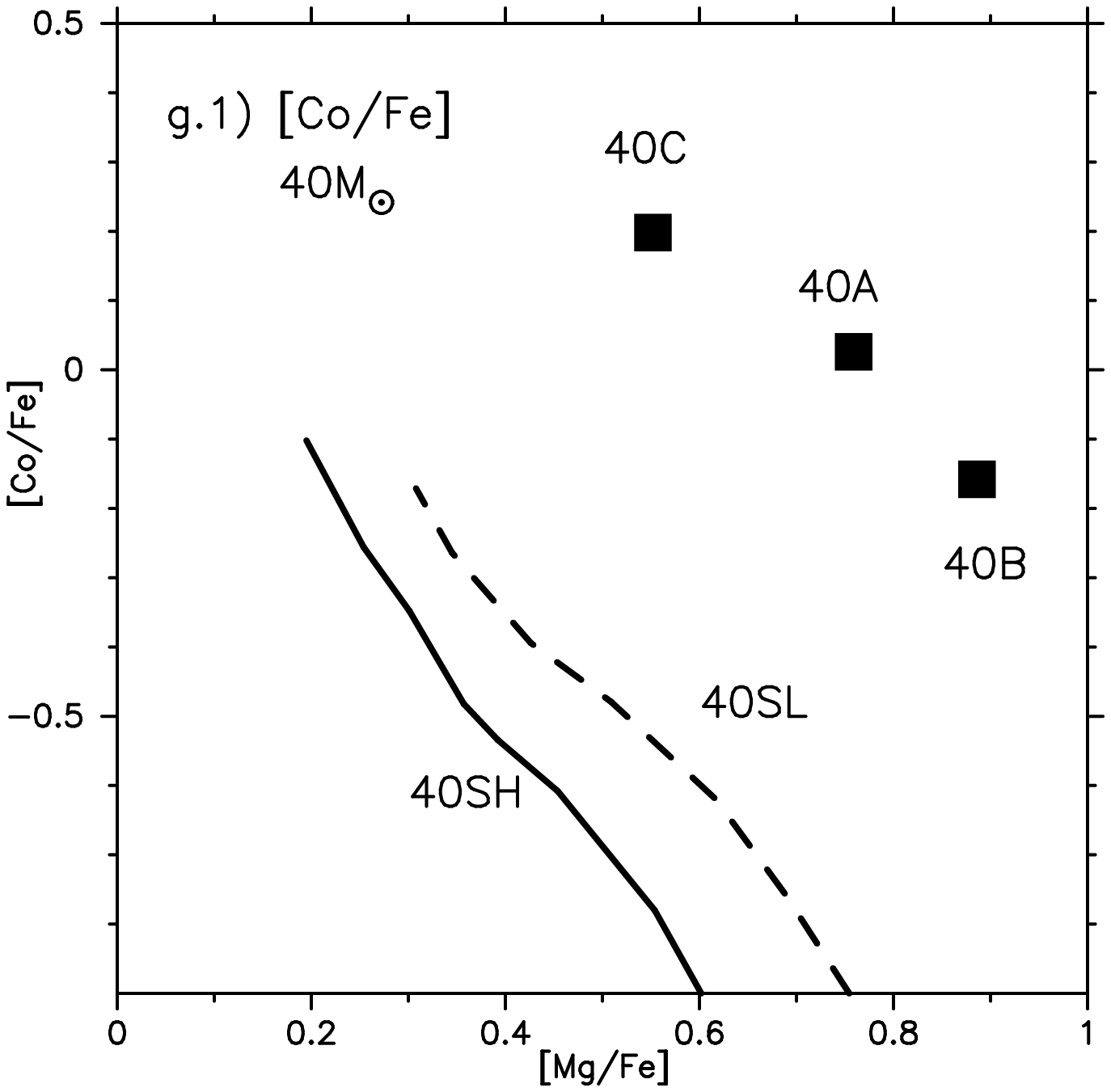}
	\end{minipage}
	\begin{minipage}[t]{0.4\textwidth}
		\epsscale{1.0}
		\plotone{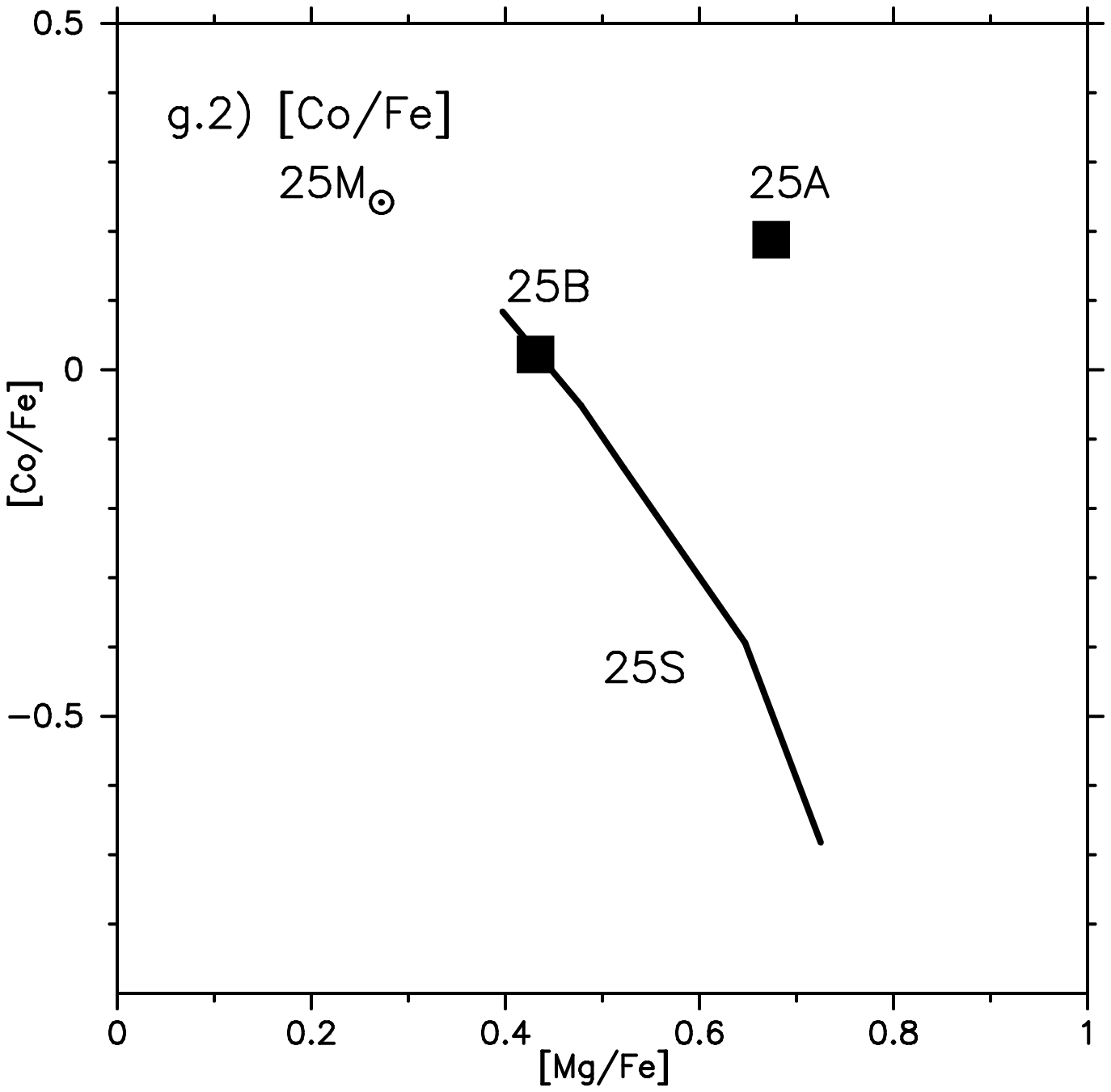}
	\end{minipage}
	\begin{minipage}[t]{0.4\textwidth}
		\epsscale{1.0}
		\plotone{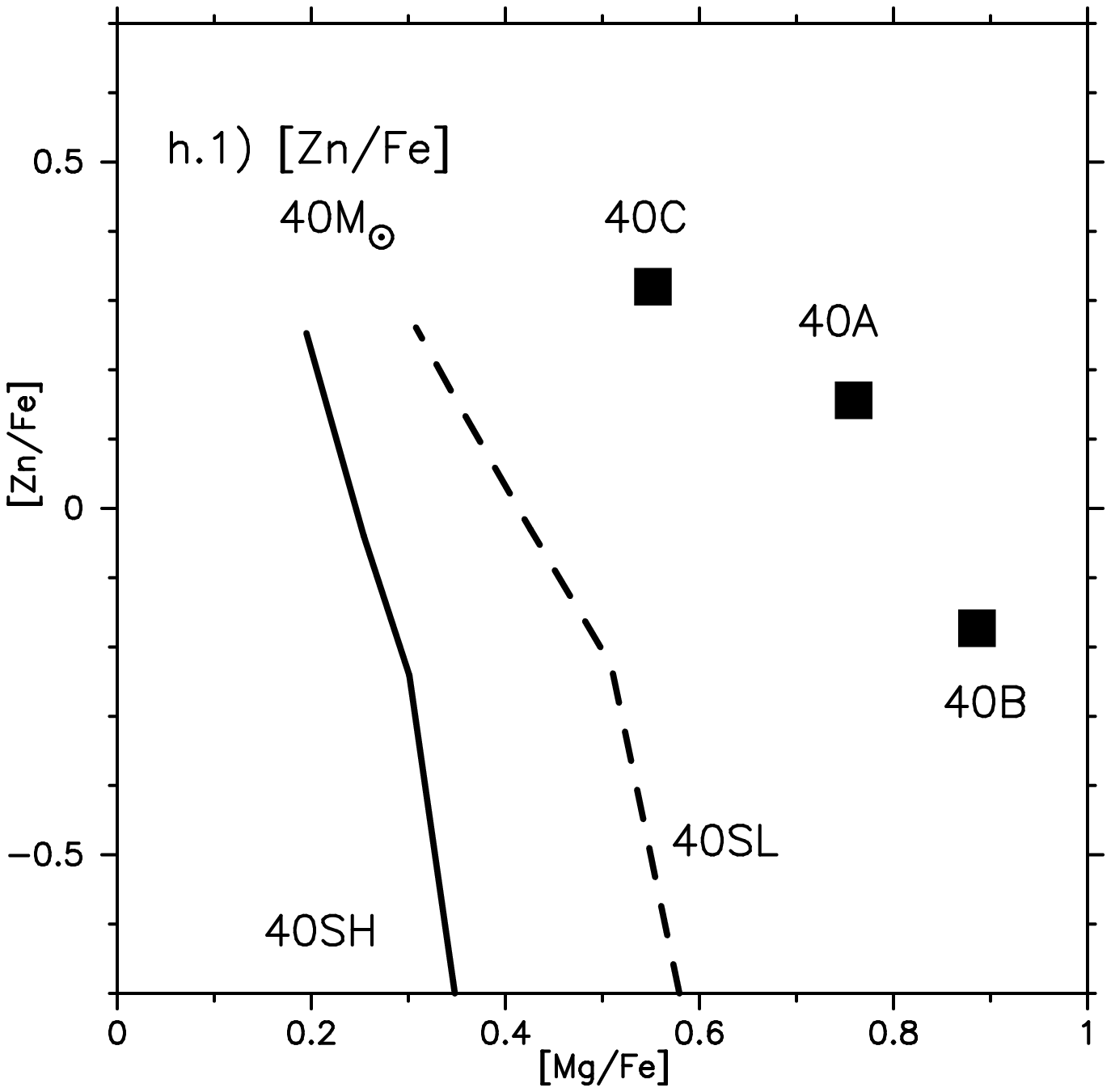}
	\end{minipage}
	\begin{minipage}[t]{0.4\textwidth}
		\epsscale{1.0}
		\plotone{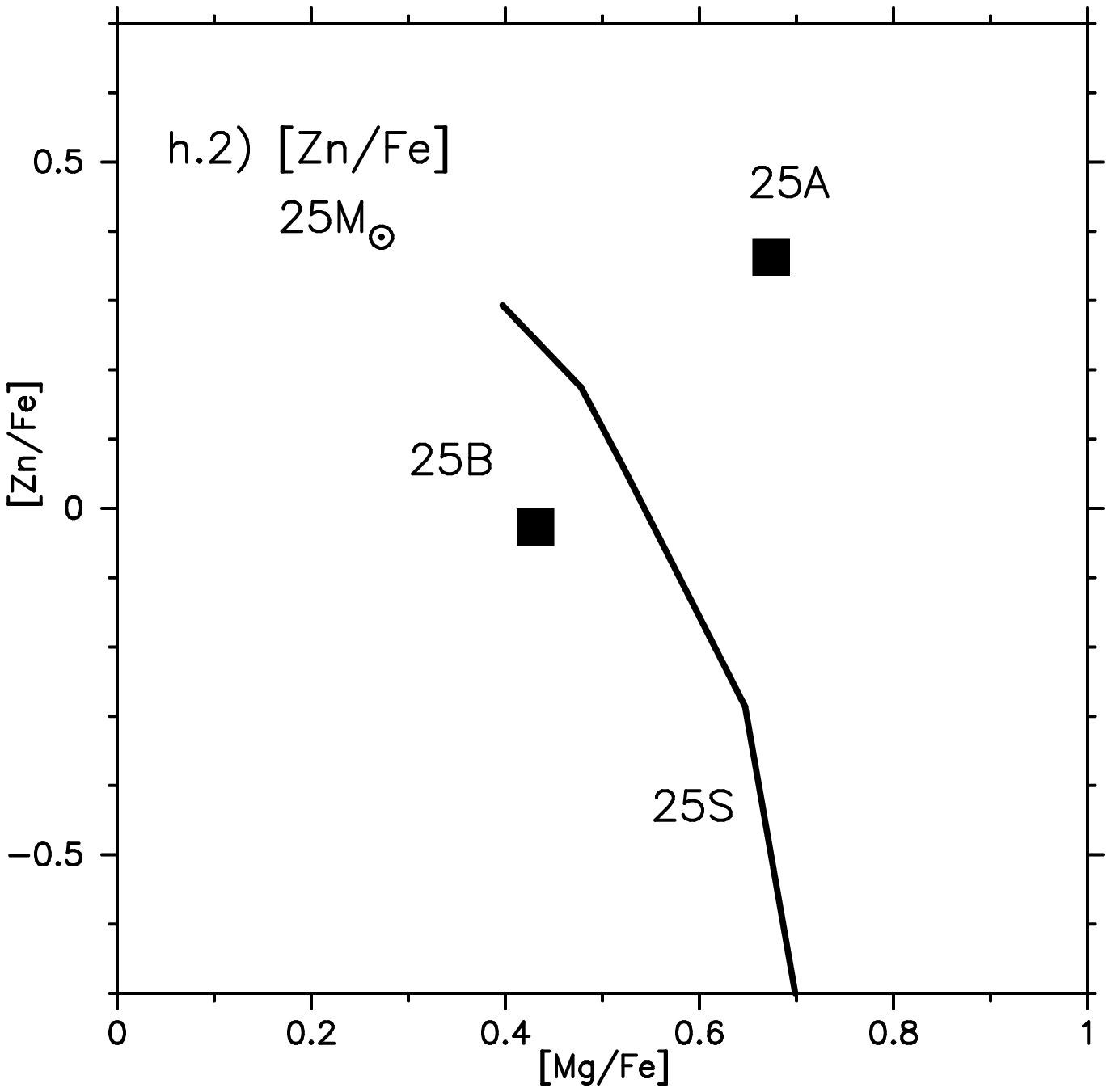}
	\end{minipage}
\end{center}
\end{figure}

\clearpage

\begin{figure}
\begin{center}
	\begin{minipage}[t]{0.4\textwidth}
		\epsscale{1.0}
		\plotone{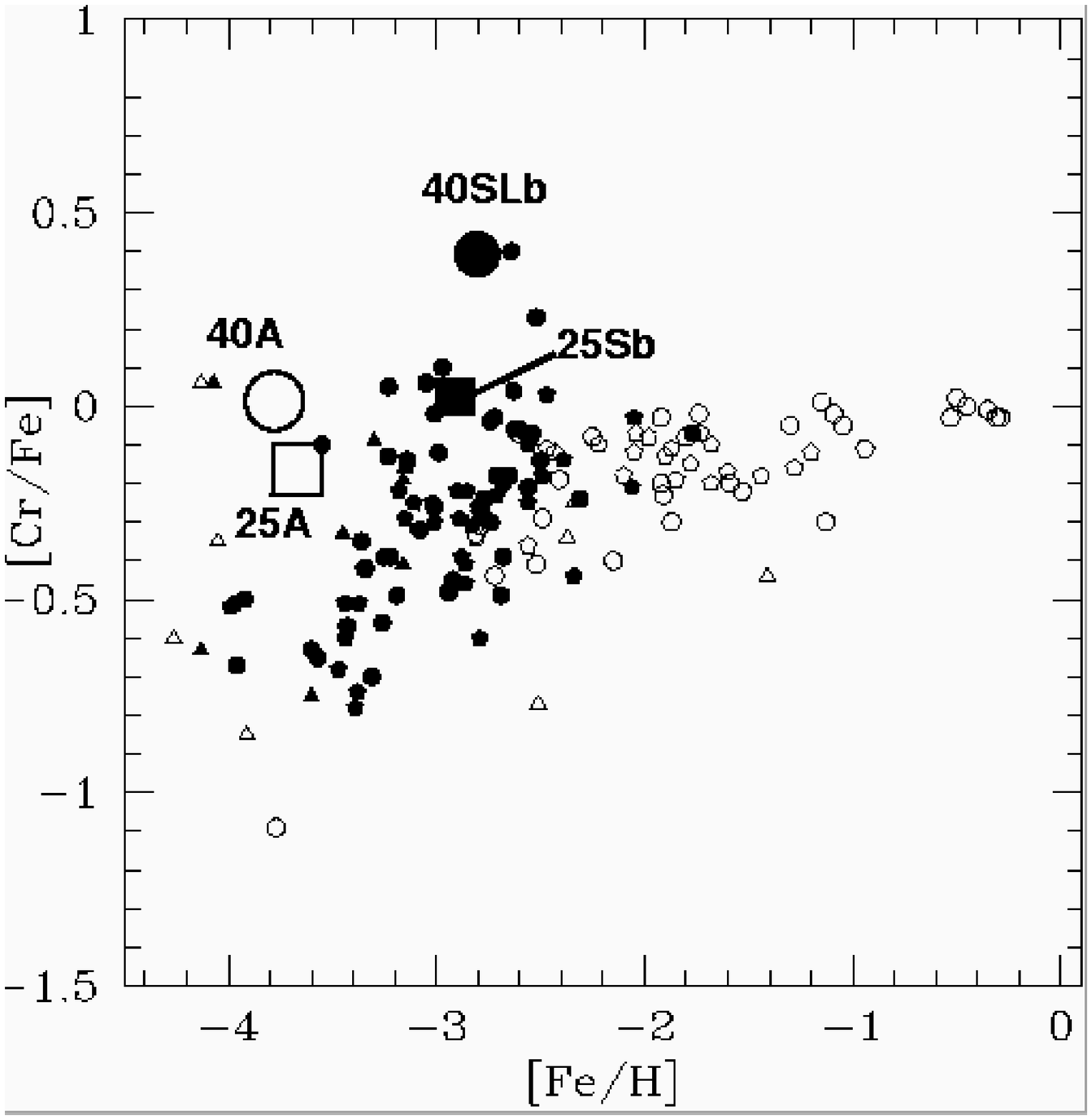}
	\end{minipage}
	\begin{minipage}[t]{0.4\textwidth}
		\epsscale{1.0}
		\plotone{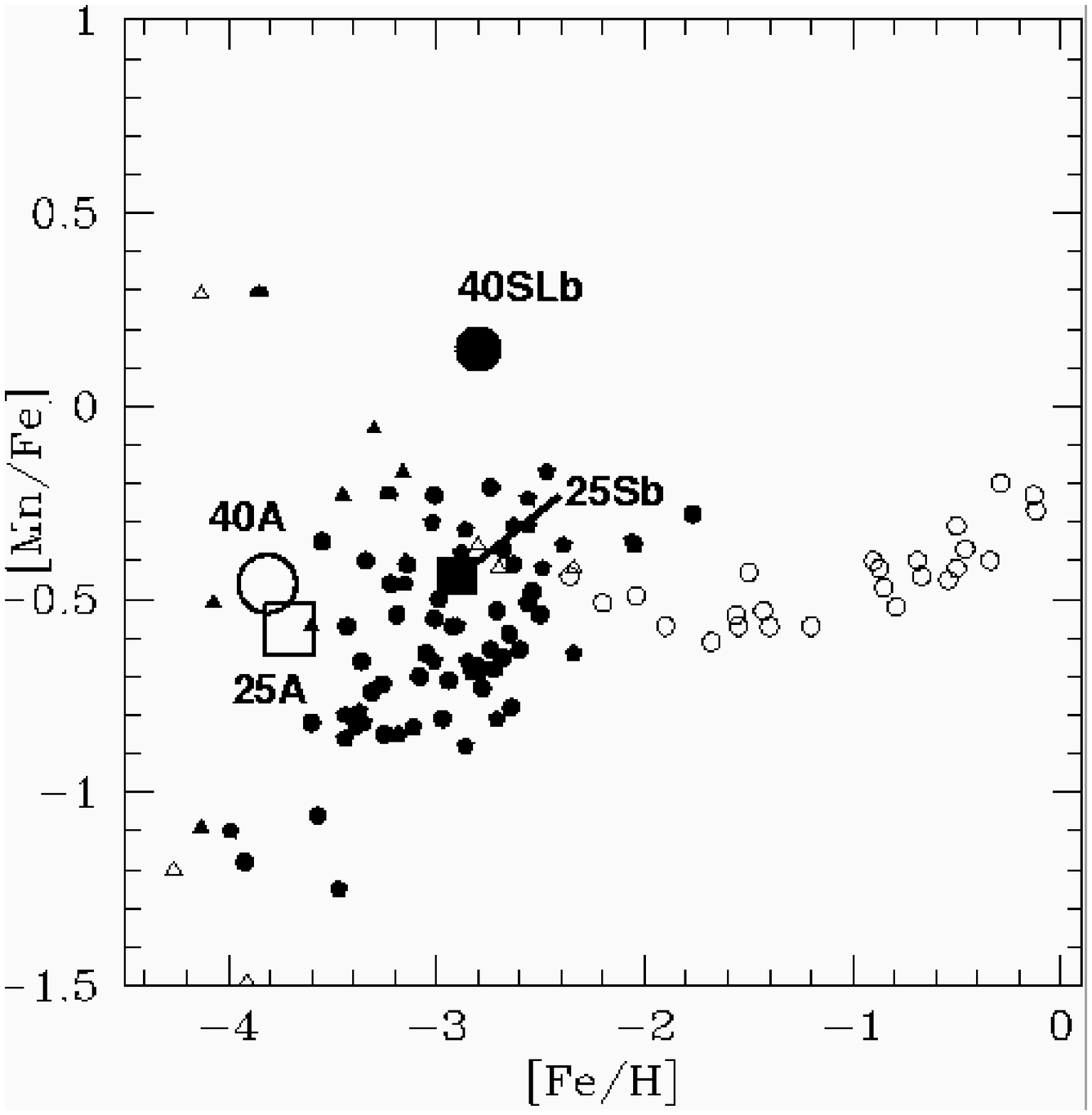}
	\end{minipage}
	\begin{minipage}[t]{0.4\textwidth}
		\epsscale{1.0}
		\plotone{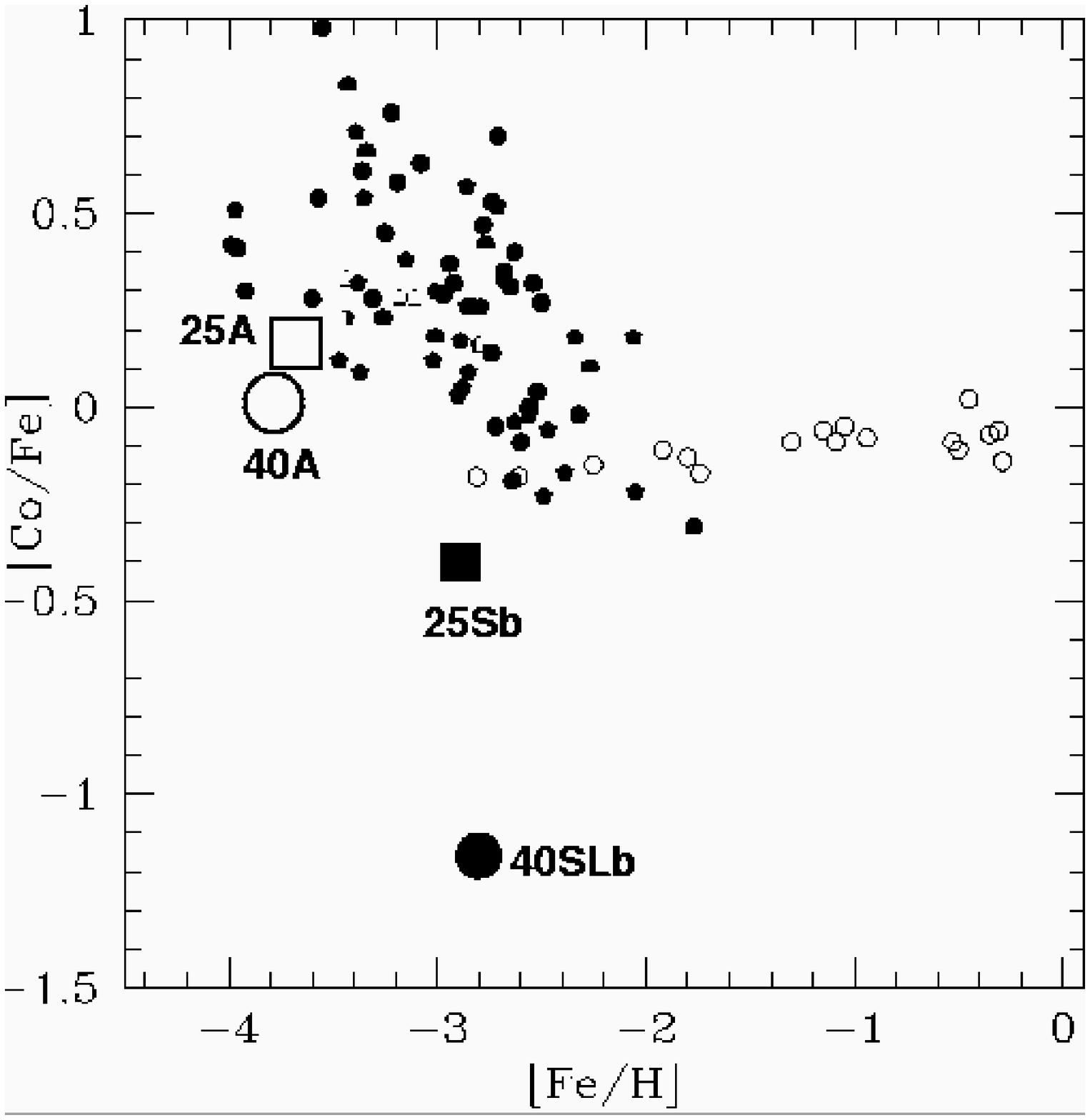}
	\end{minipage}
	\begin{minipage}[t]{0.4\textwidth}
		\epsscale{1.0}
		\plotone{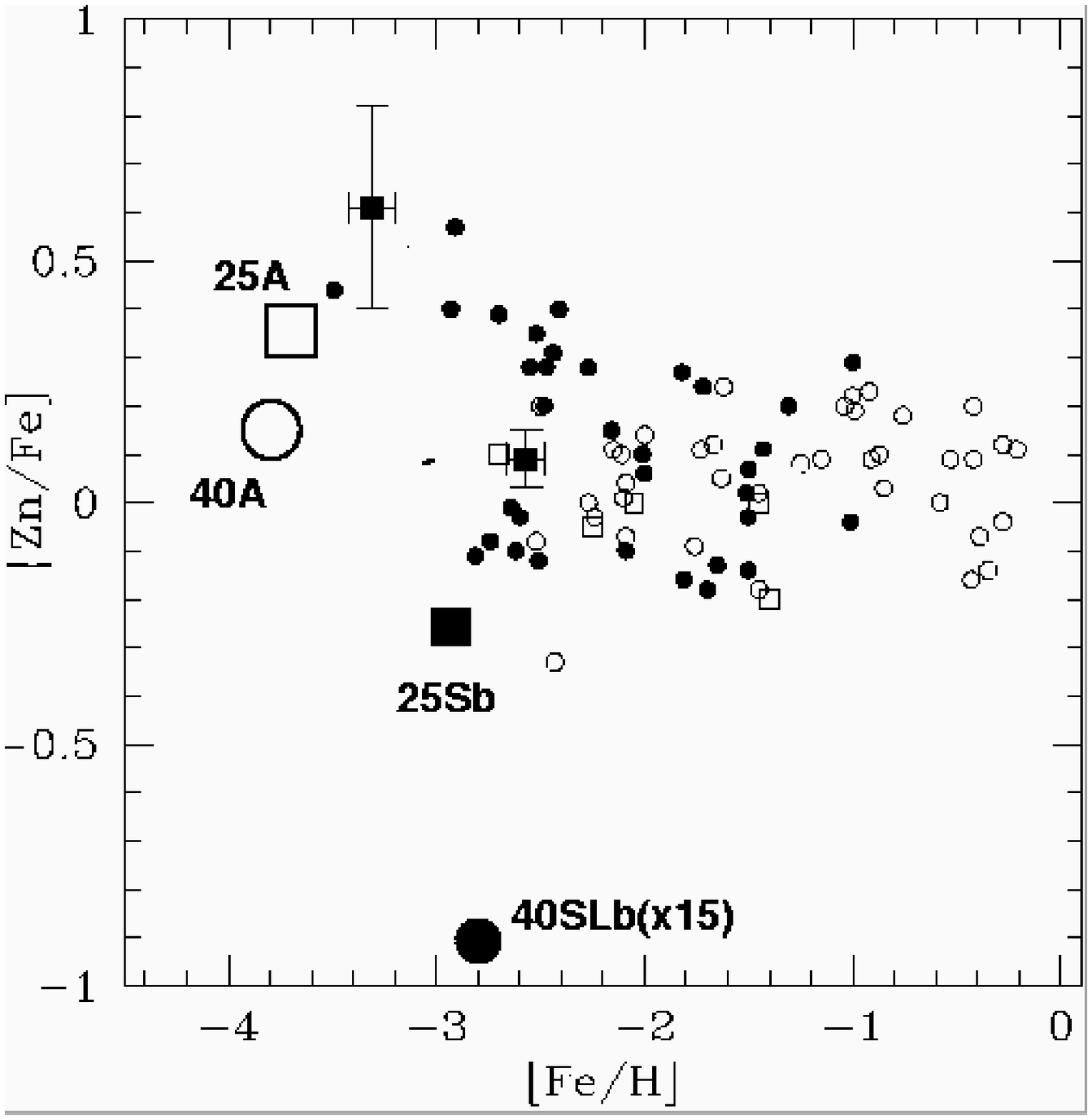}
	\end{minipage}
\end{center}
\caption{Observed abundance ratios of [Zn, Co, Mn, Cr/Fe] 
(McWilliam et al. 1995; Ryan et al. 1996; Primas et al. 2000; 
Blake et al. 2001), 
and the theoretical abundance patterns for 
normal supernovae (25Sb, 40SLb) 
and the bipolar models (25A, 40A). 
\label{f18}} 
\end{figure}

\clearpage

%\append.Figs.

\begin{figure}
\begin{center}
	\begin{minipage}[t]{0.4\textwidth}
		\epsscale{1.0}
		\plotone{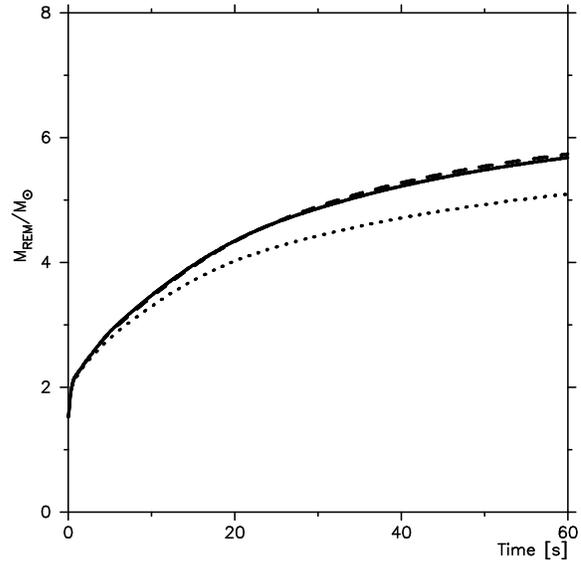}
	\end{minipage}
\end{center}
\caption{The effect of the numerical resolution on 
the growth of the mass of the central remnant ($M_{\rm REM}$) for 
model 40A. The numbers of meshes are $100 \times 30$ (solid), 
$200 \times 60$ (dashed), and $70 \times 21$ (dotted). 
\label{f19}}
\end{figure}

\clearpage

\begin{figure}
\begin{center}
	\begin{minipage}[t]{0.4\textwidth}
		\epsscale{1.0}
		\plotone{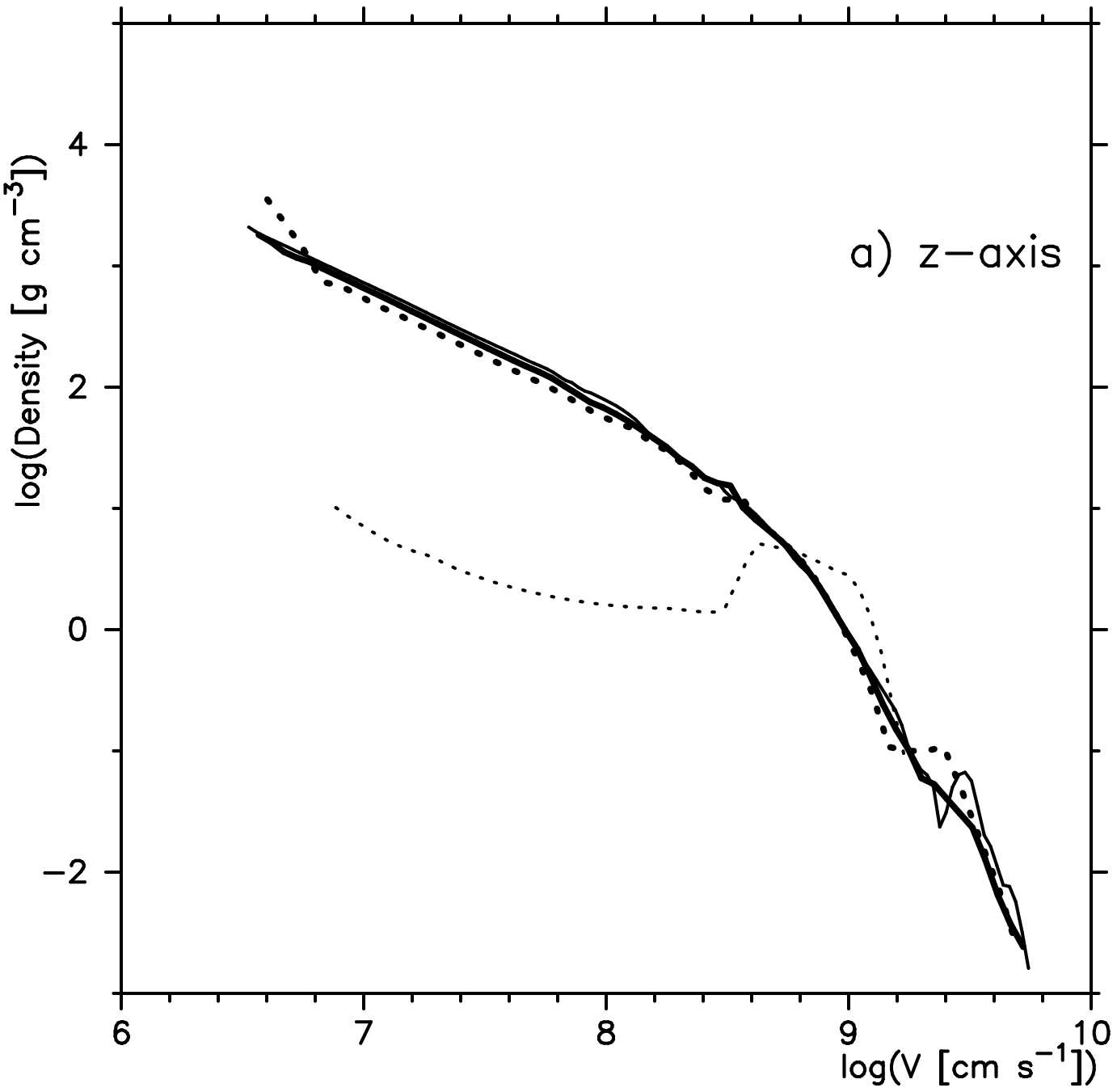}
	\end{minipage}
	\begin{minipage}[t]{0.4\textwidth}
		\epsscale{1.0}
		\plotone{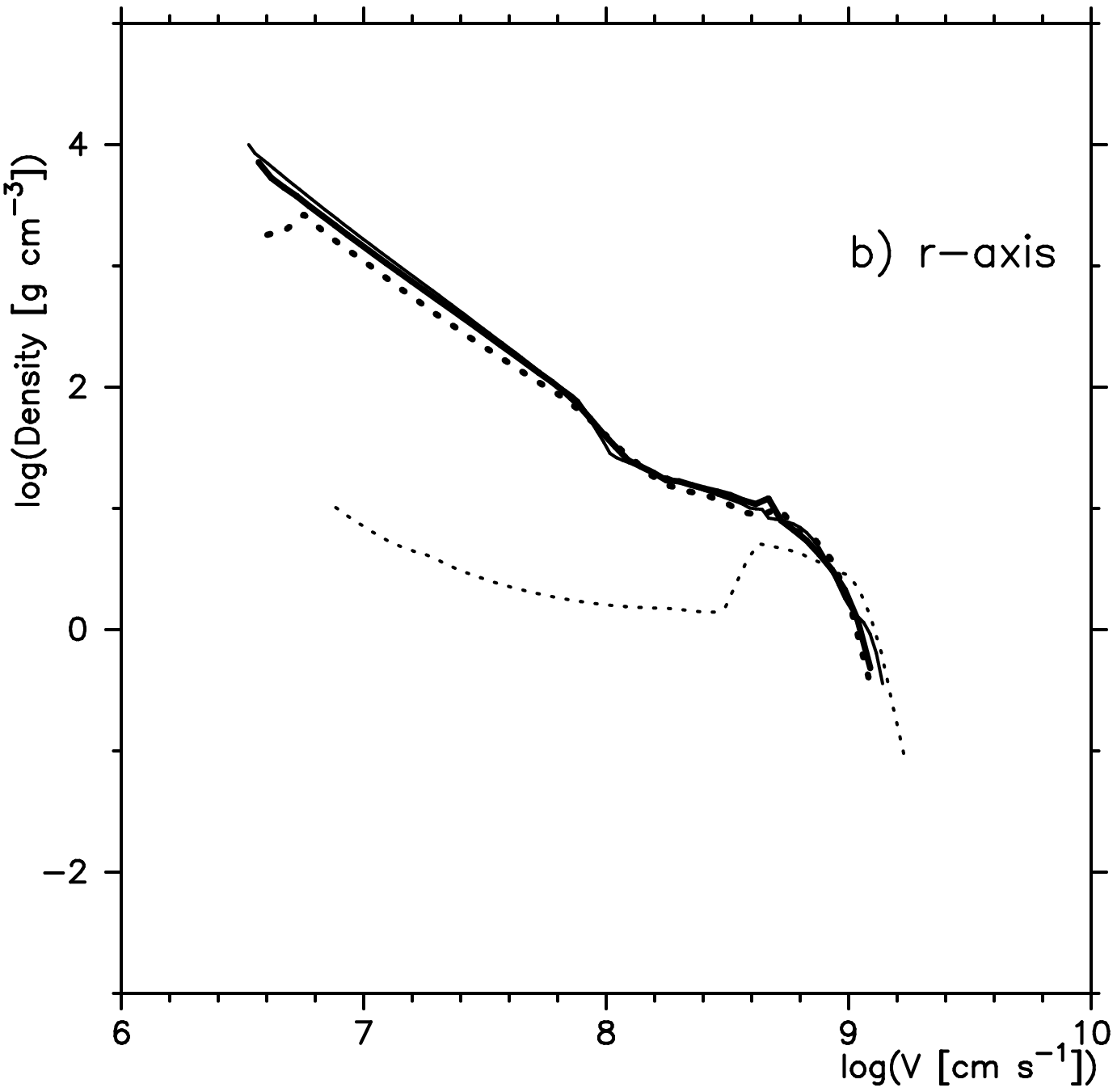}
	\end{minipage}
\end{center}
\caption{The effect of the numerical resolution on 
the density structure (at 100 sec). 
(a) Along the $z$-axis of 40A.
The numbers of meshes are $100 \times 30$ (thick-solid), 
$200 \times 60$ (thin-solid), and $70 \times 21$ (thick-dotted).
Also shown is model 40SH with $100 \times 30$ (thin-dotted). 
(b) Along the $r$-axis.
\label{f20}}
\end{figure}

\clearpage

\begin{figure}
\begin{center}
	\begin{minipage}[t]{0.4\textwidth}
		\epsscale{1.0}
		\plotone{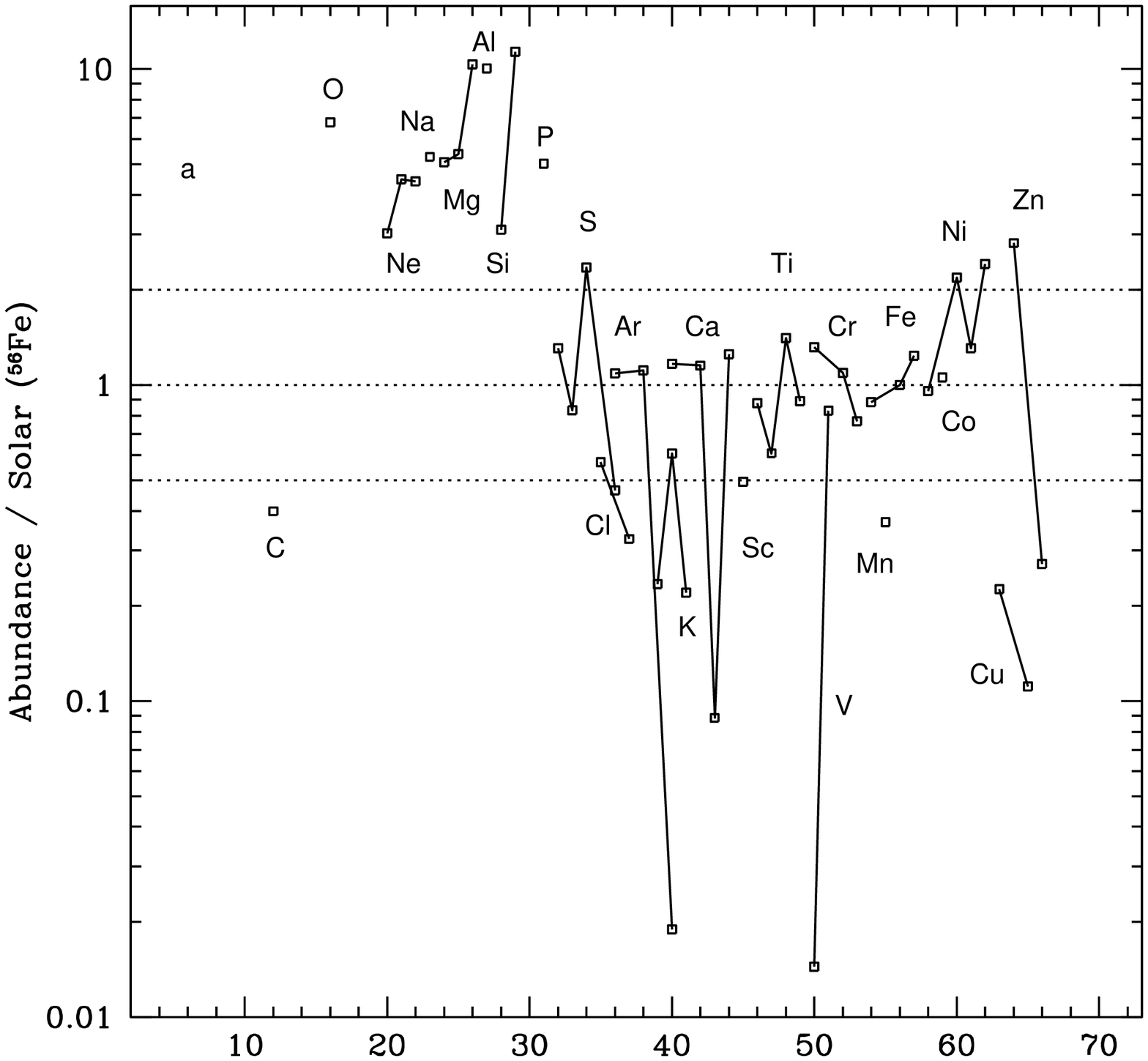}
	\end{minipage}
	\begin{minipage}[t]{0.4\textwidth}
		\epsscale{1.0}
		\plotone{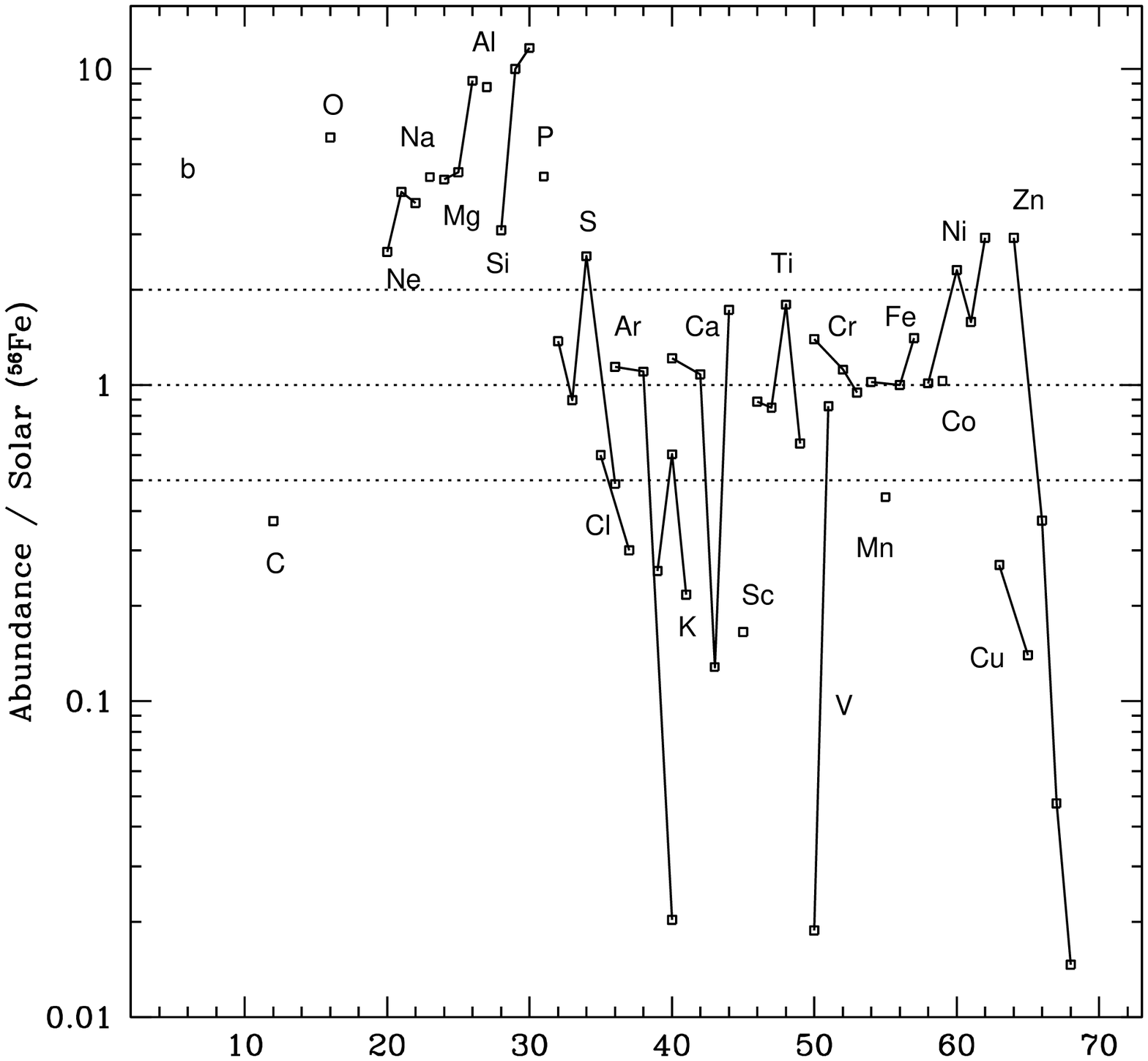}
	\end{minipage}
\end{center}
\caption{The effect of the numerical resolution on 
the isotopic yields (of model 40A; after decay of radioactive species). 
(a) The result of the run with $100 \times 30$ numerical grids. 
(b) The reslut of the run with $200 \times 60$ numerical grids.
\label{f21}}
\end{figure}

\clearpage

%\begin{figure}
%\plottwo{f2a.eps}{f2b.eps}
%\caption{This is an example of a multipart figure with a long figure caption 
%that must be set as a paragraph.  The processor has to buffer the text of the 
%caption, so it is good not to be too wordy, but that would make for 

%poor communication as well.\label{fig2}} 

%\end{figure} % 
%% If you are not including electonic art with your submission, you may 

%% mark up your captions using the \figcaption command. See the 

%% so if you have more than seven captions, insert a \clearpage 

%% Tables should be submitted one per page, so put a \clearpage before 

%% Two options are available to the author for producing tables:  the 
%% deluxetable environment provided by the AASTeX package or the LaTeX 
%% table environment.  Use of deluxetable is preferred. 
%%

%% Three table samples follow, two marked up in the deluxetable environment, 
%% one marked up as a LaTeX table.

%% In this first example, note that the \tabletypesize{}
%% command has been used to reduce the font size of the table. 
%% Note also that the \label command needs to be placed 

\begin{table}
\begin{center}
\scriptsize

\begin{tabular}{cccccccc}\hline\hline
{\bf Model} & 
{\bf $M$($\msun$)} &
{\bf$M_{\rm REM0}$($\msun$)}   & 
{\bf $\alpha$} &
{\bf $\theta_{\rm jet}$(degree)} & 
{\bf $E_{\rm total}$($10^{51}$ergs)} &
{\bf $M_{\rm REM}$($\msun$)} &
{\bf $M$($^{56}$Ni)($\msun$)}\\\hline

40A & 40 & 1.5 & 0.01 & 15 & 10.9 & 5.9 & 1.07E-1\\
40B & 40 & 1.5 & 0.01 & 45 & 1.2 & 6.8 & 8.11E-2\\
40C & 40 & 1.5 & 0.05 & 15 & 32.4 & 2.9 & 2.40E-1\\
40D & 40 & 3.0 & 0.01 & 15 & 8.5 &10.5 & 6.28E-8\\
\hline
40SHa & 40 & & & & 10.0 & 2.0 & 5.40E-1\\
40SHb & 40 & & & & 10.0 & 2.7 & 1.00E-1\\
40SLa & 40 & & & & 1.0 & 2.0 & 4.38E-1\\
40SLb & 40 & & & & 1.0 & 2.5 & 1.00E-1\\
\hline
25A & 25 & 1.0 & 0.01 & 15 & 6.7 & 1.9 & 7.81E-2\\
25B & 25 & 1.0 & 0.01 & 45 & 0.6 & 1.5 & 1.51E-1\\
\hline
25Sa & 25 & & & & 1.0 & 1.5 & 1.64E-1\\
25Sb & 25 & & & & 1.0 & 1.6 & 1.00E-1\\\hline
\end{tabular}
\caption{Models
\label{tab1}}
\end{center}
\end{table}

\clearpage

\begin{table}
\begin{center}
\scriptsize

\begin{tabular}{cccccccccc}\hline\hline
{\bf Species} & 
{\bf 40A} &
{\bf 40B} & 
{\bf 40C} &
{\bf 40D} & 
{\bf Species} &
{\bf 40A} &
{\bf 40B} &
{\bf 40C} &
{\bf 40D}\\
{\bf $M$($^{56}$Ni)} &
{1.1E-01} &
{8.1E-02} &
{2.4E-01} &
{6.3E-08} &
{} &
{1.1E-01} &
{8.1E-02} &
{2.4E-01} &
{6.3E-08}\\\hline
$^{4}$He & 1.88E+00& 1.85E+00& 1.99E+00& 1.81E+00&
$^{43}$Ca& 7.30E-07& 3.90E-08& 1.43E-06& 1.15E-08\\
$^{12}$C & 1.11E-01& 1.13E-01& 1.29E-01& 8.44E-02&
$^{44}$Ca& 1.64E-04& 1.50E-05& 4.26E-04& 8.51E-08\\
$^{13}$C & 2.76E-08& 3.21E-09& 8.95E-08& 1.58E-08&
$^{46}$Ca& 3.05E-10& 8.62E-11& 7.92E-10& 1.52E-10\\
$^{14}$N & 6.83E-05& 6.83E-05& 6.84E-05& 8.80E-05&
$^{45}$Sc& 1.77E-06& 9.08E-08& 1.04E-05& 1.81E-08\\
$^{15}$N & 8.45E-09& 2.45E-09& 2.63E-08& 5.52E-09&
$^{46}$Ti& 1.80E-05& 1.28E-05& 5.05E-05& 2.76E-06\\
$^{16}$O & 5.99E+00& 5.97E+00& 8.29E+00& 2.61E+00&
$^{47}$Ti& 1.16E-05& 1.15E-06& 1.81E-05& 5.65E-08\\
\hline
\end{tabular}
\caption{Detailed yields ($\msun$) after decay of radioactive species: 
For models 40A, 40B, 40C, and 40D.
\label{tab2}}
\end{center}
\end{table}

\clearpage

\begin{table}
\begin{center}
\scriptsize

\begin{tabular}{cccccccccc}\hline\hline 

{\bf Species} & 
{\bf 40SHa} &
{\bf 40SHb} & 
{\bf 40SLa} &
{\bf 40SLb} & 
{\bf Species} &
{\bf 40SHa} &
{\bf 40SHb} &
{\bf 40SLa} &
{\bf 40SLb}\\
{\bf $M$($^{56}$Ni)} &
{5.4E-01} &
{1.0E-01} &
{4.4E-01} &
{1.0E-01} &
{} &
{5.4E-01} &
{1.0E-01} &
{4.4E-01} &
{1.0E-01}\\\hline
$^{4}$He & 2.01E+00& 1.79E+00& 1.90E+00& 1.83E+00&
$^{43}$Ca& 3.85E-06& 8.75E-08& 1.14E-06& 8.94E-08\\
$^{12}$C & 1.46E-01& 1.46E-01& 1.51E-01& 1.49E-01&
$^{44}$Ca& 1.18E-03& 6.76E-06& 6.52E-04& 6.90E-06\\
$^{13}$C & 1.83E-08& 1.83E-08& 1.53E-08& 1.87E-08&
$^{46}$Ca& 1.09E-09& 1.09E-09& 5.30E-10& 1.11E-09\\
$^{14}$N & 8.19E-05& 8.19E-05& 8.19E-05& 8.36E-05&
$^{45}$Sc& 6.42E-07& 2.45E-07& 5.33E-07& 2.51E-07\\
$^{15}$N & 1.46E-08& 1.46E-08& 1.11E-08& 1.49E-08&
$^{46}$Ti& 2.78E-05& 2.20E-05& 8.62E-05& 2.24E-05\\
$^{16}$O & 8.82E+00& 8.82E+00& 8.94E+00& 9.01E+00&
$^{47}$Ti& 5.36E-05& 5.85E-07& 5.14E-05& 5.98E-07\\
\hline
\end{tabular}
\caption{Detailed yields ($\msun$) after decay of radioactive species: 
For models 40SH and 40SL (with different $M_{\rm cut}$).
\label{tab3}}
\end{center}
\end{table}

\clearpage

\begin{table}
\begin{center}
\scriptsize

\begin{tabular}{cccccccccc}\hline\hline 

{\bf Species} & 
{\bf 25A} &
{\bf 25B} & 
{\bf 25Sa} &
{\bf 25Sb} & 
{\bf Species} &
{\bf 25A} &
{\bf 25B} &
{\bf 25Sa} &
{\bf 25Sb}\\
{\bf $M$($^{56}$Ni)} &
{7.8E-02} &
{1.5E-01} &
{1.6E-01} &
{1.0E-01} &
{} &
{7.8E-02} &
{1.5E-01} &
{1.6E-01} &
{1.0E-01}\\\hline
$^{4}$He & 1.82E+00& 1.80E+00& 1.82E+00& 1.79E+00&
$^{43}$Ca& 7.33E-07& 8.98E-08& 7.83E-07& 4.47E-07\\
$^{12}$C & 1.35E-01& 1.38E-01& 1.41E-01& 1.41E-01&
$^{44}$Ca& 1.31E-04& 3.67E-05& 2.14E-04& 5.73E-05\\
$^{13}$C & 1.56E-07& 2.13E-09& 2.22E-09& 2.23E-09&
$^{46}$Ca& 1.31E-10& 9.74E-11& 1.20E-10& 1.20E-10\\
$^{14}$N & 1.02E-03& 1.02E-03& 9.57E-04& 9.57E-04&
$^{45}$Sc& 2.65E-06& 5.09E-08& 7.26E-08& 6.28E-08\\
$^{15}$N & 1.40E-08& 1.88E-09& 2.04E-09& 2.04E-09&
$^{46}$Ti& 8.03E-06& 6.66E-06& 1.32E-05& 6.30E-06\\
$^{16}$O & 2.66E+00& 2.86E+00& 2.91E+00& 2.91E+00&
$^{47}$Ti& 8.44E-06& 9.42E-06& 2.50E-05& 6.20E-06\\
\hline
\end{tabular}
\caption{Detailed yields ($\msun$) after decay of radioactive species: 
For models 25A, 25B, and 25S.
\label{tab4}}
\end{center}
\end{table}

\clearpage

\begin{table}
\begin{center}
\scriptsize

\begin{tabular}{ccccc}\hline\hline 

{\bf Species} & 
{\bf 40A} &
{\bf 40B} & 
{\bf 40C} &
{\bf 40D}\\
{\bf $M$($^{56}$Ni)} &
{1.1E-01} &
{8.1E-02} &
{2.4E-01} &
{6.3E-08}\\\hline

 $^{4}$He & 1.88E+00& 1.85E+00& 1.99E+00& 1.81E+00\\
 $^{6}$Li & 3.23E-14& 1.46E-15& 2.01E-16& 2.85E-21\\
 $^{7}$Li & 9.18E-19& 2.97E-20& 2.55E-19& 6.62E-20\\
 $^{9}$Be & 9.97E-17& 7.43E-26& 3.06E-16& 6.76E-17\\
 $^{10}$B & 3.25E-15& 2.43E-16& 2.14E-17& 1.30E-22\\
 $^{11}$B & 2.68E-14& 8.48E-15& 6.62E-14& 1.28E-14\\
\hline
\end{tabular}
\caption{Detailed yields ($\msun$) before decay of radioactive species: 
For models 40A, 40B, 40C, and 40D
\label{tab5}}
\end{center}
\end{table}

\clearpage

\begin{table}
\begin{center}
\scriptsize

\begin{tabular}{ccccc}\hline\hline 

{\bf Species} & 
{\bf 40A} &
{\bf 40B} & 
{\bf 40C} &
{\bf 40D}\\
{\bf $M$($^{56}$Ni)} &
{1.1E-01} &
{8.1E-02} &
{2.4E-01} &
{6.3E-08}\\\hline

 $^{33}$Ar& 1.17E-10& 5.41E-11& 3.81E-10& 1.44E-21\\
 $^{34}$Ar& 2.35E-06& 2.77E-08& 9.49E-06& 5.87E-20\\
 $^{35}$Ar& 3.42E-06& 1.79E-07& 1.02E-05& 2.23E-13\\
 $^{36}$Ar& 7.75E-03& 4.64E-03& 1.69E-02& 5.50E-04\\
 $^{37}$Ar& 2.13E-05& 1.62E-05& 6.14E-05& 3.85E-06\\
 $^{38}$Ar& 1.57E-03& 1.24E-03& 4.92E-03& 4.03E-04\\
\hline
\end{tabular}
\end{center}
\end{table}

\clearpage

\begin{table}
\begin{center}
\scriptsize

\begin{tabular}{ccccc}\hline\hline 

{\bf Species} & 
{\bf 40A} &
{\bf 40B} & 
{\bf 40C} &
{\bf 40D}\\
{\bf $M$($^{56}$Ni)} &
{1.1E-01} &
{8.1E-02} &
{2.4E-01} &
{6.3E-08}\\\hline

 $^{51}$Mn& 2.65E-05& 7.45E-06& 4.84E-05& 3.92E-09\\
 $^{52}$Mn& 4.55E-07& 3.24E-07& 8.52E-07& 2.17E-08\\
 $^{53}$Mn& 1.03E-05& 7.51E-06& 2.35E-05& 1.48E-06\\
 $^{54}$Mn& 5.20E-08& 3.12E-08& 1.23E-07& 7.52E-09\\
 $^{55}$Mn& 1.81E-08& 9.76E-09& 4.32E-08& 2.49E-09\\
 $^{56}$Mn& 1.03E-13& 2.71E-14& 2.10E-13& 8.74E-15\\
\hline
\end{tabular}
\end{center}
\end{table}

\clearpage

\begin{table}
\begin{center}
\scriptsize

\begin{tabular}{ccccc}\hline\hline 
{\bf Species} & 
{\bf 40SHa} &
{\bf 40SHb} & 
{\bf 40SLa} &
{\bf 40SLb}\\
{\bf $M$($^{56}$Ni)} &
{5.4E-01} &
{9.0E-02} &
{4.4E-01} &
{1.1E-01}\\\hline
 $^{4}$He & 2.01E+00& 1.79E+00& 1.90E+00& 1.80E+00\\
 $^{6}$Li & 2.10E-20& 2.10E-20& 1.80E-20& 1.80E-20\\
 $^{7}$Li & 4.19E-19& 4.19E-19& 2.22E-19& 2.22E-19\\
 $^{9}$Be & 0.00E+00& 0.00E+00& 0.00E+00& 0.00E+00\\
 $^{10}$B & 9.84E-22& 9.84E-22& 7.15E-22& 7.15E-22\\
 $^{11}$B & 4.79E-14& 4.79E-14& 3.73E-14& 3.73E-14\\
\hline
\end{tabular}
\caption{Detailed yields ($\msun$) before decay of radioactive species: 
For models 40SH and 40SL. 
\label{tab6}}
\end{center}
\end{table}

\clearpage

\begin{table}
\begin{center}
\scriptsize

\begin{tabular}{ccccc}\hline\hline 
{\bf Species} & 
{\bf 40SHa} &
{\bf 40SHb} & 
{\bf 40SLa} &
{\bf 40SLb}\\
{\bf $M$($^{56}$Ni)} &
{5.4E-01} &
{9.0E-02} &
{4.4E-01} &
{1.1E-01}\\\hline

 $^{33}$Ar& 8.46E-17& 5.74E-25& 1.37E-22& 2.45E-24\\
 $^{34}$Ar& 2.49E-06& 4.58E-16& 1.43E-07& 7.39E-16\\
 $^{35}$Ar& 1.21E-05& 3.99E-09& 2.35E-06& 1.30E-09\\
 $^{36}$Ar& 2.63E-02& 2.58E-02& 2.88E-02& 2.86E-02\\
 $^{37}$Ar& 3.26E-05& 3.22E-05& 8.58E-05& 8.58E-05\\
 $^{38}$Ar& 4.04E-03& 4.03E-03& 8.21E-03& 8.20E-03\\
\hline
\end{tabular}
\end{center}
\end{table}

\clearpage

\begin{table}
\begin{center}
\scriptsize

\begin{tabular}{ccccc}\hline\hline 
{\bf Species} & 
{\bf 40SHa} &
{\bf 40SHb} & 
{\bf 40SLa} &
{\bf 40SLb}\\
{\bf $M$($^{56}$Ni)} &
{5.4E-01} &
{9.0E-02} &
{4.4E-01} &
{1.1E-01}\\\hline

 $^{51}$Mn& 1.12E-04& 4.74E-05& 9.66E-05& 3.73E-05\\
 $^{52}$Mn& 2.04E-06& 1.71E-06& 2.34E-06& 2.12E-06\\
 $^{53}$Mn& 3.07E-05& 3.01E-05& 4.82E-05& 4.82E-05\\
 $^{54}$Mn& 8.85E-08& 8.84E-08& 2.10E-07& 2.10E-07\\
 $^{55}$Mn& 2.19E-08& 2.19E-08& 6.95E-08& 6.95E-08\\
 $^{56}$Mn& 2.54E-13& 6.02E-14& 2.38E-13& 1.70E-13\\
\hline
\end{tabular}
\end{center}
\end{table}

\clearpage

\begin{table}
\begin{center}
\scriptsize

\begin{tabular}{ccccc}\hline\hline 
{\bf Species} & 
{\bf 25A} &
{\bf 25B} & 
{\bf 25Sa} &
{\bf 25Sb}\\
{\bf $M$($^{56}$Ni)} &
{7.8E-02} &
{1.5E-01} &
{1.6E-01} &
{9.1E-02}\\\hline
 $^{4}$He & 1.82E+00& 1.80E+00& 1.82E+00& 1.79E+00\\
 $^{6}$Li & 3.23E-14& 2.88E-20& 2.12E-20& 2.12E-20\\
 $^{7}$Li & 3.13E-19& 7.26E-20& 9.73E-20& 9.73E-20\\
 $^{9}$Be & 9.86E-16& 0.00E+00& 0.00E+00& 0.00E+00\\
 $^{10}$B & 2.97E-15& 6.92E-22& 4.37E-22& 4.37E-22\\
 $^{11}$B & 1.23E-14& 4.92E-15& 5.11E-15& 5.11E-15\\
\hline
\end{tabular}
\caption{Detailed yields ($\msun$) before decay of radioactive species: 
For models 25A, 25B and 25S.
\label{tab7}}
\end{center}
\end{table}

\clearpage

\begin{table}
\begin{center}
\scriptsize

\begin{tabular}{ccccc}\hline\hline 
{\bf Species} & 
{\bf 25A} &
{\bf 25B} & 
{\bf 25Sa} &
{\bf 25Sb}\\
{\bf $M$($^{56}$Ni)} &
{7.8E-02} &
{1.5E-01} &
{1.6E-01} &
{9.1E-02}\\\hline

 $^{33}$Ar& 6.61E-11& 2.24E-15& 2.95E-17& 9.83E-18\\
 $^{34}$Ar& 4.90E-06& 7.80E-07& 1.59E-06& 2.74E-07\\
 $^{35}$Ar& 3.85E-06& 1.18E-06& 3.00E-06& 4.45E-07\\
 $^{36}$Ar& 3.48E-03& 4.31E-03& 4.80E-03& 4.76E-03\\
 $^{37}$Ar& 9.18E-06& 6.52E-06& 9.35E-06& 9.34E-06\\
 $^{38}$Ar& 4.61E-04& 3.15E-04& 4.83E-04& 4.83E-04\\
\hline
\end{tabular}
\end{center}
\end{table}

\begin{table}
\begin{center}
\scriptsize

\begin{tabular}{ccccc}\hline\hline 
{\bf Species} & 
{\bf 25A} &
{\bf 25B} & 
{\bf 25Sa} &
{\bf 25Sb}\\
{\bf $M$($^{56}$Ni)} &
{7.8E-02} &
{1.5E-01} &
{1.6E-01} &
{9.1E-02}\\\hline

 $^{51}$Mn& 1.84E-05& 2.06E-05& 4.96E-05& 1.47E-05\\
 $^{52}$Mn& 2.10E-07& 2.29E-07& 2.72E-07& 2.50E-07\\
 $^{53}$Mn& 3.25E-06& 2.08E-06& 3.38E-06& 3.38E-06\\
 $^{54}$Mn& 1.46E-08& 6.83E-09& 1.13E-08& 1.13E-08\\
 $^{55}$Mn& 4.18E-09& 1.85E-09& 3.06E-09& 3.06E-09\\
 $^{56}$Mn& 4.69E-14& 6.82E-15& 4.16E-14& 4.06E-14\\
\hline
\end{tabular}
\end{center}
\end{table}

\clearpage

\begin{table}
\begin{center}
\scriptsize

\begin{tabular}{ccccc}\hline\hline 
{\bf Species} &
{\bf Models} &
{} &
{} &
{}\\\hline
& {\bf 40A} & {\bf 40B}   & {\bf 40C}  & {\bf 40D}\\
$^{26}$Al & 3.39E-05 & 3.10E-05 & 5.45E-05 & 1.40E-05\\
$^{41}$Ca & 5.30E-06 & 4.05E-06 & 1.50E-05 & 9.05E-07\\
$^{44}$Ti & 1.64E-04 & 1.50E-05 & 4.26E-04 & 7.52E-08\\
$^{56}$Ni & 1.07E-01 & 8.11E-02 & 2.40E-01 & 6.28E-08\\
$^{57}$Ni & 3.25E-03 & 1.45E-03 & 7.68E-03 & 6.90E-08\\\hline
& {\bf 40SHa}    & {\bf 40SHb}    & {\bf 40SLa}    & {\bf 40SLb}\\
$^{26}$Al & 5.70E-05 & 5.52E-05 & 5.36E-05 & 5.35E-05\\
$^{41}$Ca & 8.65E-06 & 7.40E-06 & 2.27E-05 & 2.24E-05\\
$^{44}$Ti & 1.18E-03 & 6.67E-06 & 6.52E-04 & 8.72E-06\\
$^{56}$Ni & 5.44E-01 & 9.01E-02 & 4.38E-01 & 1.09E-01\\
$^{57}$Ni & 1.81E-02 & 1.39E-03 & 1.18E-02 & 2.16E-03\\\hline
& {\bf 25A}      & {\bf 25B}      & {\bf 25Sa}     & {\bf 25Sb}\\
$^{26}$Al & 1.10E-02 & 1.23E-02 & 1.22E-02 & 1.22E-02\\
$^{41}$Ca & 1.95E-06 & 1.41E-06 & 2.06E-06 & 2.06E-06\\
$^{44}$Ti & 1.31E-04 & 3.66E-05 & 2.14E-04 & 5.73E-05\\
$^{56}$Ni & 7.81E-02 & 1.51E-01 & 1.64E-01 & 9.07E-02\\
$^{57}$Ni & 2.12E-03 & 2.65E-03 & 4.18E-03 & 2.55E-03\\\hline
\end{tabular}
\caption{Major Radioact isotopes ($\msun$)
\label{tab8}}
\end{center}
\end{table}

\clearpage

\begin{table}
\begin{center}
\scriptsize

\begin{tabular}{cccccccc}\hline\hline 
{\bf Model} & 
{\bf $M$($^{56}$Ni)} &
{\bf [Fe/H]} & 
{\bf [O/Fe]} &
{\bf [Mg/Fe]} & 
{\bf [Si/Fe]} &
{\bf [S/Fe]} & 
{\bf [Sc/Fe]}\\
{} & 
{} &
{\bf [Ti/Fe]} & 
{\bf [Cr/Fe]} &
{\bf [Mn/Fe]} & 
{\bf [Co/Fe]} &
{\bf [Zn/Fe]} & 
{}\\\hline
40A & 1.1E-01 & -3.8 & 8.32E-01 & 7.59E-01 & 5.84E-01 & 1.26E-01 &-3.04E-01\\ 
&         & 8.24E-02 & 2.08E-02 &-4.32E-01 & 2.59E-02 & 1.56E-01 & \\
40B & 8.1E-01 & -3.1 & 9.55E-01 & 8.86E-01 & 5.95E-01 & 5.83E-02 &-1.47E+00\\ 
&         &-5.09E-01 &-2.91E-01 &-5.08E-01 &-1.58E-01 &-1.73E-01 & \\
40C & 2.4E-01 & -4.0 & 6.25E-01 & 5.52E-01 & 4.26E-01 & 1.31E-01 & 1.16E-01\\ 
&         & 1.06E-01 &-6.90E-02 &-4.77E-01 & 1.98E-01 & 3.20E-01 & \\
\hline
\end{tabular}
\caption{[X/Fe]. [Fe/H] is estimated with equation 
 (6) and (7) in the text. 
\label{tab9}}
\end{center}
\end{table}

%% Text for table notes should follow after the \enddata but before 
%% the \end{deluxetable}. Make sure there is at least one \tablenotemark 
%% in the table for each \tablenotetext.

%\tablenotetext{a}{Sample footnote for table~\ref{tbl-1} that was generated 
%with the deluxetable environment} 
%\tablenotetext{b}{Another sample footnote for table~\ref{tbl-1}}

%\tablecomments{Occasionally, authors wish to append a short 
%paragraph of explanatory notes that pertain to the entire table, but 
%which are different than the caption.  Such notes should be placed in 

%a {\tt tablecomments} command like this.}

%\end{deluxetable}

%% If you use the table environment, please indicate horizontal rules using 
%% Do not put multiple tabular environments within a single table. 
%% The optional \label should appear inside the \caption command.

\end{document}